\documentclass[fleqn,usenatbib]{mnras}

\usepackage{newtxtext,newtxmath}
\usepackage{tabularx}
\usepackage{siunitx}
\usepackage[T1]{fontenc}
\usepackage{tikz}\usepackage{tikz}
\DeclareRobustCommand{\VAN}[3]{#2}
\let\VANthebibliography\thebibliography
\def\thebibliography{\DeclareRobustCommand{\VAN}[3]{##3}\VANthebibliography}

\usepackage{graphicx}	
\usepackage{amsmath}	
\usepackage{xcolor} 
\usepackage{tikz}
\usepackage{float} 
\usepackage{booktabs}
\usepackage{lipsum}
\usepackage{makecell}
\usepackage{placeins}
\usetikzlibrary{positioning} 
\usepackage{subcaption}
\definecolor{box1color}{HTML}{EEF1FF} 
\definecolor{box2color}{HTML}{EEF1FF} 
\definecolor{box3color}{HTML}{EEF1FF} 
\definecolor{box4color}{HTML}{EEF1FF} 

\definecolor{bordergreen}{HTML}{719FB0} 
\DeclareSIUnit\arcsec{arcsec}

\renewcommand{\arraystretch}{1.2} 

\title[Non-uniform metal distribution in M86]{A non-uniform metal distribution in the ram-pressure stripped circumgalactic medium of M86}

\author[S. Kara et al.]{
Sinancan Kara,$^{1}$\thanks{E-mail: kara.sinancan@gmail.com}
François Mernier,$^{2}$
Norbert Werner$^{3}$
and E. Nihal Ercan$^{1}$
\\
$^{1}$Department of Physics, Bo\u{g}aziçi University, Bebek, 34342 Istanbul, Turkey\\
$^{2}$IRAP, CNRS, Universit\'{e} de Toulouse, CNES, UT3-UPS, Toulouse, France\\
$^{3}$Department of Theoretical Physics and Astrophysics, Faculty of Science, Kotlá\v{r}ská 2, Masaryk University, Brno, 611 37, Czech Republic
}

\date{Accepted XXX. Received YYY; in original form ZZZ}

\pubyear{2025}

\begin{document}
\label{firstpage}
\pagerange{\pageref{firstpage}--\pageref{lastpage}}
\maketitle

\begin{abstract}
The chemical enrichment of X-ray-emitting hot halos has primarily been studied in closed-box galaxy clusters. Investigating the metal content of lower-mass, open systems can serve as a valuable tracer for understanding their dynamical history and the extent of chemical enrichment mechanisms in the Universe. In this context, we use an 85.6 ks \textit{XMM-Newton} observation to study the spatial distribution of the abundance ratios of Mg, Si, and S with respect to Fe in the hot gas of the ram-pressure-stripped M86, which has undergone morphological transformations. We report that the chemical composition in the M86 galaxy core is more similar to the rest of the hot gaseous content of the Universe than to its stellar population. This result indicates that even supersonic ram-pressure is insufficient to strip the inner part of a galaxy of its hot atmosphere. Comparison with other galaxies undergoing ram-pressure stripping suggests that stripping the "primordial" atmosphere of a galaxy requires a combination of ram-pressure stripping and strong radio-mechanical AGN activity. The X-ray emission structures within M86, the plume and the tail, are found to be relatively isothermal. We observe that the Mg/Fe ratio in the plume is $3.3\sigma$ higher than in the M86 galaxy core and is consistent with that in the M86 group outskirts and the Virgo ICM, suggesting that the plume might originate from the low-entropy outer gas rather than from the recent ram-pressure stripping of the dense galaxy core.  

\end{abstract}

\begin{keywords}
galaxies: abundances -- galaxies: clusters: intracluster medium -- X-rays: galaxies --  galaxies: ISM -- supernovae: general \end{keywords}



\section{Introduction}
\label{sec:intro}

Most of the ordinary matter in the Universe resides in highly ionized, diffuse, hot, X-ray-emitting plasma that pervades massive galaxies, groups, clusters of galaxies, and cosmic filaments. The hot ionized gas in the inner parts of massive galaxies, known as the interstellar medium (ISM); the gas beyond their stellar components, called the circumgalactic medium (CGM); the gas permeating galaxy groups, referred to as the intergroup medium (IGrM); and the gas within galaxy clusters, known as the intracluster medium (ICM), are all found to be remarkably similar in chemical composition with uniform contributions of core-collapse supernovae (SNcc) and Type Ia supernovae (SNIa) products \citep[for a recent review, see e.g.,][]{Mernier2022}.


Since the first discovery of K-shell Fe emission lines \citep{Mitchell1976,Serlemitsos1977}, the chemical origin of the hot halos has been debated.   Metal abundances are discovered to be independent of the cluster mass \citep{dePlaa2017, Mernier2018m, Truong2019}, indicating that cluster member galaxies are not the primary factor in the metal enrichment. Moreover, the metal abundance in the ICM is spatially uniform \citep{Werner2013,Urban2017}; and the chemical composition of the ICM is constant radially, even beyond its virial radius with uniform contributions of SNIa and SNcc products \citep{Simionescu_2015,Mernier_2017}. Last but not least, the abundance ratios with respect to Fe in the central regions of clusters, groups and massive ellipticals, are very similar to our Solar system (i.e. X/Fe ratio is $\sim$1 Solar) \citep{Mernier2018b, Simionescu2019b}. In line with these observational discoveries, simulations (see, e.g., \citet{Biffi2017}) suggest that the intergalactic medium was enriched with metals before clusters assembled at $z \sim 2-3$ ($10-12$ billion years ago), at the peak era of star formation and active galactic nuclei (AGN) activity \citep{MadauDickinson2014, Hickox2018}. Then, the enriched hot gas has been accreted externally and heated by galaxies, groups and clusters of galaxies to form the present "hot atmospheres" with Solar abundance ratios \citep[for a recent review, see e.g.,][]{Werner_2020}.

Given that the chemical composition of hot gaseous content of the Universe is uniform, any deviations from homogeneity is a good tracer of dynamical histories of systems. Recently, \citet{Kara_2024} showed that M89, an AGN-hosting, ram-pressure stripped galaxy in the Virgo Cluster, has super-Solar ratios of $\alpha$ elements with respect to Fe, resembling its stellar component rather than the "Universal" hot gas content, suggesting a ram-pressure induced accretion cut-off from the surroundings and AGN activity might lead to loss and replenishment of the hot gas in an elliptical galaxy. To investigate further the extent of such replenishment, the massive galaxy M86 is an excellent candidate. Investigating the abundance ratios and chemical compositions of M86 is particularly interesting and important, as it also provides us the opportunity to infer the dynamical history of such a disturbed system.

M86 is a blue-shifted giant elliptical galaxy falling into the Virgo Cluster from the far side with a relative velocity $\sim\!$1500 km $\text{s}^{-1}$ with respect to Virgo Cluster central galaxy M87 \citep{Finoguenov_2004}. Furthermore, the highly extended hot X-ray emitting halo of M86 has a shallower X-ray surface brightness profile and larger gas mass fraction, which resembles a group of galaxies (see, e.g., \cite{Bohringer_1994} and references therein). Additionally, \cite{Binggeli_1993} has shown the blue-shifted tail of M86 consists dwarf ellipticals, which implies M86 galaxy is in fact at the core of a larger structure, the M86 group of galaxies. Within the M86 group, there is an X-ray emitting plume at 3 arcmin north of the M86 galaxy and extending more than projected 10 arcmin to northwest, first reported by \cite{Forman_1979} with \textit{Einstein} observations. Furthermore, using \textit{ROSAT} PSPC and \textit{HRI} data, \cite{Rangarajan_1995} identified another X-ray structure, a north-eastern arm extending from the M86 galaxy core to at least 5 arcmin to north-east. \textit{Chandra} analysis by \cite{Randall_2008} further revealed a long tail of emission, extending from the plume to northwest with a true length of 380 kpc. In that study they also identified two temperature structures, a cooler component of $\sim\!$0.8 keV that can be associated with the M86 galaxy halo, and a hotter component with $\sim\!$1.2 keV associated with the M86 group halo. The temperature maps of M86 show smooth gradients with no discontinuities at the apparent boundaries \citep{Finoguenov_2004,Randall_2008}. Therefore, these structures are not foreground or background objects, but are truly bound to the same system.

M86 is a very weak radio source \citep{Dressel_Condon_1978,Hummel_1980,Fabbiano_1989}, therefore the AGN activity cannot be the sole mechanism responsible for these emission irregularities within M86. The plume and other emission structures are therefore considered to be originate from ram-pressure stripping due to the interaction of the Virgo ICM with the hot gas of the M86 group \citep[see e.g.,][]{Forman_1979, Fabian_1980,Takeda_1984, White_1991, Rangarajan_1995, Randall_2008}. M86 has a close distance of $0.4 \pm 0.8$ Mpc to M87 \citep{Mei_2007}, and it travels with nearly twice of the sound speed \citep{Randall_2008}. Therefore it experiences supersonic ram-pressure stripping and the ram-pressure stripping is capable of displacing a significant fraction of ISM in a single blob \citep{Takeda_1984}. \cite{Randall_2008} shows that the stripping is capable of displacing such blob, given that the gravitational well of M86 is asymmetric. However, using \textit{XMM-Newton}, \cite{Finoguenov_2004} identified a possible shock at north-western of M86 with Mach number of $\sim\!$1.4 which is not associated with emission structures like plume and the north-eastern arm, which they found to have low entropy. They argue that, instead of ram-pressure stripping, smaller scale disruptions, such as galaxy-galaxy interactions, are the main cause of the X-ray emission structures. The existence of the unperturbed gas halo --other than the plume and other structures-- extending to almost 1 degree away from M86 \citep{Bohringer_1994}, which is not present in other ram-pressure stripped galaxies gives weight to that idea. Furthermore, galaxy-galaxy collision scenario, has crucial evidences.  \cite{Kenney_2008} showed a spectacular H$\alpha$ complex connecting M86 with the neighbor disturbed spiral NGC\,4438, which is 120 kpc projected away. The filaments show a smooth velocity gradient between M86 and NGC\,4438, strongly supporting a collision scenario. Moreover, there are dust components spatially coincident with the H$\alpha$ emissions \citep{Gomez_2010}.  Furthermore, \citep{Ehlert_2013} showed that there is a $\sim$0.6 keV X-ray emitting gas bridge between M86 and NGC\,4438, spatially coinciding with the H$\alpha$ filaments. 


Therefore, although M86 does experience strong ram-pressure stripping with the Virgo ICM, the X-ray structures with low entropy might originate from the collision with NGC\,4438 in the first place, and dragged by the gravitational well of M86 while being heated and stripped through their voyage in Virgo Cluster. These features make M86 an excellent candidate to investigate (i) if the accretion cut-off due to ram-pressure leads to stellar-like hot halo in a galaxy group, and (ii) the origins of the X-ray emission structures in M86 via their chemical compositions. Moreover, studies aiming to constrain supernova (SN) models based on chemical abundance patterns in hot gas have mostly been conducted using samples of galaxy clusters \citep[e.g.,][]{dePlaa_2007,Mernier_SNe}. A detailed investigation of the abundance pattern in M86 provides a unique opportunity to robustly compare and constrain different SN models based on the halo of an elliptical galaxy. In this study, we measure the elemental abundances and abundance ratios with respect to Fe in the hot gas of M86 with \textit{XMM-Newton}/EPIC and RGS data. With a particular attention to carefully modelling the spectra, we investigate the distribution of metals in M86 group, the galaxy core and other aforementioned emission structures. 

The structure of this paper is as follows. Sect. \ref{sec:observation} describes the data reduction and spectral analysis. We present the abundance and temperature measurement results in Sect. \ref{sec:results} and discuss the chemical enrichment history of M86 group in Sect. \ref{sec:Discussion}. The conclusion of this study is presented in Sect. \ref{sec:conclusion}. Throughout the paper we assume the standard $\Lambda$CDM cosmology with $H_0 = 70\;$km$\,$s$^{-1}\,$Mpc$^{-1}$, $\Lambda_0 = 0.73$ and $q_0 = 0$. All the abundances are expressed using the proto-Solar values of \citet{Lodders2009}, and for simplicity, the values are referred to as `Solar' throughout the paper. Unless stated otherwise, all uncertainties are expressed in the $1\sigma$ confidence. The parameter uncertainties were estimated from posterior distributions obtained from Markov chain Monte Carlo (MCMC) simulations. We use optimal binning method of \citet{Kaastra_2016}, and following \citet{Kaastra_2017} all fits are performed with C-statistics \citep{Cash_1979} unless stated otherwise.

\section{Observation and data analysis}
\label{sec:observation}
M86 was observed with a single \textit{XMM-Newton} observation on 1 July 2002, with ObsID 0108260201. The total exposure time is 85.6 ks and the net exposure time for each instrument is presented in Table \ref{table:obs}. For data reduction, we use the \textit{XMM-Newton} Science Analysis System (SAS) version 21.0.0, along with the integrated Extended Source Analysis Software (\texttt{ESAS} v9.0\footnote{\href{https://www.cosmos.esa.int/web/xmm-newton/xmm-esas}{https://www.cosmos.esa.int/web/xmm-newton/xmm-esas}}) package. For point source detection, we use \textit{Chandra} \texttt{CIAO} 4.15.

\begin{table}
\centering
\renewcommand{\arraystretch}{1.0} 
\caption{\textit{XMM-Newton}/EPIC (MOS1, MOS2 and pn) and RGS net exposure times of the M86, with observation date 2002-07-01 and ObsID 0108260201.}
\begin{tabularx}{\columnwidth}{X
  c
  c}
\hline
\hline
 {Instrument}& {Raw exposure (ks)} & {Clean exposure (ks)} \\ \hline
{MOS1} & 85.4 & 62.0 \\ 
{MOS2}  & 85.4 & 63.8 \\ 
{pn}  & 82.0 & 38.6 \\ 
{RGS1}  & 85.6 & 61.2  \\
{RGS2}  & 85.6 & 61.0 \\ \hline
\end{tabularx}
\label{table:obs}
\end{table}

\begin{figure*}
    \centering
    \makebox[\textwidth][c]{
        \includegraphics[height=6.9cm]{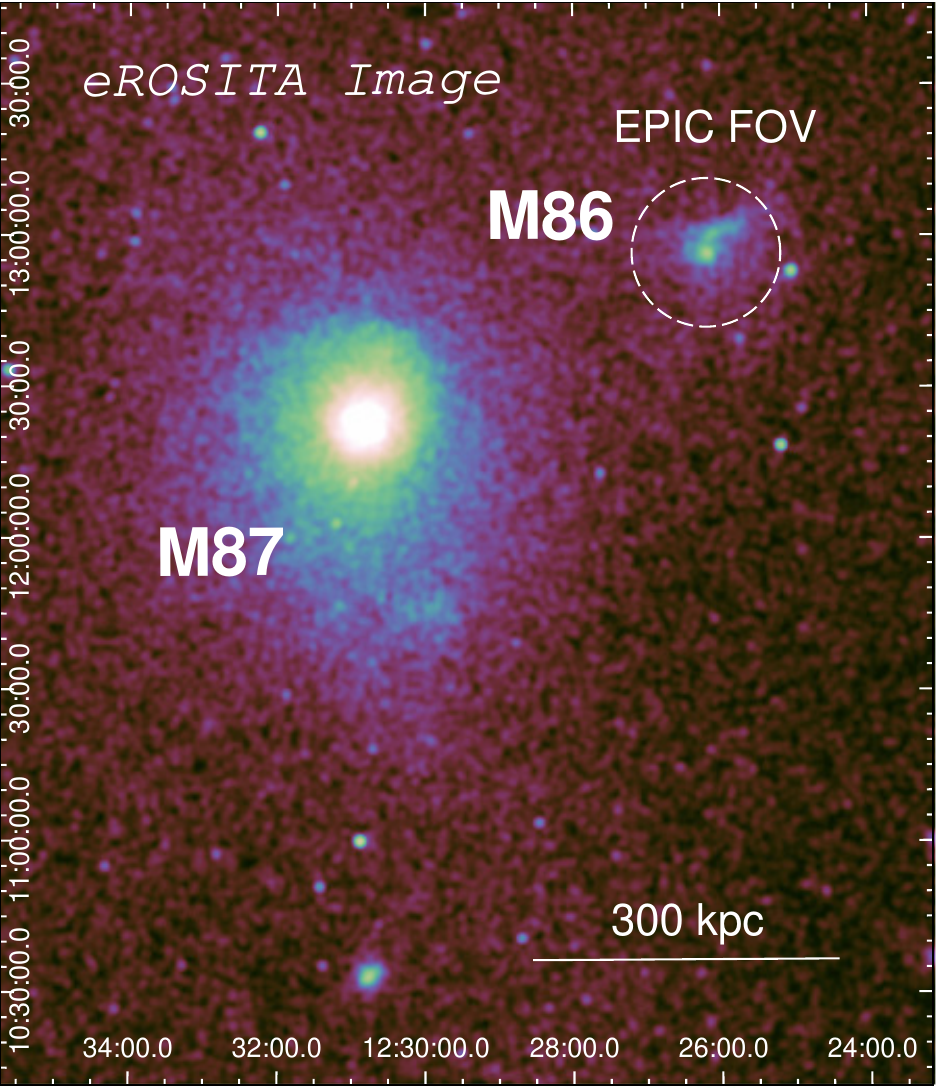}%
        \hspace{2.5mm} 
        \includegraphics[height=6.9cm]{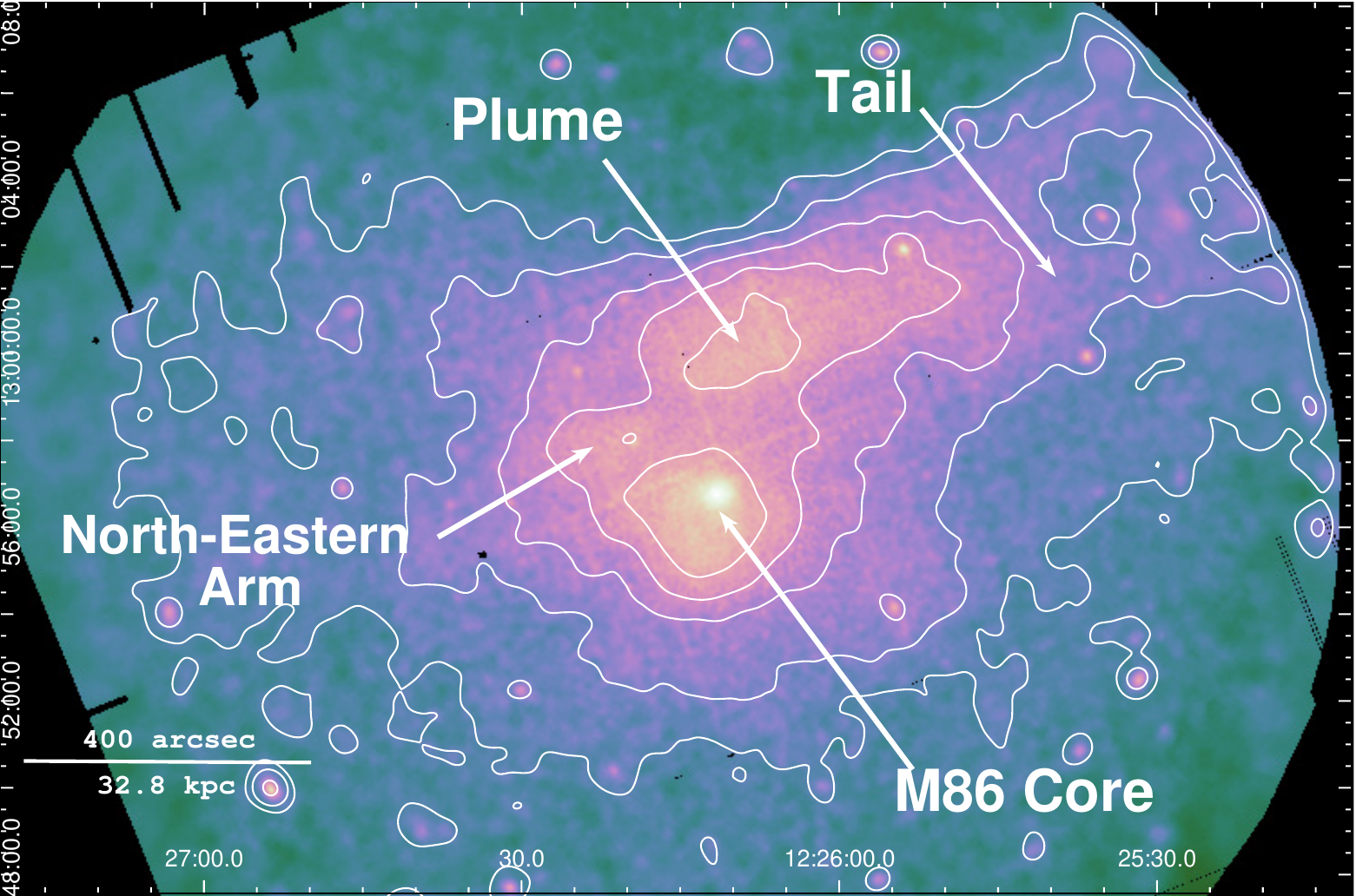}
    }
    \\[2.5mm]
    \makebox[\textwidth][c]{
        \includegraphics[height=7cm]{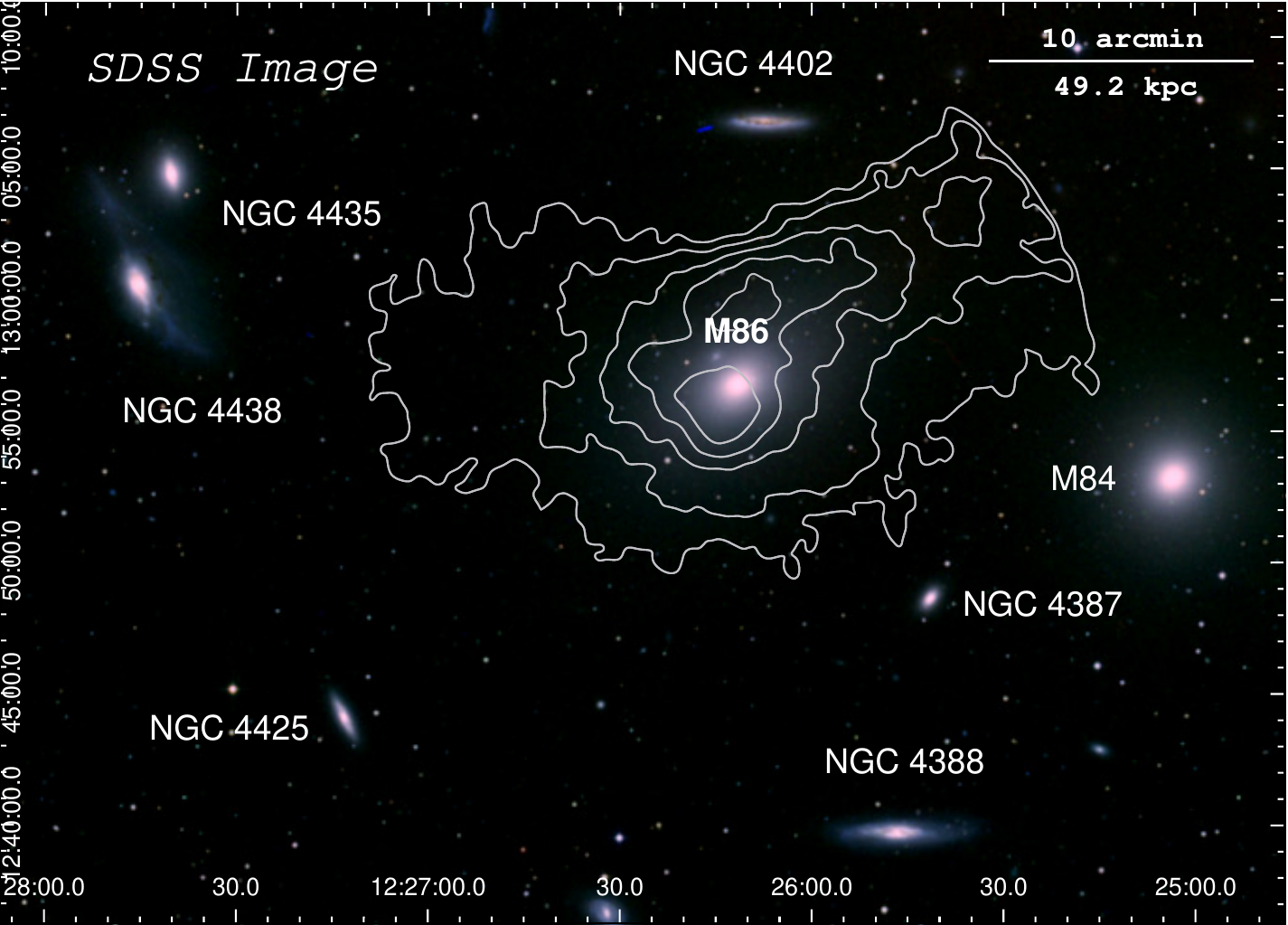}%
        \hspace{2.5mm} 
        \includegraphics[height=7cm]{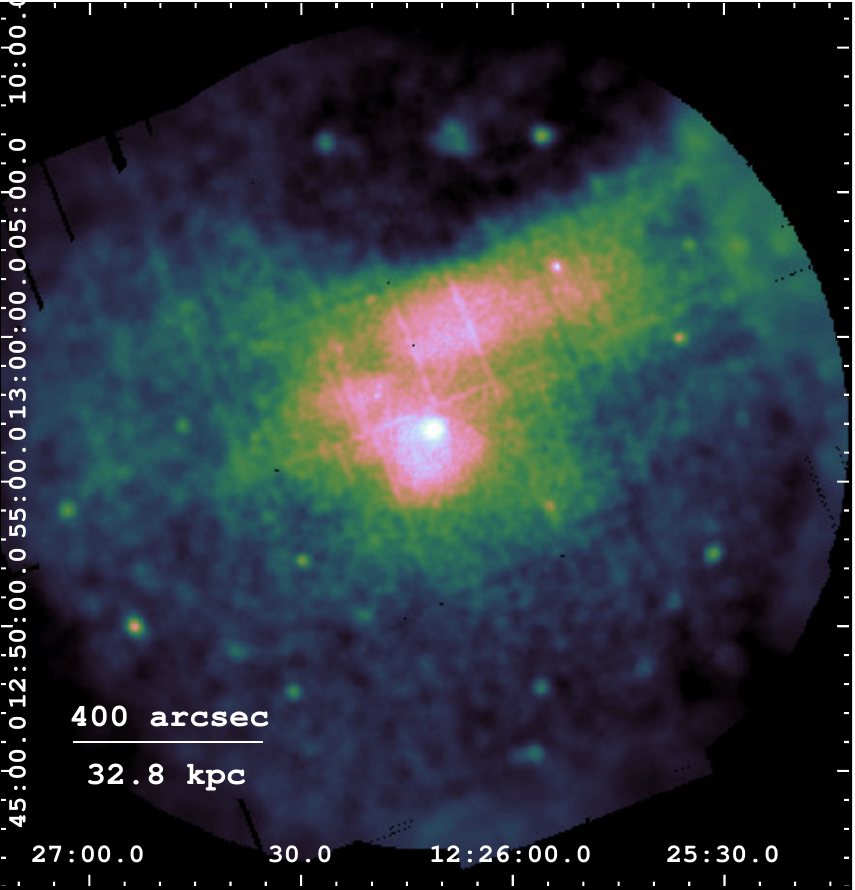}
    }
    \caption{\textit{Top Left:} Soft band eROSITA image of Virgo cluster. The dashed white line shows the \textit{XMM-Newton}/EPIC FOV. \textit{Top Right:} Background-subtracted and exposure corrected \textit{XMM-Newton}/EPIC count image of M86 in 0.3-2.0 keV band with the corresponding surface brightness contours. Core, Plume and the Tail regions are indicated. \textit{Bottom Left}: RGB image using SDSS filters: red channel from i-band, green channel from r-band, and blue channel from g-band. \textit{XMM-Newton}/EPIC surface brightness contours are overlaid in white. \textit{Bottom Right:} Exposure corrected \textit{XMM-Newton}/EPIC FOV count image of M86 in 0.3-2.0 keV band without surface brightness contours.
  }
    \label{fig:phoinddex}
\end{figure*}

\subsection{EPIC}

The calibrated photon event files for MOS and pn data are created using \texttt{emchain} and \texttt{epchain}, respectively. Further, we filter the event files to exclude soft-proton (SP) flares by building Good Time Interval (GTI) files using the routine \texttt{espfilt}, where we create light curves for the full-band spectrum in 100s bins. After calculating the best-fitting mean count rate for the Gaussian distribution, $\mu$, and the standard deviation, $\sigma$, we reject time bins outside the interval $\mu \pm 2\sigma$. Point source detection and exclusion are performed using the CIAO algorithm \texttt{wavdetect}. We process the \textit{XMM–Newton}/EPIC data using \texttt{mosspectra} and \texttt{pnspectra}, which call the \texttt{evselect}, \texttt{rmfgen}, and \texttt{arfgen} tasks to extract spectra and generate RMFs and ARFs, respectively. We keep single, double, triple, and quadruple events in the MOS data (pattern$\leq$12) and single events (pattern=0) for pn.

\subsubsection{Background modelling approach}
\label{subsection:background}

We carefully model the background components to obtain more reliable measurements compared to subtracting a local background spectrum \citep[see relevant discussions in, e.g.,][]{Mernier_2015, dePlaa__2017, Zhang_2018}. The background components can be divided into two main categories. The first category is the instrumental background, which includes the Hard Particle (HP) and residual Soft Proton (SP) backgrounds. These instrumental components are not folded by the ARF of the instrument. The second category is the Astrophysical X-ray Background (AXB), which represents the total astrophysical X-ray emission. It consists of local (i.e., unabsorbed) components, including the Local Hot Bubble (LHB) and Solar Wind Charge Exchange (SWCX), as well as absorbed AXB components, which are more distant and include the Galactic Thermal Emission (GTE) and Unresolved Point Sources (UPS). In the case of M86, the absorbed AXB also includes the X-ray emission from the Virgo Cluster.

The X-ray emission from M86 covers the entire EPIC Field of View (FOV); therefore, it is not possible to measure the background model parameters from a local background region. Instead, we fit the entire EPIC FOV spectra simultaneously with the ROSAT All-Sky Survey (RASS) spectra to estimate the background parameters, except for the HP components, for which we use Filter Wheel Closed (FWC) data. In the FOV fit, which serves to estimate the background components, both the background components and M86 emission are modelled. The M86 emission in the FOV fit represents the integrated sum of all M86 galaxy and group halo emission.

We model the hot gas of M86 with a two-temperature thermal emission model (\texttt{vapec + vapec}), where the abundance ratios of N, O, and Ne with respect to Fe are fixed to the values obtained from the RGS results (See Sect. \ref{subsec:core_res}). The temperatures and the abundance values of Mg, Si, S, and Fe are left free. In this EPIC FOV + RASS fit for determining the background parameters, we find that the dominant cooler central gas of M86 is at $0.93\pm0.01$ keV, while the extended halo is at $1.37_{-0.02}^{+0.03}$ keV, both in line with previous measurements \citep{Randall_2008, Mernier2016a}. The methods for estimating all background components and the use of the RASS data are discussed in detail in the following subsections.

\subsubsection{Hard Particle Background}

High-energy particles hit the EPIC detectors from all directions and are recorded as photon events. These cosmic rays, which reach the CCD chips even when the filter wheel is closed, create a continuum and also generate instrumental X-ray fluorescence lines. We estimate the HP background components using FWC data taken from the date closest to the M86 observation. For MOS1 and MOS2, we use data from 18 June 2002, with an observation time of 24.4 ks. For pn, we take FWC data from 5 October 2002, with an exposure time of 11.1 ks.

We model the HP continuum with a broken power-law model (\texttt{bknpower}). Moreover, incorrect subtraction of fluorescence lines can lead to inaccurate abundance measurements for elemental lines at similar energies, as well as inaccurate temperature measurements. Therefore, instead of subtracting them, we model the instrumental lines with Gaussians (\texttt{gaussian}), using instrumental line energy values from \cite{Mernier_2015}. At this stage, we determine the continuum parameters (i.e., the broken power-law model), which will later be fixed to the determined values in the spectral fitting of M86. The instrumental fluorescence emissions, on the other hand, vary more across the detector and over time; therefore, the Gaussian model normalizations of the lines are left free during the fitting stage. The results for HP background parameters are listed in Table \ref{table:hp}.

\begin{table}
\centering
\renewcommand{\arraystretch}{1.2}
\caption{Best fit parameters of the HP continuum component, estimated from the full FOV of FWC observations. The equal sign (=) means that MOS2 parameters are linked with the MOS1 parameters.}
\label{table:hp}
\begin{tabular*}{\columnwidth}{@{\extracolsep{\fill}} l 
    S[table-format=2.2(2)] 
    c 
    S[table-format=2.2(2)]}
\toprule
\toprule
 & \multicolumn{1}{c}{HP MOS1} & \multicolumn{1}{c}{HP MOS2} & \multicolumn{1}{c}{HP pn} \\ 
\midrule
Norm (\(10^{-2}~\mathrm{cm}^{-5}\)) & 3.30 \pm 0.20 &$ 1.60 \pm 0.20$ & 19.95 \pm 0.40 \\ 
$\Gamma$ & 1.38 \pm 0.02 & = & 0.95 \pm 0.21 \\ 
$\Delta\Gamma$ & -1.25 \pm 0.02 & = & -0.34 \pm 0.13 \\ 
$E_{\mathrm{break}}$ (keV) & 2.14 \pm 0.01 & = & 1.15 \pm 0.33 \\ 
\bottomrule
\end{tabular*}
\end{table}

\subsubsection{Residual Soft-Proton Background}

Even after building GTI files and clipping the event list, a residual SP background continuum is expected in the filtered spectra. For the SP background and other background components discussed in the following sections, we fit the full FOV spectra of M86 using the fixed HP background components derived from the FWC data. This approach has been used in recent enrichment studies \citep[e.g.,][]{Urdampilleta_2019}.

The SP background can be modeled with a power-law component (\texttt{powerlaw}). We allow the power-law indices to vary freely for each instrument, within the range 0.1–1.4, as reported in \citet{Snowden_Kuntz_2013}. The best-fit parameters for the SP background are presented in Table \ref{table:sp}.

\begin{table}
\caption{Best fit parameters of the SP background, measured from the full FOV M86 spectra.}
\begin{minipage}{\columnwidth} 
\centering
\renewcommand{\arraystretch}{1.2} 
\begin{tabularx}{\columnwidth}{X
  S[table-format=1.2(2), separate-uncertainty]
  S[table-format=1.2(2), separate-uncertainty]
  S[table-format=1.2(2), separate-uncertainty]} 
\hline
\hline
 & {SP MOS1} & {SP MOS2} & {SP pn} \\ \hline
{Norm ($10^{-2} \mathrm{cm^{-5}}$)} & 5.95 \pm 0.10 & 7.76 \pm 0.30 & 4.15 \pm 0.18 \\ 
{$\Gamma$} & 0.28 \pm 0.02 & 0.11 \pm 0.01 & 0.13 \pm 0.03 \\  \hline
\end{tabularx}
\end{minipage}
\label{table:sp}
\end{table}

\subsubsection{Solar Wind Charge Exchange}
The first local (i.e., unabsorbed and not redshifted) soft X-ray foreground component is SWCX emission, which occurs within the heliosphere when highly ionized solar wind ions interact with neutral atoms in Earth's magnetosheath \citep[see, e.g.,][]{Cravens_2001,Snowden_2009}. Although the ionized portion of the interstellar medium (ISM) cannot penetrate the heliosphere, neutral ISM atoms can pass through the Solar System \citep{Cox_1998}. When solar wind ions interact with these neutral atoms—whether in the magnetospheric (also referred to as geocoronal, describing the same near-Earth region) or heliospheric environment—they capture an electron into an excited state, which then decays while emitting X-rays \citep[for a review, see, e.g.,][]{Kuntz_2019}.

We observe the 0.56 keV O {\footnotesize VII} triplet from $n = 2 \rightarrow 1$ at the soft X-ray background. The SWCX emission is modelled using Gaussian components with widths set to zero. We calculate the line fluxes in units of photons $\text{cm}^{-2} \, \text{s}^{-1} \, \text{sr}^{-1}$ (hereafter LU, for "line units"), which are presented in Table\,\ref{table:swcx}. We see that these values are in the range reported in other observations \citep{Koutroumpa_2011, Bulbul_2012}.

\begin{table}
\caption{Line Surface Brightness of O {\scriptsize VII} due to SWCX.}
\begin{minipage}{\columnwidth} 
\centering
\renewcommand{\arraystretch}{1.2} 
\begin{tabularx}{\columnwidth}{X
  S[table-format=1.2(2), separate-uncertainty]
  S[table-format=1.2(2), separate-uncertainty]
  S[table-format=1.2(2), separate-uncertainty]} 
\hline
\hline
 & {MOS1} & {MOS2} & {pn} \\ \hline
{Line} & \multicolumn{1}{l}{O {\scriptsize VII} (0.56 keV)} & \multicolumn{1}{c}{=} & \multicolumn{1}{c}{=} \\ 
{Line Surface Brightness (LU)} & 3.96 \pm 0.62 & \multicolumn{1}{c}{=} & 3.73 \pm 0.71 \\  \hline
\end{tabularx}
\end{minipage}
\label{table:swcx}
\end{table}




\subsubsection{Local Hot Bubble, Galactic Thermal Emission and Unresolved Point Sources}

\begin{table}
\caption{Best fit parameters for the astrophysical background components. The background components obtained from the \textit{XMM-Newton} M86 FOV spectrum and the RASS spectrum are indicated with the dagger symbol ($\dag$).}
\begin{minipage}{\columnwidth} 
\centering
\renewcommand{\arraystretch}{1.2} 
\begin{tabularx}{\columnwidth}{l
>{\centering\arraybackslash}p{0.3\columnwidth}
>{\centering\arraybackslash}X
>{\centering\arraybackslash}X}
\toprule
\toprule
         & Norm ($10^{-2}\mathrm{cm^{-5}}$) & kT (keV) & $\Gamma$ \\ \midrule
$\mathrm{LHB^{\dag} }$ & $0.030 \pm 0.002$                          & $0.11 \pm 0.01$   &                  \\
$\mathrm{GTE^{\dag} }$ & $0.037 \pm 0.002$                  & $0.56 \pm 0.03$   &                  \\
$\mathrm{UPS^{\dag} }$ & $0.057 \pm 0.004$                    &                   & 1.41 (fixed)     \\
\addlinespace 
\addlinespace 

Virgo ICM              & $0.39 \pm 0.05$                    & 2.3 (fixed)       &                  \\ \bottomrule
\end{tabularx}
\end{minipage}
\label{table:axb}
\end{table}

The Milky Way originated X-ray foregrounds include two primary components. The first is the Local Hot Bubble (LHB), a local (i.e., unabsorbed) shock-heated region between the heliosphere and the local interstellar medium \citep{Kuntz_2008}. The LHB is modeled using an unabsorbed single-temperature thermal emission model (\texttt{apec}) with Solar abundances, and its expected temperature is approximately $\sim0.1\, \text{keV}$. The second Galactic foreground component is the Galactic Thermal Emission (GTE), which originates from the X-ray thermal emission of the Milky Way halo. The GTE lies beyond most of the absorption caused by the Galactic ISM and is therefore represented using an absorbed thermal emission model (\texttt{phabs*apec}) with Solar abundances. The hydrogen column density for the GTE absorption is set to $N_{\text{H}} = 1.26\times10^{20} \text{cm}^{-2}$ \citep{Mernier_2015}. The temperature of the GTE component is expected to vary between $0.1 - 0.7\,\text{keV}$ across the sky \citep{Sanskriti_2019}. Although the mean GTE temperature is approximately $\text{kT} \sim0.2\, \text{keV}$ \citep{McCammon_2002}, recent studies have reported higher temperatures in the sky vicinity of M86 (e.g., 0.48 keV for the GTE in the Virgo Cluster; \citealp{Gatuzz_2023}). Therefore, we expect a higher GTE temperature. Another background component is the Unresolved Point Sources (UPS), which corresponds to the integrated X-ray emission from active galactic nuclei (AGNs), dominant in the hard band, and galaxies, which contribute most of the UPS counts in the soft band \citep{Lehmer_2012}. We model the UPS emission using a power-law model with a photon index of $\Gamma=1.41$ \citep{deLuca_2004}.

To estimate the background components mentioned above, we fit the FOV EPIC spectra combined with RASS spectra, which is more sensitive to the soft band. Because the M86 X-ray emission is highly extended and there are X-ray bright galaxies in the vicinity of M86, we were unable to extract a RASS spectra from an annulus surrounding M86. Therefore we moved our extraction pointing 1\textdegree\, to North and 1\textdegree\, to West from the M86 center and extracted a circular region with 1\textdegree\,radius in Galactic coordinates. Based on previous studies reporting a significant drop in the Virgo ICM surface brightness profile at distances similar to our RASS extraction region \citep{Simionescu_2017}, we assume the Virgo ICM contribution is negligible in the RASS spectra. To confirm that our RASS extraction region is not affected by any contamination, we carefully checked the diffuse X-ray maps from HEASARC X-Ray Background Tool\footnote{\href{https://heasarc.gsfc.nasa.gov/cgi-bin/Tools/xraybg/xraybg.pl}{https://heasarc.gsfc.nasa.gov/cgi-bin/Tools/xraybg/xraybg.pl}}. In the fitting, we couple all the LHB, GTE and UPS model parameters in EPIC with their counterpart in RASS, with a free scaling parameter. We find LHB temperature to be $\text{kT}_{\text{LHB}} = 0.11\pm0.01$ keV, and the GTE temperature to be $\text{kT}_{\text{GTE}} = 0.56\pm0.03$ keV, presented in Table\,\ref{table:axb}.


\subsubsection{Virgo Cluster Emission}
M86 is embedded in the ICM of Virgo Cluster. In line with \citet{Ehlert_2013} we model the Virgo ICM with a thermal emission with a fixed temperature, $\text{kT}_{\text{Virgo}}$ = 2.3 keV taken from the measurements of the Virgo ICM at the similar distances from M87 \citep{Urban_2011} and metal abundance $\text{Z}_{\text{Virgo}}$ = 0.25 Solar. We find that the \texttt{XSPEC} normalization of the Virgo ICM component is approximately $\sim$1 order of magnitude larger than the other astrophysical fore- and background components, presented in Table\,\ref{table:axb}.

\subsubsection{EPIC region selection and spectral modelling}

\begin{figure}
    \centering
    \includegraphics[width=1\columnwidth]{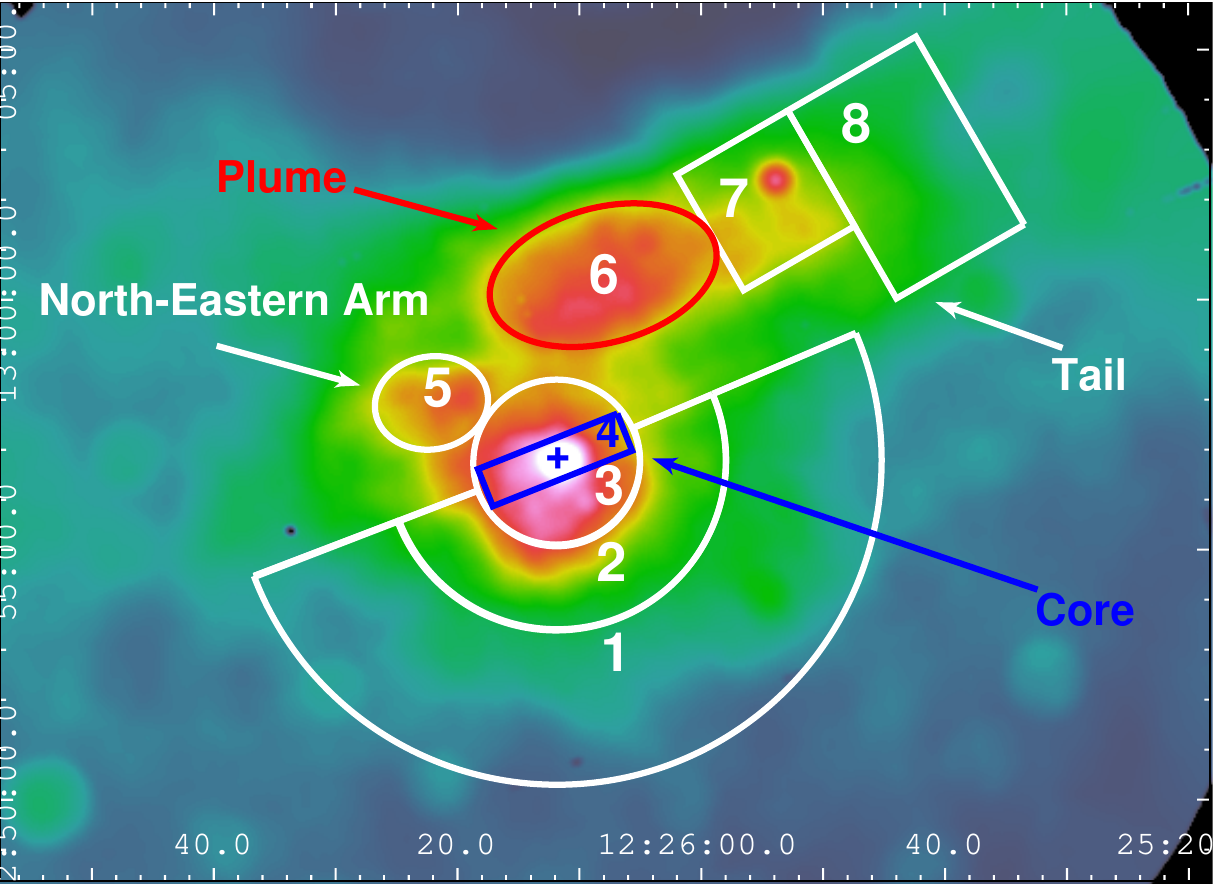}
    \caption{Exposure corrected EPIC image of M86 in 0.3-10 keV band. The extracted regions are presented.}
    \label{fig:regions}
\end{figure}

\begin{figure*}
    \centering
    \makebox[\textwidth][c]{
        \includegraphics[height=7.8cm]{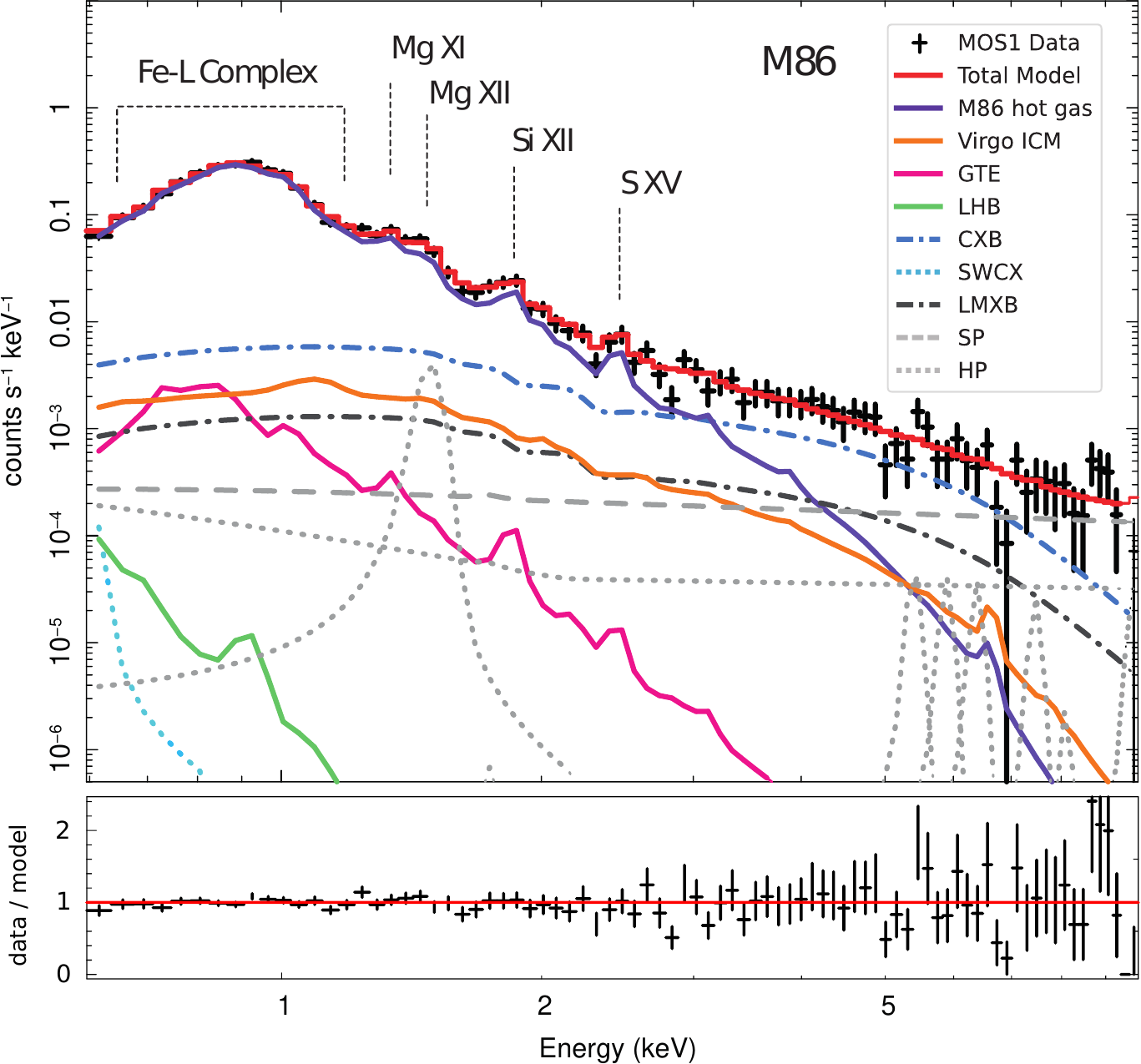}%
        \label{fig:mos-spectrum}
        \hspace{6.5mm} 
        \includegraphics[height=7.8cm]{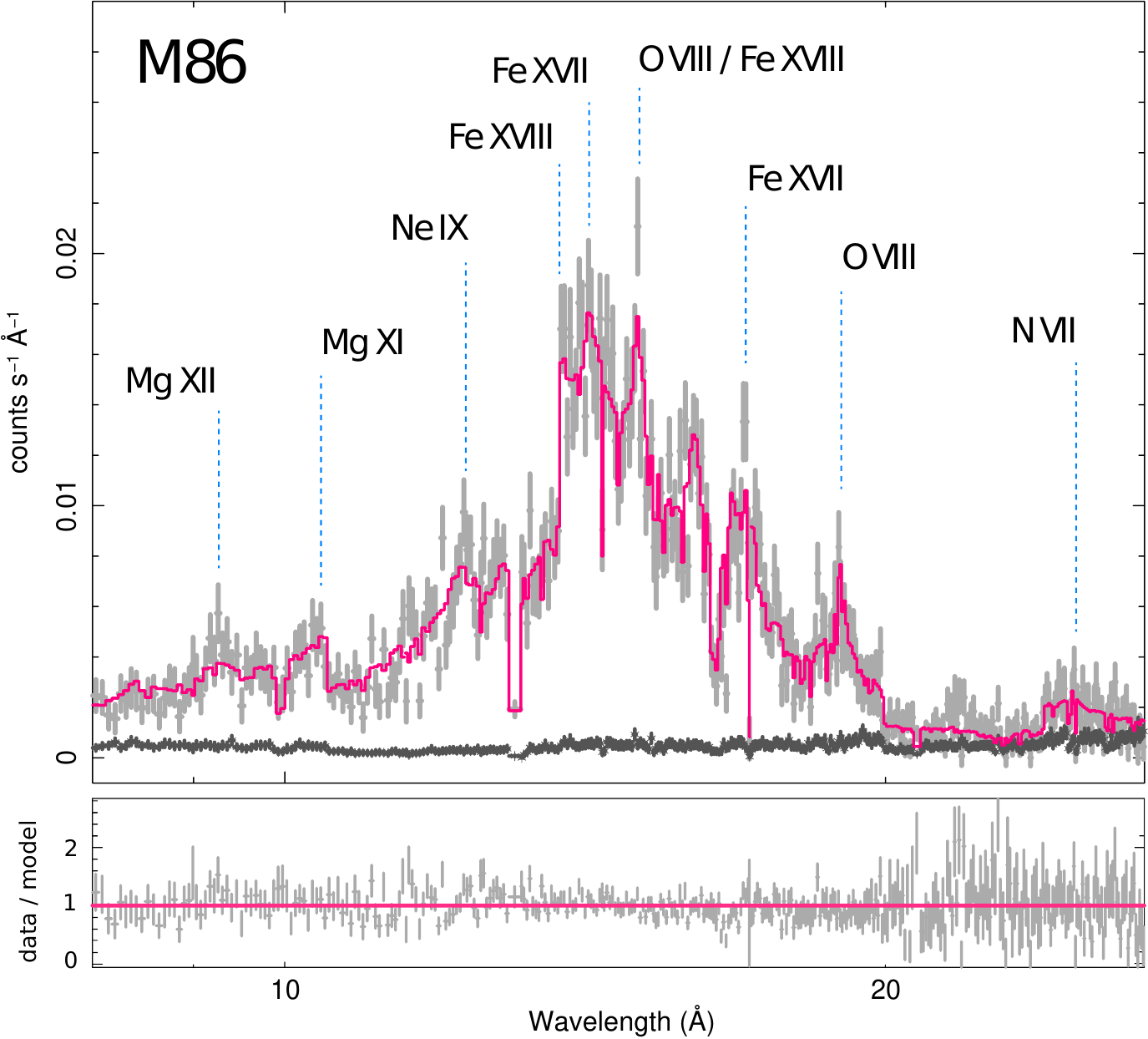}
        \label{fig:rgs-spectrum}
    }
    \caption{\textit{Left:} \textit{XMM-Newton}/EPIC fit of the M86. For visualization purposes we present the MOS1 fit only. The background components are also shown. See the text for details. \textit{Right:} RGS spectra in light gray, the template background in dark gray, and the total model in magenta. For visualization purposes, we present first-order combined spectra.}
  
    \label{fig:spectra}
\end{figure*}

To investigate the spatial distribution of metals, we extract spectra from eight individual regions, as shown in Figure \ref{fig:regions}. Specifically, we examine (i) the radial distribution of metals in the M86 group, (ii) the chemical composition of the M86 galaxy core, and (iii) the metal content in individual emission structures of M86.

For the (i) radial distribution, we first extract a spectrum from a 100-arcsec circular region surrounding but excluding the core, followed by spectra from 200- and 400-arcsec half-annuli. These regions are numbered 1, 2, and 3, from the outermost to the innermost.

To study the (ii) M86 galaxy core, we extract the spectrum from a 0.8-arcmin rectangular region that spatially overlaps with the RGS extraction, corresponding to region number 4.

Lastly, for (iii) the prominent structures, we extract spectra from elliptical or rectangular regions corresponding to the Eastern Structure\footnote{Throughout the text, we specifically refer to our extraction regions when the name of the emission structure begins with a capital letter.} (region number 5), the Plume (region number 6), and the Tail (regions number 7 and 8 -- Tail 1 and Tail 2).

In fitting each extracted spectrum, we model all background components with fixed parameters obtained from the FOV fit, including their normalizations. Afterward, we scale down the background normalizations for each region and corrected for vignetting effects using exposure maps for each instrument (MOS1, MOS2, pn) and region. 

For the region number 4 and 3, which corresponds to the galaxy core and the surrounding circular region; and in the North-Eastern Arm, which is region number 5, we observe an additional background component corresponding to the integrated X-ray emission from Low-Mass X-ray Binaries (LMXBs) in M86. To model this component, we use a redshifted power-law model with a fixed photon index $\Gamma = 1.56$ \citep{Irwin_2003, Su_2017}. Initially, we fix the normalization ratio of this component to that of the core hot gas to the value we observed in \cite{Kara_2024}. Later, we allow its normalization to vary. We note that LMXB component cannot be constrained in the FOV fit and other regions. In the regions with the prominent LMXB emission, we also allow the normalization of the other astrophysical power-law model component accounting for the UPS to vary. However, we observe that the change is not significant and causes very little changes in abundance ratios (less than 5\%), so we fix the parameter to the original scaled value. We note that the background parameters except LMXB are fixed for all fits across all regions, and the normalizations for each region are scaled down separately. 

The temperature structure of a collisional-ionization equilibrium (CIE) plasma can be estimated using either single-temperature or multi-temperature models. Abundance measurements in cool systems ($kT \lesssim 2-3$ keV) are known to be affected by the so-called `Fe bias' \citep{Buote1998, Buote2000, Gastaldello2021}. This is an underestimation of Fe abundance when multiple temperature components of a spectrum are estimated with a single-temperature model. Because multi-temperature components are inevitable in the spectrum of M86 hot gas due to issues such as projection effects and cooling in the core, we do not use a single-temperature (1T) fit in our modelling. As for multi-temperature models, it is possible to use a two-temperature (2T) model where the hot plasma is assumed to have one hotter and one cooler components. The more complex approach is the multi-temperature model with Gaussian distribution of emission. To model such a distribution, we use \texttt{vgadem} model in our fittings. Additionally, we fit the M86 core spectra using a 2T model for testing purposes. We find that the absolute abundances and abundance ratios are in good agreement within 1$\sigma$ uncertainties for 2T and \texttt{vgadem} models. 

In our EPIC modelling, we measured the Mg, Si, S, and Fe abundances, which were left free during the fitting. For consistency, we coupled the N, O, and Ne abundance parameters to Fe with a fixed ratio obtained from our RGS measurement of these elements with respect to Fe, presented in Sect. \ref{subsec:core_res}. We note that coupling these elements directly to Fe without using our RGS ratios resulted in virtually identical abundances for Mg, Si, and S (differences of $\sim$5\% in abundances and ratios). We present the spectral fit of the Core (4) region in Fig. \ref{fig:spectra}. Additionally, we also measure abundances from the MOS and pn spectra of the Core (4) region separately, with applying both broad-band and narrow-band fits, explained in detail in Section \ref{subsec:core_res}.

\subsection{RGS}
\label{subsection:RGS}

We use the SAS task \texttt{rgsproc} to process the RGS data. Since the RGS and MOS cameras are co-aligned on the satellite, the RGS 1 and RGS 2 data are filtered for flaring events using the same GTIs applied to MOS 1 and MOS 2, respectively. The RGS spectra are extracted using 90\% of the pulse height distribution.



\subsubsection{Selecting the RGS region size}
\label{subsub:rgs_size}
First, we extract the spectrum from (i) 90\% of the point spread function (PSF) using \texttt{rgsproc} (xpsfincl=90), which roughly corresponds to a 0.8 arcmin region. Additionally, we extract a broader region from (ii) 99\% of the PSF (xpsfincl=99), corresponding to a 3.4 arcmin region. Due to its larger size, the second spectrum is expected to exhibit more non-intrinsic line broadening and to include emissions from a larger area.
\subsubsection{Selecting the RGS background spectra}
\label{subsub:rgs_spectra}
We first create the (i) model background spectrum using the standard \texttt{rgsbkgmodel} routine. This background template file is based on the count rate in CCD 9 of the RGS, which is expected to contain no source counts. Additionally, we extract (ii) a background spectrum from outside 98\% of the PSF. This second background spectrum includes source counts from outside the galaxy core.

\begin{figure*}
    \centering
    \includegraphics[width=0.95\textwidth]{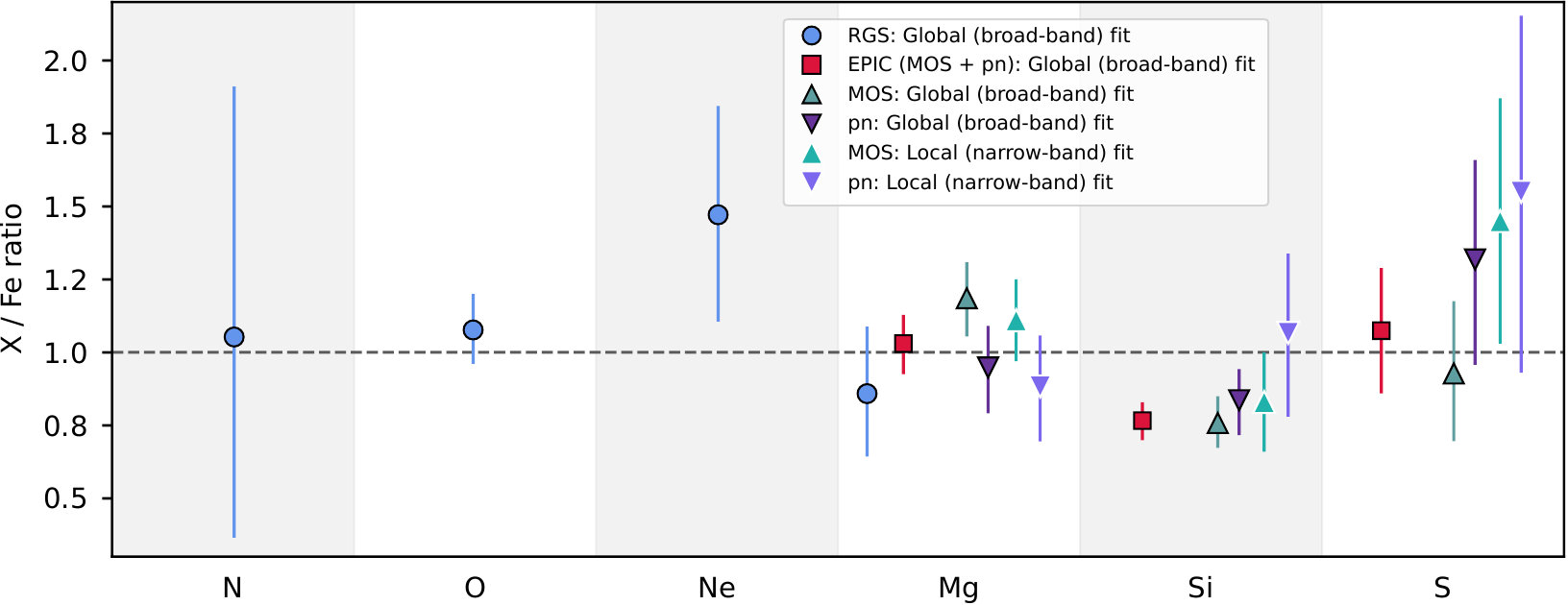}
    \caption{Global and local fits of RGS (blue circle), EPIC (red square), MOS (blue triangle-up) and pn (purple triangle-down) spectra for the galaxy core (Region number 4), obtained from \texttt{vgadem} model. }
    \label{fig:agreement}
\end{figure*}

\subsubsection{RGS spectral modelling}
\label{sub:rgs-spectral}

The RGS spectrometers are slitless, therefore, the spatially extended sources in the dispersion direction broaden the spectrum by

\begin{equation}
    \Delta\lambda \approx \frac{0.139}{\textit{m}}\Delta\theta\r{A}
    	\label{eq:rgs_broadening}
\end{equation}

where $m$ is the spectral order and $\theta$ is the half energy width of the source in arcmin\footnote{See the \textit{XMM-Newton} Users Handbook.}. The broadening is corrected with the \texttt{XSPEC} convolution model \texttt{rgsxsrc} at the spectral fitting stage. We generate MOS1 image in the 0.46-1.55 keV ($\sim$8--28 \r{A}) band and the suffice brightness profile in the dispersion direction. We use first and second-order RGS spectra in 8--28 \r{A} without combining R1 and R2 spectra. Similar to our EPIC analysis, we focus on \texttt{vgadem} model to estimate the multi-temperature structure of hot gas. The RGS spectral fit is presented in Fig. \ref{fig:spectra}, where we combine R1 and R2 first-order spectra using \texttt{rgscombine} for visualization purposes. Additionally, to test the "O/Fe bias" observed in \cite{Kara_2024}, which is the discrepancy of O/Fe ratio between 1T and multi-temperature models, we also fit the RGS spectra with 1T (\texttt{vapec}) and 2T (\texttt{vapec + vapec}) models.
 

We also fit our 0.8 arcmin spectra with the template background using 1T, 2T and \texttt{vgadem} models to test the measurement differences between different temperature models. For all 2T modelings, we assume that the hotter and colder components of the hot plasma have the same elemental abundance values. The measurement differences between these variations are briefly discussed in Sect. \ref{subsec:core_res} and presented in detail in Appendix \ref{subsection:RGS-systematics}.



\begin{table*}
\centering
\caption{Best-fitting results for the core of M86. The box region for the core is chosen to coincide the RGS extraction region. The Global fits refer to 0.55--10 keV fit, while Local fits refer to the energy interval where the line emission of the measured element is prominent. The luminosity results are obtained via the \texttt{XSPEC} convolution model \texttt{clumin}. $^{a}$The local fits are performed for each element separately for the energy interval described in the text. For Fe abundance, the Fe abundance obtained from the Global fit of that instrument (MOS or pn) is applied.}
\begin{tabular}{l
                c
                c
                c
                c
                c
                c} 
\toprule
\toprule
 & \multicolumn{1}{c}{RGS} & \multicolumn{1}{c}{EPIC (MOS + pn)} & \multicolumn{1}{c}{MOS Global} & \multicolumn{1}{c}{MOS Local$^{a}$} & \multicolumn{1}{c}{pn Global} & \multicolumn{1}{c}{pn Local$^{a}$} \\ 
\midrule
$\text{Luminosity}_{\text{Gas}}$ (\si{10^{39} \, \mathrm{erg} \, \mathrm{s}^{-1}}) & $4.03 \pm 0.05$ & $1.08 \pm 0.02$ & $1.20 \pm 0.05$ & {0.76--1.06} & $1.04 \pm 0.02$ & {1.02--1.08} \\
$\text{Luminosity}_{\text{ICM}}$ (\si{10^{39} \, \mathrm{erg} \, \mathrm{s}^{-1}}) & {--} & {< 0.13} & {< 0.11} & {(MOS Global)} & {< 0.45} & {(pn Global)} \\
$\text{Luminosity}_{\text{LMXB}}$ (\si{10^{39} \, \mathrm{erg} \, \mathrm{s}^{-1}}) & {--} & $0.18 \pm 0.03$ & $0.19 \pm 0.02$ & {(MOS Global)} & $0.18_{-0.03}^{+0.04}$ & {(pn Global)} \\ 
\midrule
$\text{kT}_{\text{mean}}$ (\si{\kilo\electronvolt}) & $0.88 \pm 0.02$ & $0.83 \pm 0.01$ & $0.84 \pm 0.01$ & {(MOS Global)} & $0.83 \pm 0.01$ & {(pn Global)} \\
$\sigma_{\text{kT}}$ (\si{\kilo\electronvolt}) & $0.38 \pm 0.04$ & $0.18_{-0.01}^{+0.02}$ & $0.15 \pm 0.02$ & {(MOS Global)} & $0.21_{-0.03}^{+0.02}$ & {(pn Global)} \\
\midrule
N & $0.47_{-0.32}^{+0.43}$ & {--} & {--} & {--} & {--} & {--} \\
O & $0.48_{-0.06}^{+0.08}$ & {--} & {--} & {--} & {--} & {--} \\
Ne & $0.66_{-0.16}^{+0.17}$ & {--} & {--} & {--} & {--} & {--} \\
Mg & $0.39_{-0.11}^{+0.12}$ & $0.68_{-0.09}^{+0.10}$ & $0.98_{-0.13}^{+0.14}$ & $0.92_{-0.08}^{+0.09}$ & $0.56_{-0.10}^{+0.12}$ & $0.53_{-0.09}^{+0.10}$ \\
Si & {--} & $0.50_{-0.06}^{+0.08}$ & $0.62_{-0.11}^{+0.08}$ & $0.69_{-0.11}^{+0.13}$ & $0.50_{-0.08}^{+0.07}$ & $0.64_{-0.15}^{+0.16}$ \\
S & {--} & $0.71_{-0.15}^{+0.17}$ & $0.75_{-0.22}^{+0.17}$ & $1.20_{-0.34}^{+0.33}$ & $0.79_{-0.20}^{+0.26}$ & $0.93_{-0.33}^{+0.36}$ \\
Fe & $0.45_{-0.05}^{+0.06}$ & $0.66 \pm 0.04$ & $0.83 \pm 0.07$ & {(MOS Global)} & $0.60 \pm 0.06$ & {(pn Global)} \\ 
\bottomrule
\end{tabular}
\label{table:parameters}
\end{table*}

\section{Results}
\label{sec:results}

\subsection{Galaxy core}
\label{subsec:core_res}

We measure the elemental abundances in the hot halo pervading the galaxy core of M86 using both RGS and EPIC data. To improve consistency between the RGS and EPIC measurements, we extract our core spectra from a rectangular region overlapping the RGS extraction band with a 0.8 arcsec width (Region 4), following a similar approach to \citet{Mernier2022b}. The elemental abundances in a CIE plasma are derived from the equivalent width (EW), which is proportional to the ratio of line flux to continuum flux. Due to imperfections in the effective areas of EPIC instruments, the continuum measurement may be biased \citep{Mernier_2015, Simionescu2019b}. Consequently, line fluxes may compensate for under- or overestimation of the continuum flux, leading to corresponding over- or underestimation of elemental abundances. We first perform a global fit using the \texttt{vgadem} model on the full-band EPIC spectra (0.55 keV–10 keV). Then, we repeat the global fitting separately for MOS (MOS1 + MOS2) and pn spectra. Additionally, we perform local fits of MOS and pn for narrow-band spectra corresponding to well-resolved elemental lines: 1–1.7 keV for Mg, 1.6–2.3 keV for Si, and 2.2–2.8 keV for S. In these more conservative local fits of MOS and pn, all parameters are frozen to the values derived from the global fit for that specific instrument, except for the normalization of the CIE model (\texttt{vgadem}) and the relevant elemental abundances, which are left free.

After the fittings, we observe that all global and local fittings obtained from MOS and pn spectra result in abundance ratios that agree within 1$\sigma$ uncertainties for all elements. These results are also consistent with global EPIC (MOS + pn) measurements. For the rest of the analysis, we use global EPIC measurements for all elements and regions.



As for the RGS, we fit the spectra with different methods described in Sect. \ref{subsection:RGS}. The best statistics between different temperature models is obtained with \texttt{vgadem} model. We observe that different temperature models result in discrepancies in Fe absolute abundance measurements due to the "Fe-bias". Additionally, we observe that O/Fe and Ne/Fe show significant differences between different temperature models. The O/Fe ratio differs more than 2$\sigma$ between single-temperature and \texttt{vgadem} models. Similarly, the Ne/Fe is overestimated to unrealistically high values ($\sim4$) for a single-temperature model, while the \texttt{vgadem} model provides physically realistic values for Ne/Fe. We find that different extraction regions (0.8 arcmin or 3.4 arcmin) yield abundance ratios that agree well. However, the background methods lead to discrepancies in Mg/Fe. Specifically, the background extracted between 95-100\% of the PSF results in a high Mg/Fe ratio ($\sim$2) and lower statistics, which is inconsistent with the ratio obtained using the template background within $1\sigma$. Here, we present our RGS results based on 0.8 arcmin extraction size and template background from CCD 9, which gives the best statistics. Although the X-ray data itself is Poisson, the background template is Gaussian. Therefore we fit our RGS spectra using a modified version of C-statistics (\texttt{pgstat}\footnote{\href{https://heasarc.gsfc.nasa.gov/xanadu/xspec/manual/XSappendixStatistics.html}{https://heasarc.gsfc.nasa.gov/xanadu/xspec/manual/XSappendixStatistics.html}}), which combines the Poisson likelihood for the source data with a Gaussian likelihood for the background, allowing us to properly account for the uncertainties in the background model. The results obtained from other methods and the comparison of the systematics in the RGS data of low-temperature plasma are further discussed in detail in Appendix \ref{subsection:RGS-systematics}. The differences in abundance ratios with respect to different temperature models are presented in Fig. \ref{fig:rgs_syst}, and the biases related to the extraction regions and background methods are presented in Fig. \ref{fig:rgs_syst_2}.

For the RGS and EPIC results of the galaxy core, we see that the overall abundance ratios are Solar. We find Solar ratios for N/Fe, O/Fe, Mg/Fe, and S/Fe. The Ne/Fe ratio is super-Solar within the 1$\sigma$ interval, while Si/Fe is sub-Solar with a significance of 3.8$\sigma$. The Mg/Fe ratios of EPIC (both local and global) and RGS data agrees within 1$\sigma$. The results are presented in Figure \ref{fig:agreement} and Table \ref{table:parameters}.

We see that, although M86 experiences strong ram-pressure stripping and consequently an accretion cut-off from its surroundings, its abundance ratios with respect to Fe are similar to rest of the gaseous content of the Universe. The abundance ratios in M86 core is discussed in detail in Sect. \ref{sec:core}.

\subsection{Radial distribution of metals}
\label{sec:radial}

We measure abundances and abundance ratios of 8 regions, covering a projected region roughly 15 arcmin ($\sim$80 kpc). For all regions, we use a multi-temperature \texttt{vgadem} model. Additionally, throughout this analysis we consider the Virgo ICM X-ray background to be constant across the FOV. To test the validity of this assumption, we evaluate the most extreme scenario by setting the ICM flux to zero and refitting the outermost region, Tail 2 (Region number 8). To perform this test, we fix the normalization of the ICM \texttt{apec} model to zero and let the other astrophysical scaled-down background normalizations to vary as a whole (with a multiplicative \texttt{constant} model). We see that for the case without the ICM model, the astrophysical background normalizations increase by $\sim$20\%, while the abundance ratios change roughly by $\sim$5\%. Since this fit provides poorer C-statistics, we decide to model the ICM emission for all regions with a scaled-down fixed normalizations according to the vignetting-corrected size of that region, as described in Sect \ref{subsection:background}.

The temperature and metal distribution is presented in Table \ref{table:radial}. We found $\sigma_{\text{kT}}$ is roughly $\sim\!5$\% of the $\text{kT}_{\text{mean}}$ at the Plume and Tail regions (Tail 1 and Tail 2), meaning a steeper Gaussian distribution compared to other regions. This indicates that the temperature structure of the Plume and Tail regions shows more isothermal features and are closer to a single-temperature structure. 

\begin{figure*}
    \centering
    \includegraphics[width=0.85\textwidth]{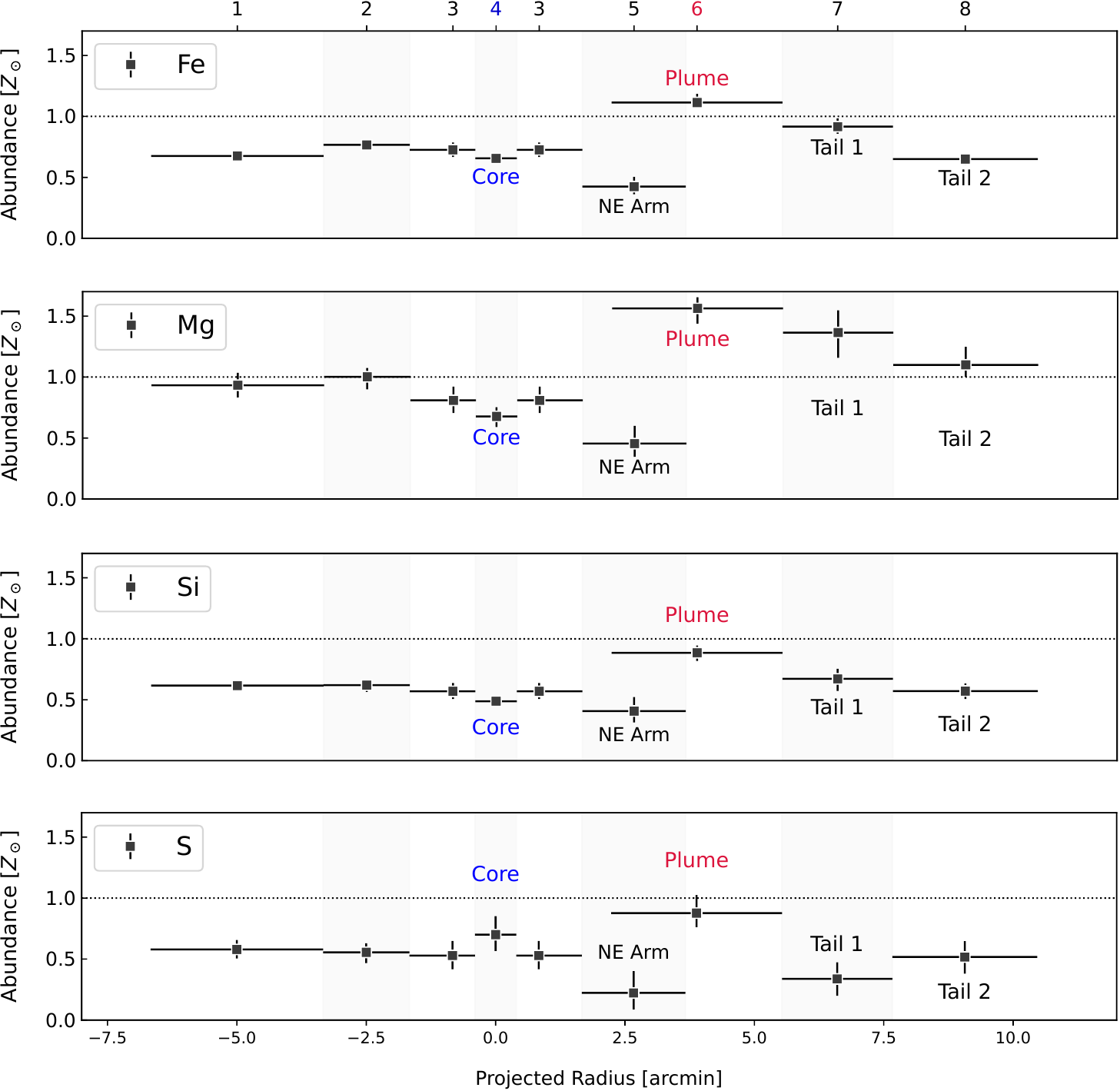} 
     \caption{Absolute abundance distribution in the hot atmosphere of M86 in Solar units for each region (upper x-axis). Region 3, which is plotted twice, corresponds to the circular region surrounding but excluding the core. The projected distances in arcminutes from the galaxy core are indicated. The emission structures are also labeled accordingly.}
     \label{fig:abundances}
    
\end{figure*}

The changes in absolute abundances are shown in Fig. \ref{fig:abundances}, and each ratio is further discussed in the following sections. We observe an abundance drop at the galaxy core for Mg and Si; while Fe and S abundances increase in the core. Similarly, at the North-Eastern Arm, we observe an significant abundance drop. Additionally, we observe that the Plume has the highest  abundance for all elements. Furthermore, the abundance ratio gradients of Mg/Fe, Si/Fe and S/Fe between regions do not show a similar trend. Specifically, Mg/Fe appears to have super-Solar ratios outside of the core, and becomes Solar for the regions close to the galaxy core. Si/Fe and S/Fe however, shows sub-Solar values except the core where S/Fe reaches a Solar ratio value. 

Additionally, we compare the abundance ratios within M86 to those of the ambient medium. The projected distance between the core of M87 and M86 is $\sim$75 arcmin, corresponding to $\sim$340 kpc. \citet{Simionescu_2015} measured the abundance ratios in the hot gas of the Virgo Cluster with \textit{Suzaku} observations beyond the virial radius by binning their spectral data into four radial regions. Their results for the region between 300--500 kpc from M87, which encompasses M86, are $1.39\pm0.07$ for Mg/Fe, $0.82\pm0.04$ for Si/Fe, and $1.17\pm0.07$ for S/Fe.

In the following sections, we compare our results for each abundance ratio with the average ratio in hot gas obtained from the CHEERS (CHEmical Evolution RGS Sample) sample \citep{Mernier2018b}, the ratios in the stars of elliptical galaxies \citep{Conroy_2013}, and the Virgo ICM ratios from \citet{Simionescu_2015}, which represent the chemical composition of the ambient medium through which M86 travels.



\subsubsection{Mg/Fe increases to super-Solar towards outskirts and tail}

\begin{table*}
\centering
\caption{Best-fitting results for all regions. The region numbers are indicated in parenthesis. All abundances and abundance ratios are obtained from the Global fits using EPIC measurements.}
\begin{tabular}{l
                c
                c
                c
                c
                c
                c
                c
                c
                c
                c
                c
                c
                c} 
\toprule
\toprule
 Region (arcsec) & 400--200 (1) & 200--100 (2) & 100--24 (3) & 0--24 (4) & 100--220 (5) & 132--332 (6) & 332--460 (7) & 460--628 (8) \\ 
 &  &  &  & (Core) & (N.E. Arm) & (Plume) & (Tail 1) & (Tail 2) \\ 
\midrule


\makecell[l]{\texttt{XSPEC} Norm\textsubscript{Gas} per \\ \si{\arcsec\squared} (\( \times 10^{-7} \))} &
$0.10\pm0.02$ & $0.18\pm0.03$ & $0.35\pm0.05$ & $0.63\pm0.06$ & $0.25\pm0.04$ & $0.18\pm0.04$ & $0.12\pm0.02$ & $0.13\pm0.02$ \\

\makecell[l]{\texttt{XSPEC} Norm\textsubscript{LMXB} per \\ \si{\arcsec\squared} (\( \times 10^{-7} \))} & 
-- & -- & $< 0.004$ & $0.002\pm0.001$ & $< 0.002$ & -- & -- & -- \\

\midrule

$\text{kT}_{\text{mean}}$ (\si{\kilo\electronvolt}) & $1.17\pm0.01$ & $0.89\pm0.01$ & $0.87_{-0.01}^{+0.02}$ &  $0.83\pm0.02$& $0.77\pm0.02$  &  $0.88\pm0.01$ & $0.97\pm0.01$ & $1.03\pm0.01$ \\
 $\sigma_\text{kT}$ (\si{\kilo\electronvolt}) & $0.19_{-0.01}^{+0.02}$ & $0.11_{-0.01}^{+0.02}$  & $0.20_{-0.01}^{+0.02}$ &  $0.18_{-0.01}^{+0.02}$  &   $0.13_{-0.04}^{+0.03}$& $0.04_{-0.03}^{+0.02}$ & $0.04\pm0.03$ & $0.05_{-0.03}^{+0.04}$  \\ \midrule

Fe & $0.68 \pm 0.04$ & $0.77 \pm 0.04$ & $0.73 \pm 0.06$ & $0.66 \pm 0.04$ & $0.42_{-0.06}^{+0.08}$ & $1.11_{-0.04}^{+0.07}$ & $0.92_{-0.06}^{+0.07}$ & $0.65_{-0.05}^{+0.04}$ \\
Mg/Fe & $1.38_{-0.11}^{+0.12}$ & $1.30_{-0.08}^{+0.07}$ & $1.12 \pm 0.09$ & $1.03 \pm 0.09$ & $1.07_{-0.17}^{+0.18}$ & $1.39_{-0.07}^{+0.05}$ & $1.48_{-0.16}^{+0.15}$ & $1.72_{-0.13}^{+0.14}$ \\
Si/Fe & $0.91 \pm 0.04$ & $0.81_{-0.05}^{+0.04}$ & $0.80_{-0.06}^{+0.06}$ & $0.77 \pm 0.06$ & $0.96_{-0.14}^{+0.15}$ & $0.79_{-0.04}^{+0.05}$ & $0.73_{-0.08}^{+0.07}$ & $0.88_{-0.07}^{+0.09}$ \\
S/Fe & $0.86 \pm 0.10$ & $0.72 \pm 0.11$ & $0.73 \pm 0.15$ & $1.08_{-0.21}^{+0.20}$ & $0.52_{-0.31}^{+0.40}$ & $0.78_{-0.09}^{+0.12}$ & $0.37_{-0.16}^{+0.14}$ & $0.81_{-0.20}^{+0.18}$ \\

\bottomrule
\end{tabular}
\label{table:radial}
\end{table*}

The average Mg/Fe ratio in clusters, groups and ellipticals is $0.94\pm0.07$ \citep{Mernier2018b}. The Mg/Fe ratio in the radial region of 300-500 kpc from M87, which we assume the Mg/Fe in the ambient medium of M86, is $1.39\pm0.07$ \citep{Simionescu_2015}. We see that the Mg/Fe ratio in Plume ($1.39_{-0.07}^{+0.05},$), the group outskirts ($1.38_{-0.11}^{+0.12}$), and the Virgo ICM ($1.39\pm0.07$) are in excellent agreement.

\begin{figure}
    \centering
        \includegraphics[width=\columnwidth]{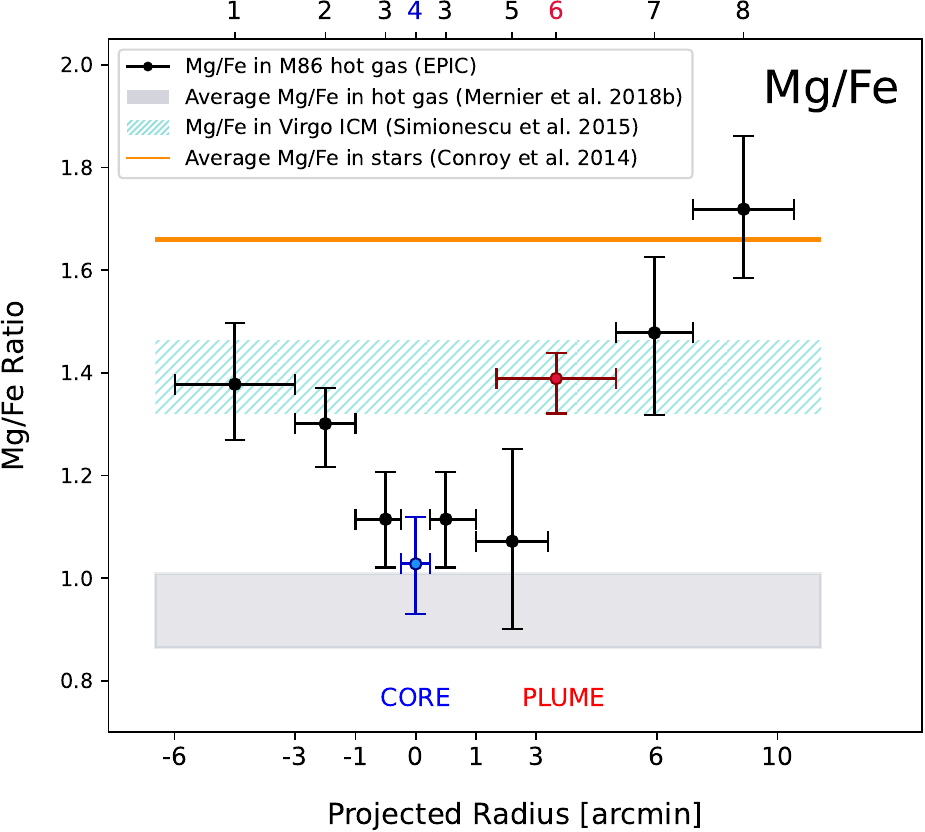}
        \caption{Mg/Fe ratio distribution in the hot gas of M86. The dashed blue band is the Mg/Fe ratio in the Virgo Cluster with comparable altitude. The gray band refers to the average Mg/Fe ratios in the hot gas of galaxy clusters, groups and elliptical galaxies. The yellow line shows the average Mg/Fe ratio in the stellar populations of early-type galaxies.}
        \label{fig:mg_fe}

\end{figure}

 The Mg/Fe ratio drops to the Solar abundance ratio values towards the galaxy core, in which Mg/Fe ratio is the lowest of all regions with the Solar value $1.03\pm0.09$. North-Eastern Structure shows a similar Solar ratio, while the Plume has super-Solar ratio of $1.39_{-0.07}^{+0.05}$, in agreement with the ambient ICM. Interestingly, in the X-ray tail, we observe that the Mg/Fe ratio exceeds the average value of Mg/Fe ratio in the stellar populations of elliptical galaxies \citep{Conroy_2013} with $1.72_{-0.13}^{+0.14}$. Our result shows a clear trend where the dense core exhibits Solar Mg/Fe, similar to rest of the gaseous content in the Universe, whereas the outskirts and the stripped gas shows Mg/Fe ratio similar to the Virgo ICM, which is presented in Fig. \ref{fig:mg_fe}. This distribution might indicate an efficient mixing of the less-dense hot gas of M86 in the outskirts and the stripped regions with the surrounding Virgo ICM.

\subsubsection{Si/Fe agrees with sub-Solar Virgo ICM ratio} 

We observe that, the outermost radial region (Region 1) in our analysis shows slightly sub-Solar Si/Fe ratio, in agreement with the average Si/Fe ratio in hot gas pervading the Universe \citep{Mernier2018b}. For the regions closer to the core, Si/Fe ratio drops to the Virgo ICM\footnote{In the context of this paper Virgo ICM refers to the ambient medium in Virgo Cluster where M86 voyages.} values, which is  $0.82\pm0.04$. This is in contrast with the Mg/Fe distribution, which shows a clear similarity for the outermost regions with the Virgo ICM, and dissimilarity within dense regions. For Si/Fe, we see that except the very outskirt, all other regions including the tail are in agreement with the Si/Fe ratio in Virgo ICM within 1$\sigma$ error. Additionally, we observe that the Si/Fe ratio of the dense core is slightly lower than that of Virgo ICM, although still agrees. The North-Eastern Arm shows the highest Si/Fe ratio, however due to high uncertainty it is not possible to make any physical interpretation.

\begin{figure}
        \centering
        \includegraphics[width=\columnwidth]{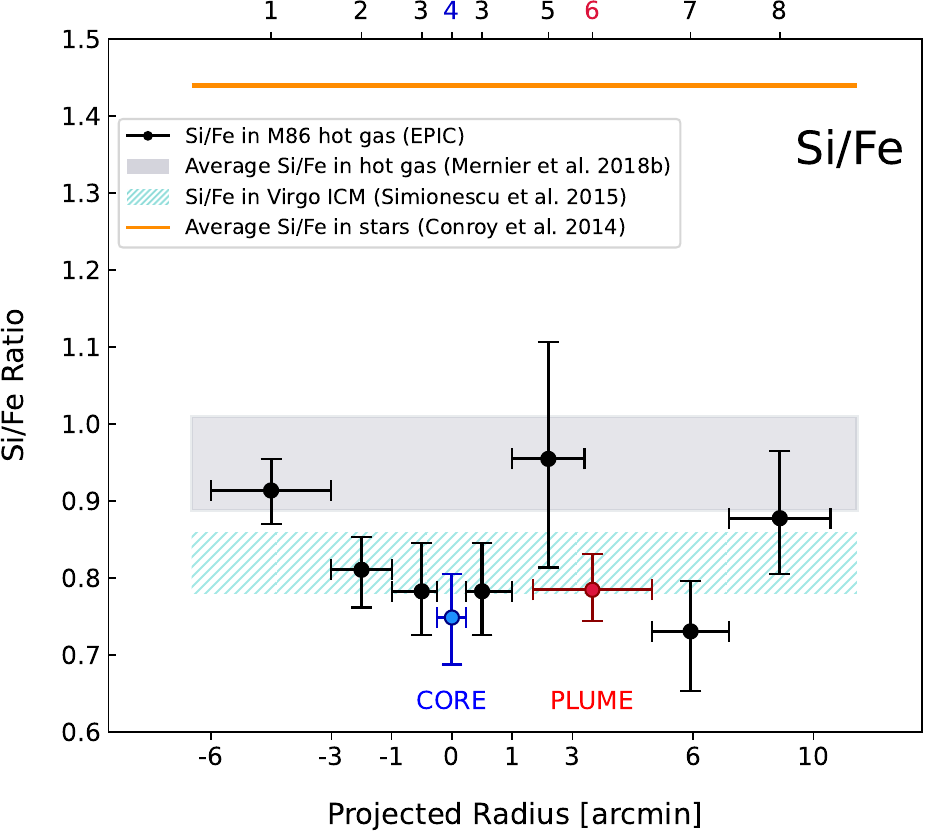}
        \caption{Si/Fe ratio distribution in the hot gas of M86. The dashed blue band is the Si/Fe ratio in the Virgo Cluster with comparable altitude. The gray band refers to the average Si/Fe ratios in the hot gas of galaxy clusters, groups and elliptical galaxies. The yellow line shows the average Si/Fe ratio in the stellar populations of early-type galaxies.}
        \label{fig:si_fe}

\end{figure}

\subsubsection{Sub-Solar S/Fe everywhere except the core} 

\begin{figure}
        \centering
        \includegraphics[width=\columnwidth]{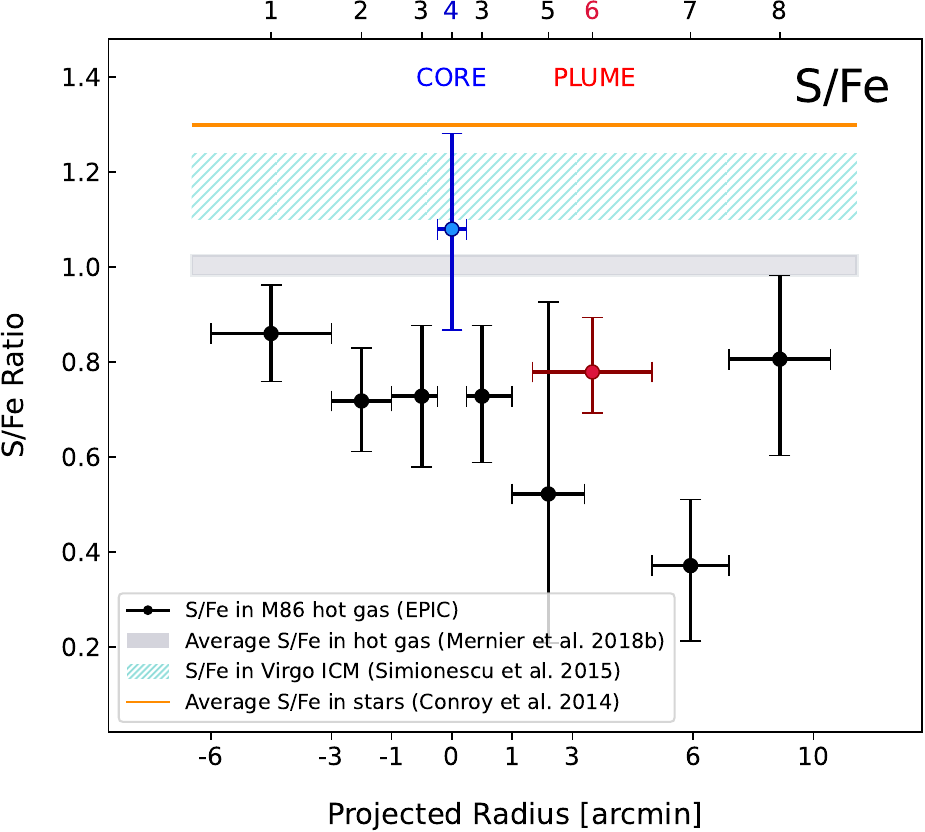}
        \caption{S/Fe ratio distribution in the hot gas of M86. The dashed blue band is the S/Fe ratio in the Virgo Cluster with comparable altitude. The gray band refers to the average S/Fe ratios in the hot gas of galaxy clusters, groups and elliptical galaxies. The yellow line shows the average S/Fe ratio in the stellar populations of early-type galaxies.}
        \label{fig:s_fe}

\end{figure}

We observe that the sub-Solar S/Fe ratio is lower in all regions except the Core compared to the average hot gas and Virgo ICM ratio of $1.17\pm0.07$. The low S/Fe ratios are in tension with the Si/Fe ratios, which are very similar to those of the Virgo ICM, and with the Mg/Fe ratios, which are comparable to the Virgo ICM except in the core. We observe a clear S/Fe discrepancy between M86 and the Virgo ICM. This is particularly interesting given that Mg, Si, and S are all $\alpha$-elements. The most outstanding region is the Tail 1, Region 7, shows very low S/Fe ratio of $0.37_{-0.16}^{+0.14}$. In the Tail 1 region, the absolute abundance of S is $0.33\pm0.11$, which can be considered as a usual value for an element in outskirts. Moreover, we observe for all other elemental abundances a gradual decrease between the three regions, Plume, Tail 1, and Tail 2 ($\text{Z}_{\text{Plume}}>\text{Z}_{\text{Tail 1}}>\text{Z}_{\text{Tail 2}}$). S, however, is more abundant in Tail 2 compared to Tail 1 (  $\text{Z}_{\text{S: Plume}}>\text{Z}_{\text{S: Tail 2}}>\text{Z}_{\text{S: Tail 1}}$). Therefore the relative abundance ratio of S is found to be quite low compared to other ratios. The Tail 2, on the other hand, shows a ratio in agreement with all other regions with $0.81_{-0.20}^{+0.18}$.

\section{Discussion}
\label{sec:Discussion}

\subsection{The chemical composition of the M86 galaxy core}
\label{sec:core}

\begin{figure*}
    \centering
    \includegraphics[width=0.9\textwidth]{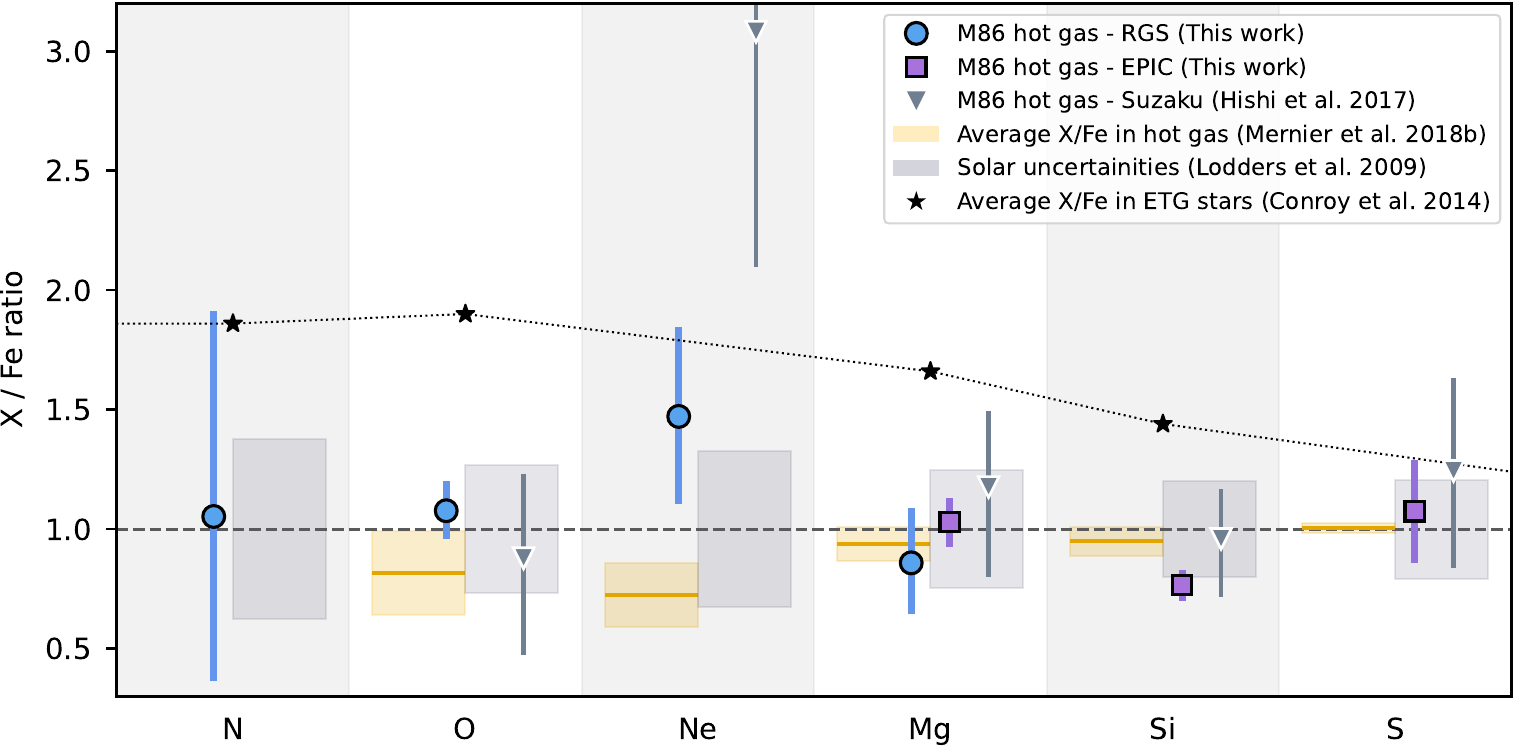} 
    \caption{Abundance ratios with respect to Fe in the galaxy core from RGS (blue circle) and EPIC (purple square) data. The gray triangles show the Suzaku measurements of the M86 galaxy core \citep{Hishi_2017}. The yellow bars represent the average abundance ratios in hot plasma \citep{Mernier2018b}, while the gray bars show the uncertainties in the Solar measurements \citep{Lodders2009}. Finally, the black stars indicate the average abundance ratios in the stellar populations of elliptical galaxies \citep{Conroy_2013}.}
    \label{fig:discussion-core}
\end{figure*}

In Fig. \ref{fig:discussion-core} we present the abundance ratios with respect to Fe in the M86 core. The blue circles represents the RGS and purple boxes are for EPIC measurements obtained from a box region with 0.8 arcmin size overlapping the RGS dispersion direction. In this plot, we also compare our results with (i) the multi-temperature \textit{Suzaku} measurements of M86 galaxy core by \citet{Hishi_2017} (gray triangles), (ii) the average elemental ratios with respect to Fe in the 44 nearby systems of the CHEERS sample by \citet{Mernier2018b} (yellow boxes), (iii) the average X/Fe ratios in the stellar content of early-type galaxies by \citet{Conroy_2013} (black stars). Finally, we also show the uncertainties in the Solar measurements \citep{Lodders2009} in gray boxes.

We compare our hot gas abundance ratios with average stellar abundance ratios to investigate whether the original hot gas in M86 has been depleted by ram-pressure stripping and replaced by a new generation of hot gas produced by stellar winds, as thermalised stellar mass loss is a continuous internal source of hot gas \citep{Mathews1990, MathewsBrighetni2003}. Chemical composition of hot gas pervading cluster member galaxies and their stellar population is considered to be decoupled. The stars in such elliptical galaxies are richer in light $\alpha$ elements (e.g. O, Ne, Mg, Si, and S), compared to heavy Fe-peak elements (e.g. Ca, Cr, Mn, Fe and, Ni) \citep{Conroy_2013}. The reason of such difference is that the light $\alpha$ elements are mostly produced in core-collapse supernovae (SNcc), while Fe-peak elements are dominantly created in Type Ia supernovae (SNIa). Elliptical (i.e., early-type) galaxies are thought to have undergone rapid star formation early in cosmic history, with a peak around $z \sim 3$ \mbox{\citep{Thomas2010}}, which is also referred as the "downsizing". Considering SNIa events occur later in cosmic time compared to SNcc explosions, the star-forming gas in the ISM has not been enriched by SNIa products as efficiently as SNcc products at $z \sim 3$. Therefore, the rapidly formed stars in ellipticals are richer in $\alpha$ elements, leading super-Solar abundance ratios of $\alpha$ elements with respect to Fe (i.e., $\alpha/$Fe > 1 in stars). In M86, because of the ram-pressure stripping, any gas inflow with Solar abundance ratios onto the galaxy is stopped \citep{Gunn_1972}. Therefore, any agreement with the stellar abundance ratios might indicate that due to ram-pressure stripping and the consequent accretion cut-off, the hot gas of M86 is replenished by stellar mass-loss products.



However, we see that O/Fe, Mg/Fe, and S/Fe are in agreement with the average ICM abundance ratio measurements of 44 nearby systems in the CHEERS sample \citep{Mernier2018b}, represented by yellow boxes. Overall, in the M86 core halo, we observe Solar and sub-Solar abundance ratios with respect to Fe, indicating that stellar winds with super-Solar ratios do not dominate the overall composition of the hot halo.

Our result is in tension with another ram-pressure stripped galaxy, M89, which shows stellar-like super-Solar abundance ratios of $\alpha$ elements with respect to Fe \citep{Kara_2024}, likely due to AGN activity facilitating the stripping of the original Solar gas. However, our results are comparable to another infalling elliptical galaxy, NGC\,1404, which also does not harbor an AGN and shows Solar abundance ratios \citep{Mernier2022b}. These comparisons suggest that, for the loss and replenishment of hot gas in a galaxy, strong AGN activity comparable to the gravitational potential might be required. The high mass of M86, together with the lack of any AGN activity might lead to the preservation of the original chemical composition of the hot atmosphere.

We found Solar N/Fe abundance ratios in the M86 core. Although the Solar abundance ratios in hot plasma are typical for heavier elements that are mostly produced in supernovae, N/Fe ratios are reported usually super-Solar. For instance, NGC\,1404 is found to have $\sim\!2.6$ Solar N/Fe \citep{Mernier2022b}, in the Centaurus cluster N/Fe is found $\sim\!3$ Solar \citep{Sanders_2008}, and N/Fe in NGC\,4636 is $\sim\!2.5$ Solar \citep{Werner_2009}. Additionally, \cite{Mao_2019} showed N/Fe ratios in 8 systems (mostly elliptical galaxies) are super-Solar, similar to \citet{Sanders_Fabian_2011} showing in a stacked RGS spectra the N/Fe is $\sim\!1.7$ Solar. One explanation of this phenomena is the fact that N is not created in supernovae explosions (unlike the other elements in X-ray band except C), but in AGB stars \citep{Molla_2006}. This suggests that the N enrichment rate is invariant to the SNIa/SNcc production ratio in an ambient medium. Therefore, we can assume an ongoing N enrichment in a system, independent of its enrichment history. Consequently, considering the hot halos are dominantly enriched by SNe products early in the cosmic time at z $\sim$ 2--3, we can expect a N surplus accumulated from $z\sim$ 2--3 to $z\sim$ 0, which leads to high N/Fe ratios. Therefore, our Solar N/Fe ratio among other super-Solar N/Fe samples is peculiar, although not unique. \cite{Sanders_2010} also reported Solar N/Fe ratios for Abell\,262, Abell\,3581 and HCG\,62. Although the reason of such deviations are unclear, we can assume that N abundance is low in M86 compared to other similar systems. We also note that the uncertainties in our N/Fe ratio is relatively high, therefore deeper observations are needed for making more robust interpretations.

Contrary to our relatively low N/Fe ratio, we find a super-Solar Ne/Fe ratio that is in tension with the average Ne/Fe ratio in the ICM, though not a strong discrepancy with a $\sim\!1.6\sigma$ difference. Such a higher Ne/Fe ratio is also reported for NGC\,1404 in \cite{Mernier2022}, which is also a relatively cool object. The difference between our super-Solar Ne/Fe ratio and the slightly sub-Solar Ne/Fe ratio in the CHEERS sample can be explained by the measurement degeneracy of Ne {\footnotesize X}. Between 12--13 \r{A}, Fe {\footnotesize XXI} and Fe {\footnotesize XXII} lines are enhanced at high temperatures, making the Ne {\footnotesize X} line indistinguishable from them. In galaxy clusters, which constitute most of the CHEERS sample, the core is expected to exhibit hotter temperature structures compared to M86. Therefore, the degeneracies caused by the blending of the Ne {\footnotesize X} line with Fe {\footnotesize XXI} and Fe {\footnotesize XXII} might lead to an underestimation of the Ne abundance in galaxy cluster cores. Thus, an increased Ne/Fe ratio in a relatively cool, relatively isothermal system can be expected in other similar systems as well, compared to those obtained from hotter galaxy clusters. Our result is in agreement with the Ne/Fe ratio in NGC\,1404 measured by \cite{Mernier2022}, which is $\sim\!1.13$. 

We observe a sub-Solar Si/Fe ratio in the galaxy core, which disagrees with the average Si/Fe ratio more than 2$\sigma$. Central drops in metal abundances in cool cores can be attributed to the cooling of hot-phase reactive metals such as Fe, Si, S, Mg, and Ca, followed by their subsequent depletion into dust grains \citep{Panagoulia_2013, Panagoulia_2015}. M86 hosts approximately $\sim10^6 \, \text{M}_{\odot}$ of dust located 10 kpc from the center toward the southeast \citep{Gomez_2010}. Therefore, a possible explanation for this low Si/Fe ratio might be a more efficient depletion of Si ions into dust. However, this would require a selectively high depletion of Si compared to Fe, whereas reactive elements have been shown to be highly correlated in their abundance drop \citep[e.g.,][]{Mernier_2017, Lakhchaura_2019, Liu_2019}. Another explanation of low Si/Fe ratio might be attributed to a fitting bias due to the complex temperature structure. However, complexities due to multi temperature structure cause an underestimation of Fe absolute abundance, the so-called "Fe bias" as discussed previously. Therefore any modelling bias is expected to result in an increased Si/Fe ratio. The last explanation of low Si/Fe ratio is that there is indeed a \textit{real} low abundance of Si enrichment compared to Fe and other $\alpha$ elements. This requires an enrichment mechanisms to explain such an abundance pattern. We also note that, although Si/Fe ratio in M86 is >2$\sigma$ lower than the average Si/Fe ratio in ICM, our result is in agreement with the Solar value when the Solar uncertainties of \cite{Lodders2009} are included (See Fig. \ref{fig:discussion-core}).

Moreover, we observe that our results of \textit{XMM-Newton}/EPIC and RGS are in agreement with the \textit{Suzaku} measurements by \citet{Hishi_2017}, except Ne/Fe, which they found an unusually high value, greater than 3 Solar. In addition to instrumental differences, the main distinctions between our study and \citet{Hishi_2017} lie in the choice of temperature models and background treatments. We adopted the \texttt{vgadem} model, while they used 2T models. In our analysis, we found that both approaches --\texttt{vgadem} and 2T-- yield consistent results. For the background, we determined the parameters from a full FOV fit and rescaled the normalizations for each region to improve stability, whereas they fixed the normalizations of all background components—except the ICM—without such rescaling. Other differences include the assumed foreground temperatures and the inclusion of an SWCX component in our model, which was not included in theirs.




\subsection{The dissimilar metal distribution in the hot gas of M86}

The metal distribution in the halo of M86 is non-uniform. We observe a clear trend in the Mg/Fe ratio: in the galaxy core it is closer to the Solar value while in the outer group gas and the extended emission structures --Plume and Tail-- the hot gas is richer in Mg, resulting in high (super-Solar) Mg/Fe ratios. The Mg/Fe is approximately 3.3$\sigma$ higher in the Plume compared to the Core. Si/Fe and S/Fe ratios, on the other hand, do not exhibit a similar trend. 

The Si/Fe ratio remains roughly uniform, while there is a decrease from the outskirts to the core, similar to the Mg/Fe trend, although less prominent because, apart from the outermost region, all ratios agree within 1$\sigma$. The S/Fe ratio, on the other hand, differs from both Mg/Fe and Si/Fe: in all regions, S/Fe is sub-Solar except in the core, where it increases to a Solar value. 

In their \textit{Suzaku} analysis, \citet{Hishi_2017} reported Solar abundance ratios everywhere within the $1\sigma$ uncertainties, except for Ne/Fe, which they found to be super-Solar throughout. The only deviation of their results from the overall Solar ratios—aside from Ne/Fe—is their Mg/Fe ratio in the plume, for which they found a super‑Solar value, consistent with ours. Their Si/Fe and S/Fe ratios for the plume are Solar, comparable with our results. Additionally, they measured a super‑Solar Mg/Fe ratio in their outer group gas region at a distance from the core comparable to our Region 1, though still consistent with the Solar value within $1\sigma$. However, for the tail, they reported Solar abundance ratios, which is inconsistent with our super-Solar Mg/Fe result, as well as our significantly sub-Solar Si/Fe and S/Fe ratios in the same tail regions. Nevertheless, our \textit{XMM-Newton} results and the \textit{Suzaku} results of \citet{Hishi_2017} agree for the plume and core, while the tail results do not.


\subsubsection{Mg/Fe ratio}



We observe that in the outskirts of the M86 group the Mg/Fe ratio is in excellent agreement with that of the ambient Virgo ICM medium (See Fig. \ref{fig:mg_fe}). Therefore, we expect that the Virgo ICM and M86 group halo have mixed efficiently through inflows and outflows. 

The Mg/Fe ratio in Plume is in agreement with that of M86 group outskirts and Virgo ICM. The Plume structure might have formed from this aforementioned mixed low-entropy group gas after the collision with NGC\,4438 and the onset of ram-pressure stripping. Since these regions extend well beyond the stellar population of the galaxy and experience an accretion cut-off from their surroundings due to ram-pressure stripping \citep{Gunn_1972}, the chemical composition in emission structures might have remained unchanged since their formation. Similarly, the chemical composition of the extended group gas (Regions 1 and 2), which is mixed efficiently with the Virgo ICM, is expected to preserve its chemical composition compared to that of the galaxy core. In contrast, in the M86 \textit{galaxy} core, SNIa events continue to enrich the medium. Therefore, SNIa products such as Fe are expected to increase in the galaxy core, leading to a decreased Mg/Fe ratio. 

 We further performed narrow-band fits in the Plume to measure the Mg/Fe abundance ratio. All the separate full-band fits of MOS and pn, as well as the narrow-band fits, are in good agreement with the global EPIC measurement. Therefore, our result indicating that Mg/Fe in the Plume is consistent with the low-entropy gas in the M86 group outskirts and Virgo ICM is robust.

Additionally, ram-pressure stripping enhances the star formation, which increases the number of massive stars and SNcc explosions. In Tail 2 (Region 8), the far end of the M86 tail in our observation, we observe the highest Mg/Fe ratio in M86. It is plausible that the dynamical disruption in the M86 tail enhanced the Mg enrichment.

\subsubsection{Si/Fe and S/Fe ratios}
\label{subsubsection:si-se}

Si/Fe ratio is relatively uniform, agreeing with the ICM value (See Fig. \ref{fig:si_fe}). The Si/Fe ratio variation from the group outskirts to the galaxy core shows a similar trend to Mg/Fe, though apart from the outermost region, Region 1, it is consistent everywhere. Notably, the M86 galaxy core, and all other emission structures have Si/Fe ratios in agreement with the Virgo ICM, which supports the argument that these structures formed from the well-mixed group gas. We also note that the North-Eastern Arm shows the highest Si/Fe ratio, although still Solar and also in agreement with the Virgo ICM. 

S/Fe ratio, on the other hand, is more peculiar as it disagrees with the Virgo ICM everywhere except the core, where there is a sharp increase of S/Fe ratio (See Fig. \ref{fig:s_fe}). Such a non-flat trend between different $\alpha$-elements has been observed before by \citet{Million_2011} where they reported a centrally peaked Si/Fe in M87 while Mg/Fe is flat; and \citet{Mernier2022b} where they reported a Si-rich arc in NGC\,1404 between $\sim2-7$ kpc from the centre.

The origins of nucleosynthesis for Mg, Si and S might explain the non-uniformity to some extend. Although Mg, Si and S are primarily produced by SNcc; Si and S also are produced by SNIa with non-negligible amounts \citep[see, e.g.,][]{Mernier_SNe,Mernier2022}. Therefore any ongoing SNIa in the galaxy core is also expected to produce Si and S, and, through time, Si/Fe and S/Fe ratios may not drop as much as Mg/Fe ratio in the galaxy core. However, the Si and S production in the core due to SNIa events is expected to be very limited compared to Fe production. 

To investigate any potential bias related to the complex temperature and abundance structure, which may lead to measurement discrepancies of Mg, Si, and S, we simulated a two-temperature spectrum with cooler and hotter components. A higher metal abundance was assigned to the cooler component than to the hotter plasma, and the spectrum was fitted using the Gaussian temperature distribution, \texttt{vgadem}. No significant bias was observed between the Mg, Si, S, and Fe abundances, with the differences between elements being less than 10\%.

\begin{figure}
    \centering
    \includegraphics[width=0.95\linewidth]{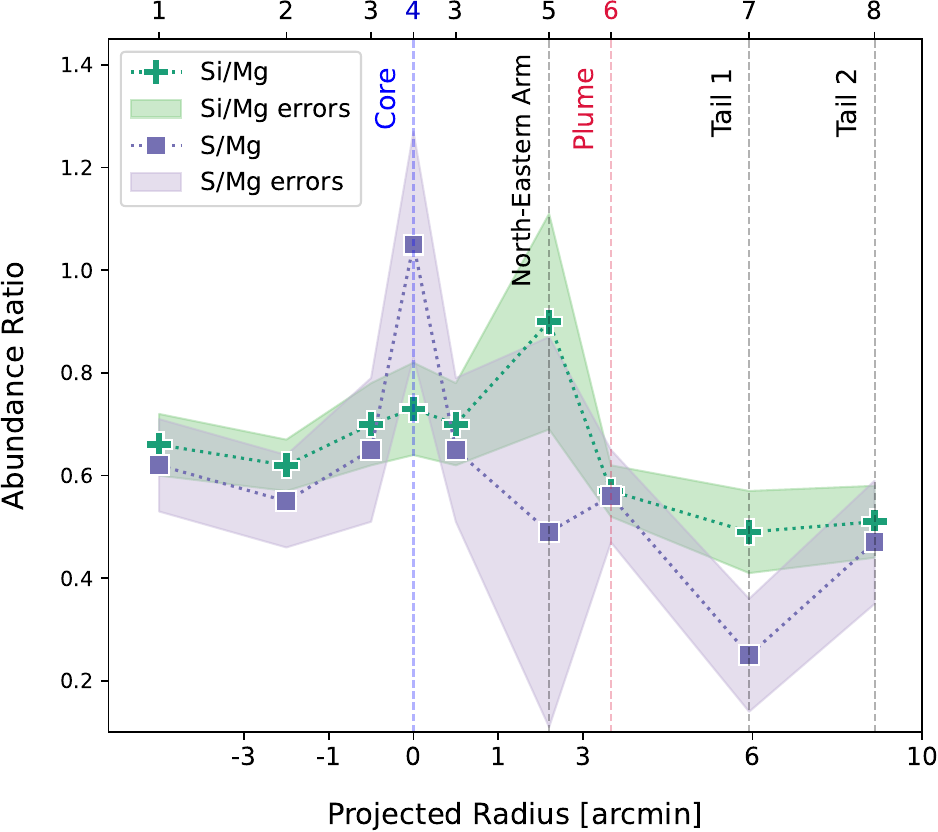}
    \caption{The radial differences between Si/Mg and S/Mg. The green plus signs are for S/Mg, while purple squares are S/Mg. The bars with same colors represent the 1$\sigma$ error bars. We observe that Si and S are correlated well apart from the Core, the North-Eastern Arm and Tail 1.}
    \label{fig:si_s}
\end{figure}

In Fig \ref{fig:si_s}, we represent the change in Si and S with respect to Mg. We observe that, Si/Mg and S/Mg ratios agree within 1$\sigma$ except the first tail region, Tail 1. In fact, the spatial variations of Si/Mg and S/Mg ratios are remarkably consistent with each other --meaning Si/S $\approx\!1$-- except the Core, the North-Eastern Arm and Tail 1. We observe that S is more abundant in the galaxy Core compared to the other parts of M86 group. Additionally, we observe that in Tail 1, the Si has a higher ratio with respect to Mg, compared to S/Mg. It is also notable that in the Plume, S and Si absolute abundances are almost the same. 

Additionally, similar to the gradient in Mg/Fe, we see $>1\sigma$ increase in Si/Fe and S/Fe ratios in Tail 2, compared to Tail 1, which may attributed to the increased star formation and the consequent increase in the SNcc enrichment due to ram-pressure stripping (See Fig. \ref{fig:si_fe} and Fig. \ref{fig:s_fe}).

\subsection{Constraining the SNe enrichment based on the abundance pattern of M86 core}
\label{subsection:SNE_head}
Metals accumulated in the hot halos of galaxies, galaxy groups and galaxy clusters provide a unique way to test SNe enrichment scenarios. Previous studies are mostly done for cluster scale \citep[e.g.,][]{Werner_2006,dePlaa_2007,Mernier_SNe,Simionescu2019b,Gatuzz_2023}, which represents the integrated yields of SNe products that are able to reach the ICM. While the closed-box ICM acts like a repository for all produced and ejected metals, individual galaxies and galaxy groups may not be closed-box systems. In order to have a complete picture of enrichment mechanisms, it is important to understand the SNe contribution variations in the lower-mass systems \citep[for a review, see, e.g.,][]{Loewenstein_2004}. The M86 group which experiences galaxy-galaxy interaction and strong ram-pressure stripping provides a unique opportunity to constrain the SNe enrichment in such a disturbed halo. 

 The chemical enrichment in M86 halo from SNe explosions can be categorized briefly with three different channels, the first one is the SNe enrichment that enriched the ICM at $z \sim 2-3$, inside and outside of M86, based on the early enrichment paradigm. The second channel is the SNe explosions during the brief star formation epoch of M86, which is at $z \sim 2$ for elliptical galaxies. The third one is the ongoing SNe explosions in M86 after that epoch. Our calculated SNe fractions in this section represents the integrated sum of all channels.



 
 In this section, we calculate and compare the yields—the mass of synthesized and ejected metals—of SNIa and SNcc in the galaxy core\footnote{We combine EPIC and RGS measurements of Mg with a weight factor $1/\sigma^2$.}, using, 

\begin{equation}
M_i = \frac{\int_{}M_i(m)m^{\alpha}dm}{\int_{}m^{\alpha}dm},
\end{equation}
where $M_i$ is the total yield (also denoted as $Y_i$) of the $i$-th element, $M_i(m)$ is the $i$-th element mass produced in a star of mass $m$, and $\alpha$ is the slope of the initial mass function (IMF). Lower and upper limits of the integration are the zero-age main-sequence mass of stars, which are different for models.

Furthermore, the number of atoms of element X is given by

\begin{equation}
N_X = \frac{M_X}{A_X m_u},
\end{equation}

where  $N_X$ is the total number of atoms of element X, $M_X$ is the total yield, $A_X$ is the atomic mass number, and $m_u$ is the atomic mass unit defined as $1.66\times 10^{-24} \, \text{g}$.

Moreover, the individual contributions of SNIa and SNcc to the number of atoms of element X can be expressed as

\begin{equation}
N_X = aN_{X,\text{SNIa}} + bN_{X,\text{SNcc}},
\end{equation}
where $N_{X,\text{SNIa}}$ and $N_{X,\text{SNcc}}$ are the number of atoms produced in a single SNIa and SNcc event, respectively, and $a$ and $b$ are the multiplicative factors which are the free variables. We can fit this linear relation to our abundance pattern of O, Ne, Mg, Si, S and F to obtain SNIa contribution, 
\begin{equation}
    \frac{\text{SNIa}}{\text{SNIa} + \text{SNcc}},
\end{equation}

which is the SNIa fraction to the total SNe responsible for the enrichment. Using the average abundance pattern in 44 systems from the CHEERS sample, which are mostly galaxy clusters, \citet{Mernier_SNe} calculated that the SNIa contribution to the total enrichment of hot halos is approximately 29--42\%. Similarly, \citet{Gatuzz2023} recently showed that the SNIa contribution to the enrichment of the Virgo ICM, through which M86 travels, is 31--47\%.

In order to find the SNIa and SNcc contribution to the total metal enrichment, we use \texttt{abunfit}\footnote{https://github.com/mernier/abunfit} code \citep{Mernier_SNe} to fit the abundance ratios in the M86 core with different sets of SNcc and SNIa models, and IMF slopes to obtain the best $\chi^2$/d.o.f. ratio. We test a wide range of possibilities for SNe progenitors to comprehensively investigate the SNe enrichment in the hot halo of M86. 

SNcc synthesis occurs when a massive star collapses onto its core, where the ejected metal yields depend on the IMF, the initial metallicity, and the equatorial rotation of the star. For SNcc, we tested (i) \cite{Nomoto_2013} (hereafter Nomoto when referring to the model) model, in which the lower and upper limits are $10 M_{\odot}$ and $50 M_{\odot}$ respectively. We tested the Nomoto model with different initial metallicities $\text{Z}_{\text{init}} = 0.0, 0.001, 0.004, 0.008, 0.02, 0.05$. The Nomoto model assumes zero initial rotational velocity for SNcc progenitors. However, any equatorial rotation induces circulations and consequent mixing of materials within the star. This mixing leads to significant differences in the ejected abundance pattern compared to non-rotating progenitors, as discussed in detail in Sect. \ref{subsect:SNcc}. Therefore, in addition to the non-rotating Nomoto models, we also tested (ii) \cite{Limongi_2018} (hereafter Limongi when reffering to the model) models which includes equatorial rotational velocity for the progenitor star. We used the Limongi models with varying initial metallicities 
$\text{Z}_{\text{init}} = 0.0, 0.001, 0.01, 0.1$ and varying rotational velocities $\text{V}_{\text{rotation}} = 0, 150, 300$\,\text{km/s}. In Limongi models the lower and upper limits are  $13 M_{\odot}$ and $120 M_{\odot}$ respectively. The Limongi models also account for stellar wind contributions to the total elemental yields before the SNcc explosions occur. The stars with $M > 25M_{\odot}$ collapse into black hole entirely, therefore their contribution to the enrichment is limited to the stellar winds.

SNIa explosions, on the other hand, occur when a white dwarf accretes material from its companion star (i.e., single-degenerate), or merge with another white dwarf (i.e., double-degenerate). The explosion can be modelled by three different phases of burning: a deflagration in which the burning propagates subsonically allowing the white dwarf to expand before the burning reaches the outermost regions; a detonation where the burning is supersonic. The third burning scenario is the delayed-detonation where an initial deflagration (slow burning) is followed by detonation (fast burning). The yields differ between these explosion mechanisms \citep[e.g.,][]{Mernier_SNe}.  Therefore we expect different mechanisms to produce different abundance patterns. In this study, we tested 3D pure deflagration models of \cite{Fink_2014} (hereafter Fink) and 3D delayed-detonation models of \cite{Seitenzahl_2012} (hereafter Seitenzahl). Both of the models use varying ignition spots within the central ignition region deep inside the white dwarf which varies as $\text{N}_{\text{s}}=1, 3, 5, 10, 20, 40, 100, 150, 200, 300, 1600$, and varying inner core density of $\rho_9=1.0, 2.9, 5.9$, with the units of $10^9 \,\text{g}/\text{cm}^3$.

As for the IMF, we used the Salpeter IMF \citep{Salpater1955} with $\alpha=-2.35$, and also tested intermediate and heavy IMFs with $\alpha=-2, -1.8, -1.5$.

We emphasize that, our analysis holds supernova parameters of IMF slope, progenitor metallicity, explosion geometry, and stellar rotation are fixed for all progenitors, even though in reality each of these properties varies for each star. To definitively identify explosion scenarios in M86 is out of scope of this paper. Our goal in this section is to compare how these SN models with fixed parameters that are previously constrained by abundance patterns in closed‐box galaxy clusters perform when applied to an open system like M86.

\begin{table}
    \centering
    \sisetup{table-format=1.2} 
    \begin{minipage}{\columnwidth}
    \captionsetup{justification=centering} 
    \captionof{table}{The 5 best-fitting results of SNe model combinations in M86 core with increasing $\chi^{2}\text{/d.o.f.}$ ratio. For SNcc yields we used Nomoto with varying initial metallicity, and Limongi models with varying initial metallicity and equatorial rotational velocity. As for SNIa yields, we used Seitenzahl and Fink models with varying number of ignition spots. See the text for the complete explanation.}
    \centering
    \sisetup{table-format=1.2} 
    \begin{tabular}{lllS[table-format=1.2]S[table-format=-1.2]S[table-format=1.2]l}
        \hline
        \hline
        SNcc & SNIa & IMF Slope & \(\frac{\text{SNIa}}{\text{SNIa} + \text{SNcc}}\) &  \hspace{4mm} \(\chi^{2}\)\text{/d.o.f.}\\  \hline

       Nomoto  & Seitenzahl & & & & \\ \hline
      Z0.02 &   N5    & -1.50 & 0.21 & 3.93 / 4 \\
      Z0.02 &  N3    &  -2.00 & 0.17 & 3.99 / 4 \\
      Z0.001 &   N10   & -1.80 & 0.25 & 4.03 / 4 \\
       Z0.02  &  N1    & -2.35 & 0.15 & 4.08 / 4 \\
     Z0.001 &   N3    &  -1.50 & 0.24 & 4.14 / 4 \\
                \hline
        Nomoto & Fink & & & \\ \hline
       Z0.001 & N10  &  -2.35 & 0.56 & 4.02 / 4 \\
      Z0.001 &  N5   &  -2.00 & 0.62 & 4.11 / 4 \\
     Z0.02  &   N1   &  -1.50 & 0.87 & 4.15 / 4 \\
      Z0.001 &   N1   & -2.00 & 0.88 & 4.18 / 4 \\
      Z0.02  &   N5   & -1.50 & 0.59 & 4.19 / 4 \\

            \hline
                        \hline

       Limongi & Seitenzahl & & & & \\ \hline
      Z0\_V0 &   N1    & -1.50 & 0.07 & 6.23 / 4 \\
      Z0\_V0 &  N3    &  -1.50 & 0.07 & 6.92 / 4 \\
      Z0\_V0 &   N1   & -1.80 & 0.08 & 6.96 / 4 \\
       Z0\_V0  &  N1    & -2.00 & 0.08 & 7.50 / 4 \\
     Z0\_V0 &   N3    &  -1.80 & 0.09 & 7.75 / 4 \\ 
                 \hline
       Limongi  &Seitenzahl  & & & & \\ \hline
      Z0\_V150 &   N1    & -1.50 & 0.08 & 25.34 / 4 \\
      Z0\_V150 &  N3    &  -1.50 & 0.09 & 25.75 / 4 \\
      Z0\_V150 &   N5   & -1.50 & 0.09 & 26.39 / 4 \\
       Z0\_V150  &  N10    & -1.50 & 0.10 & 26.92 / 4 \\
     Z0\_V150 &   N1    &  -1.80 & 0.09 & 27.63 / 4 \\ 
                 \hline
       Limongi  &Seitenzahl  & & & & \\ \hline
      Z0\_V300 &   N1    & -1.50 & 0.12 & 26.81 / 4 \\
      Z0\_V300 &  N3    &  -2.00 & 0.16 & 27.63 / 4 \\
      Z0\_V300 &   N5   & -1.50 & 0.14 & 28.82 / 4 \\
       Z0\_V300  &  N1    & -1.80 & 0.13 & 29.08 / 4 \\
     Z0\_V300 &   N10    &  -1.50 & 0.14 & 29.78 / 4 \\ 
                 \hline
                 \hline
       Limongi  & Fink & & & & \\ \hline
      Z0\_V0 &   N1    & -1.50 & 0.70 & 7.90 / 4 \\
      Z0\_V0 &  N5    &  -1.50 & 0.57 & 8.02 / 4 \\
      Z0\_V0 &   N1600   & -1.50 & 0.33 & 8.36 / 4 \\
       Z0\_V0  &  N10    & -1.50 & 0.73 & 8.64 / 4 \\
     Z0\_V0 &   N200    &  -1.50 & 0.30 & 8.74 / 4 \\ 
                 \hline
       Limongi  &Fink  & & & & \\ \hline
      Z0\_V150 &   N100H    & -1.50 & 0.21 & 24.86 / 4 \\
      Z0\_V150 &  N40    &  -1.50 & 0.22 & 25.95 / 4 \\
      Z0\_V150 &   N1   & -1.50 & 0.73 & 26.09 / 4 \\
       Z0\_V150  &  N100    & -1.50 & 0.21 & 26.20 / 4 \\
     Z0\_V150 &   N5    &  -1.50 & 0.37 & 26.25 / 4 \\ 
                 \hline
       Limongi  &Fink  & & & & \\ \hline
      Z0\_V300 &  N100H    & -1.50 & 0.29 & 26.94 / 4 \\
      Z0\_V300 &  N40    &  -1.50 & 0.31 & 28.10 / 4 \\
      Z0\_V300 &   N1600   & -1.50 & 0.28 & 28.18 / 4 \\
       Z0\_V300  &  N100    & -1.50 & 0.29 & 28.36 / 4 \\
     Z0\_V300 &   N5    &  -1.50 & 0.47 & 28.42 / 4 \\ 
        \hline
    \end{tabular}
    \label{tab:SNe_table}
    \end{minipage}
    \hfill
\end{table}

In Table \ref{tab:SNe_table}. we present the 5 best-fitting results for each selected set of SNe models and IMFs combinations with increasing $\chi^{2}\text{/d.o.f.}$ ratio. We first present the best-fit results from the \texttt{Nomoto}+\texttt{Seitenzahl} and \texttt{Nomoto}+\texttt{Fink} model combinations. Then, in the same table, we present the separated Limongi results based on the rotational velocity of the SNcc progenitor, combined with SNIa models. We show separately, the $\texttt{Limongi}_\texttt{0 km/s}+\texttt{Seitenzahl}$, $\texttt{Limongi}_\texttt{150 km/s}+\texttt{Seitenzahl}$, and $\texttt{Limongi}_\texttt{300 km/s}+\texttt{Seitenzahl}$ combinations. The "Z" in the model names refer to the initial metallicity, while "V" stands for the velocity. Finally we present the best-fitting result obtained from the $\texttt{Limongi}_\texttt{0 km/s}+\texttt{Fink}$, $\texttt{Limongi}_\texttt{150 km/s}+\texttt{Fink}$, and $\texttt{Limongi}_\texttt{300 km/s}+\texttt{Fink}$ combinations. In the SNIa models, the "N" stands for the number of ignition spots. The best-fits presented in this table use inner core density $\rho_9=2.9\times10^9 \,\text{g}/\text{cm}^3$ for all SNIa models except N100H, which has 100 ignition spots and $\rho_9=5.5\times10^9 \,\text{g}/\text{cm}^3$. 

The discussion based on these results for SNcc progenitors is presented in the next Sect. \ref{subsect:SNcc}, while the SNIa models are compared in Sect. \ref{subsect:SNIa}. Finally, the overall best-fitting results of yields of SNe and AGB stars are discussed in \ref{subsect:SNe_AGB_conclusion}.

\subsubsection{SNcc explosions: How do galaxy-galaxy interaction and ram-pressure stripping affect the SNcc enrichment?}
\label{subsect:SNcc}

  


Both galaxy-galaxy interaction \citep{SF_Larson_1978,SF_Kennicutt_1987,SF_Mihos_1996,SF_Donzelli_1997,SF_Moreno_2019} and ram-pressure stripping \citep{sf_2003,sf_2006,sf_2016,sf_2018} are known to enhance the star formation, before the quenching due to gas loss. In this section, based on our results listed in Table \ref{tab:SNe_table}, we discuss the possible effects of such mechanisms onto SNcc enrichment. 

M86 does not show significant far-ultraviolet emission \citep{Armando_2017}, indicating very little star formation. However, given that M86 experienced galaxy-galaxy interaction with NGC\,4438, and extreme supersonic ram-pressure stripping, it is possible that the resulting star-forming and turbulent environment may have triggered the formation of fast-rotating SNcc progenitors for a period, which might have significantly enriched the present hot halo of M86 galaxy core. 


In rotating SNcc progenitors, elemental abundances of $\alpha$-elements differ significantly from their non-rotating counterparts due to rotation-induced mixing in stars. The pressure difference between the poles and the equator of a rotating star, along with the resulting thermal imbalance, induces large-scale currents known as meridional circulation — or Eddington–Sweet circulation \citep{Eddington_rot,Sweet_rot}. Meridional circulation, and shear instabilities in local scales, transport chemically processed material (e.g., He, C, O) from the core-burning regions outward, while simultaneously bringing H-rich material from the envelope inward \citep{Maeder_2009}. This increase in core mass extends the duration of He-, C-, and Ne-burning. As a result, the yields of \(^{16}\text{O}\), \(^{20}\text{Ne}\), and \(^{24}\text{Mg}\), which are products of these stages, are enhanced. The enhancement of \(^{28}\text{Si}\) and \(^{32}\text{S}\) production due to the increased core mass, on the other hand, is less pronounced due to the shorter timescale of the later stages of O-burning and explosive Si-burning. Nevertheless, depending on the IMF and metallicity of the SNcc progenitor, the rotation is able to increase mass ejection from the inner shells during the explosion. This increased mass ejection could result in an even greater enhancement of Si and S abundances compared to O, Ne, and Mg, which would otherwise be trapped in the remnant. Therefore, the rotation of SNcc progenitors causes different abundance patterns compared to their non-rotating counterparts, which could be tested using the Nomoto and Limongi models.

First, we note that the abundance yields obtained from the non-rotating models of Limongi (v0 models) and the Nomoto models differ. Limongi integrates over a wider range with more massive stars (described in Sect. \ref{subsection:SNE_head}), and also uses different stellar evolution code compared to Nomoto, which produces higher Fe yields due to less fallback \citep{Nomoto_2013,Limongi_2018}. 



Second, we observe that when no equatorial rotation is introduced, both Nomoto and Limongi provide better statistics when the 3D delayed-detonation Seitenzahl models are used over 3D pure deflagration models of Fink for SNIa explosions (see Table \ref{tab:SNe_table}). However, when rotation is applied with Limongi models, the Seitenzahl and Fink models result in comparable statistics in these model combinations. The differences between SNIa models of Seitenzahl and Fink are further discussed in Sect. \ref{subsect:SNIa}.

Third, when the equatorial rotation is introduced, the best-fits provide poorer statistics in Limongi models. We see that, in both Seitenzahl and Fink combinations, the non-rotating models provide the best statistics and 150 km/s models provide slightly better statistics over 300 km/s ones. 


Last but not least, our results suggest that the best fits to the Limongi models are obtained when assuming zero initial metallicity ($\text{Z}_{\text{init}} = 0$) across all model combinations. Consistent with a zero-metallicity environment—and the resulting inefficient cooling and limited fragmentation—we observe a heavy ($\alpha=-1.5$) IMF for all SNcc progenitor stars. This result from Limongi is in tension with our findings based on the Nomoto models, which provide better statistical fits over Limongi, and instead suggest either low ($\text{Z}_{\text{init}} = 0.001$) or Solar ($\text{Z}_{\text{init}} = 0.02$) initial metallicities, along with either intermediate-heavy ($\alpha=-2, -1.8, -1.5$) or Salpeter ($\alpha=-2.35$) IMFs.

We caution the reader that our time-invariant IMF assumption is a simplification intended to compare trends between closed-box galaxy clusters and the relatively open M86 system. Similarly, the rotation and metallicity of SNcc progenitors are assumed to be uniform both spatially and temporally. In reality, however, the picture is more complex, as these parameters vary between stars and between different evolutionary episodes of M86. Our aim is not to determine the exact explosion scenario but rather to test a wide range of possibilities. By doing so, we show that SNcc enrichment in M86 is consistently high. We find that given the delayed-detonation SNIa explosions, which are preferred by the supernovae community, our SNIa contribution result for M86 from the best fits lies in the range of 8--25\%. This contribution ratio is significantly lower than the 31--47\% contribution in the Virgo Cluster \citep{Gatuzz2023}, 29--42\% of the averaged abundance sample from the CHEERS sample, consisting mostly of clusters \citep{Mernier_SNe}. The higher contribution of SNcc enrichment in the hot gas of M86 suggests an increased starburst and subsequent SNcc explosions in the galaxy. Our results indicate that the ram-pressure stripping and galaxy-galaxy interaction induced a starburst episode, which led to the increased SNcc enrichment and high SNcc contribution.

\subsubsection{SNIa explosions: Discrepancy between SNIa models based on their explosion symmetry }
\label{subsect:SNIa}

\begin{figure}
    \centering
    \includegraphics[width=0.9\columnwidth]{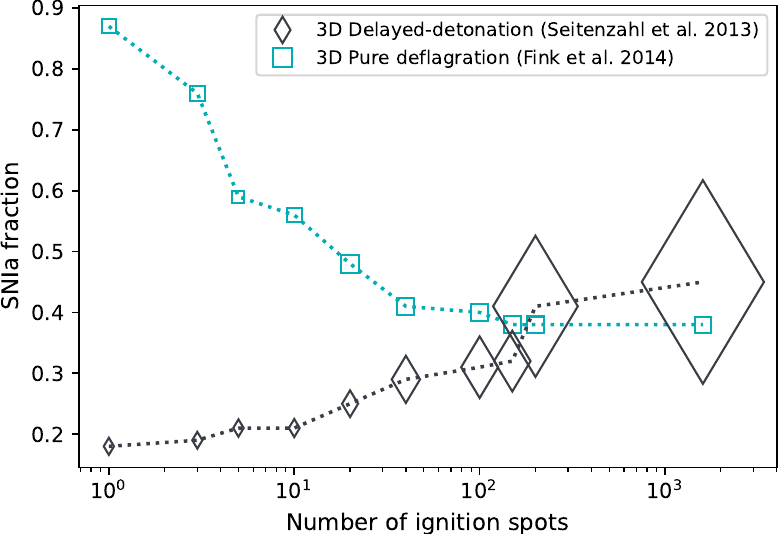}
    \caption{SNIa fraction variations in 3D pure deflagration and delayed-detonation models based on their number of ignition spots. The size of each marker is proportional to the $\chi^2/\text{d.o.f.}$ of that fit. Lower numbers of ignition spots result in geometrically asymmetric explosions, while high numbers indicate a more symmetric explosion.}
    \label{fig:SNe_compar}
\end{figure}

The best statistics for both SNIa models of pure deflagration Fink and delayed-detonation Seitenzahl are obtained when a small number of ignition spots (1, 3, 5, 10) are equipped. The physical consequence of a small number of ignition spots is that the initial deflagration burns only a small fraction of the total mass with moderate expansion, leaving a high central density at the deflagration-to-detonation transition state, which then burns into Fe-group elements abundantly during the detonation. Therefore, the abundance of intermediate-mass elements, such as $^{28} \text{Si}$ and $^{32} \text{Si}$, increases with the number of ignition spots \citep{Seitenzahl_2012}, whereas Fe-peak elements decrease. Given the low Si/Fe ratio in the hot gas of the M86 galaxy core, a small number of ignition spots better explain our observed abundance pattern. In more general terms, the explosion models with a small number of ignition spots have a more asymmetric burning geometry \citep{Seitenzahl_2012}; asymmetric burning leads to an asymmetric configuration of elements, with higher Fe-peak element production and lower yields of intermediate-mass elements like Si, while symmetric burning results in more uniform element formation. 

Our results in Table \ref{tab:SNe_table} indicate a discrepancy in the SNIa contributions between 3D pure deflagration and delayed-detonation models if the explosion is asymmetric (i.e., a low number of ignition spots). We see that, based on our 5 best-fitting results, the required SNIa contribution is between 59--88\% for Fink models, and 15--25\% for Seitenzahl models. Similarly, if we limit the IMF Slope to be only Salpeter IMF (i.e., $\alpha=-2.35$), the same discrepancy between burning mechanisms is still prominent for 5 best-fitting results obtained from a Salpeter IMF.

To investigate the discrepancy further, we fit our abundance pattern for the Fink and Seitenzahl models with all possible ignition spots, using a fixed initial metallicity of 0.001 from the Nomoto model, a Salpeter IMF, and an inner core density of $\rho_9=2.9\times10^9 \,\text{g}/\text{cm}^3$. In Fig. \ref{fig:SNe_compar}, we present the change in the SNIa fraction with the number of ignition spots ($\text{N}{\text{s}}$) for the Fink and Seitenzahl models, with marker sizes proportional to the $\chi^2/\text{d.o.f.}$ of each fit. We observe that the $\chi^2/\text{d.o.f.}$ increases with the increasing $\text{N}_{\text{s}}$ for the 3D delayed-detonation Seitenzahl models\footnote{The $\text{N}_{\text{s}}\!=\!1$ provides $\chi^2/\text{d.o.f.}\!=3\!.32/4$, while $\text{N}_{\text{s}}\!=\!1600$ results in 39.51/4}, while 3D pure deflagration Fink models are more consistent in terms of $\chi^2/\text{d.o.f.}$. Additionally, we see that the discrepancy decreases as $\text{N}{\text{s}}$ increases, reaching the best agreement between models at $\text{N}{\text{s}}=200$. The SNIa fraction for the $\text{N}_{\text{s}}=200$ fit is found to be approximately 0.40 for both models, with $\chi^2$/d.o.f. ratios of 4.45/4 for Fink and 27.38/4 for Seitenzahl, respectively. Such a high number of ignition spots ($\text{N}_{\text{s}}\gtrsim20$) in the Fink deflagration model—which results in a symmetric pure deflagration with high expansion, slow burning, and reduced $^{56}\text{Ni}$—does not resemble any observed class of SNIa \citep{Fink_2014}. However, when detonation burning is introduced, this combination successfully explains observed normal SNIa explosions \citep{Ropke_2012,Sim_2013}. A lower number of ignition spots, on the other hand, can only be explained by the brightest SNIa explosions \citep{Seitenzahl_2012}. 

Based on our SNIa contribution results over a wide range of progenitor parameters, we observe the following: (i) When SNcc progenitor parameters are allowed to vary freely, based on the best fits, the 3D delayed-detonation models --which are preferred by the supernova community-- result in consistently low SNIa and high SNcc contributions, for both Nomoto and Limongi. (ii) When SNcc progenitor parameters are fixed and only the SNIa model parameters are varied, the delayed-detonation and deflagration models converge at an SNcc contribution of $\sim\!60\%$. However, the fit quality is poor in this case for the delayed-detonation model, and acceptable fits are only achieved when the SNcc contribution exceeds $\sim\!70\%$.

\subsubsection{SNe and AGB yields in the M86 core}
\label{subsect:SNe_AGB_conclusion}

\begin{figure}
    \centering
    \includegraphics[width=0.9\columnwidth]{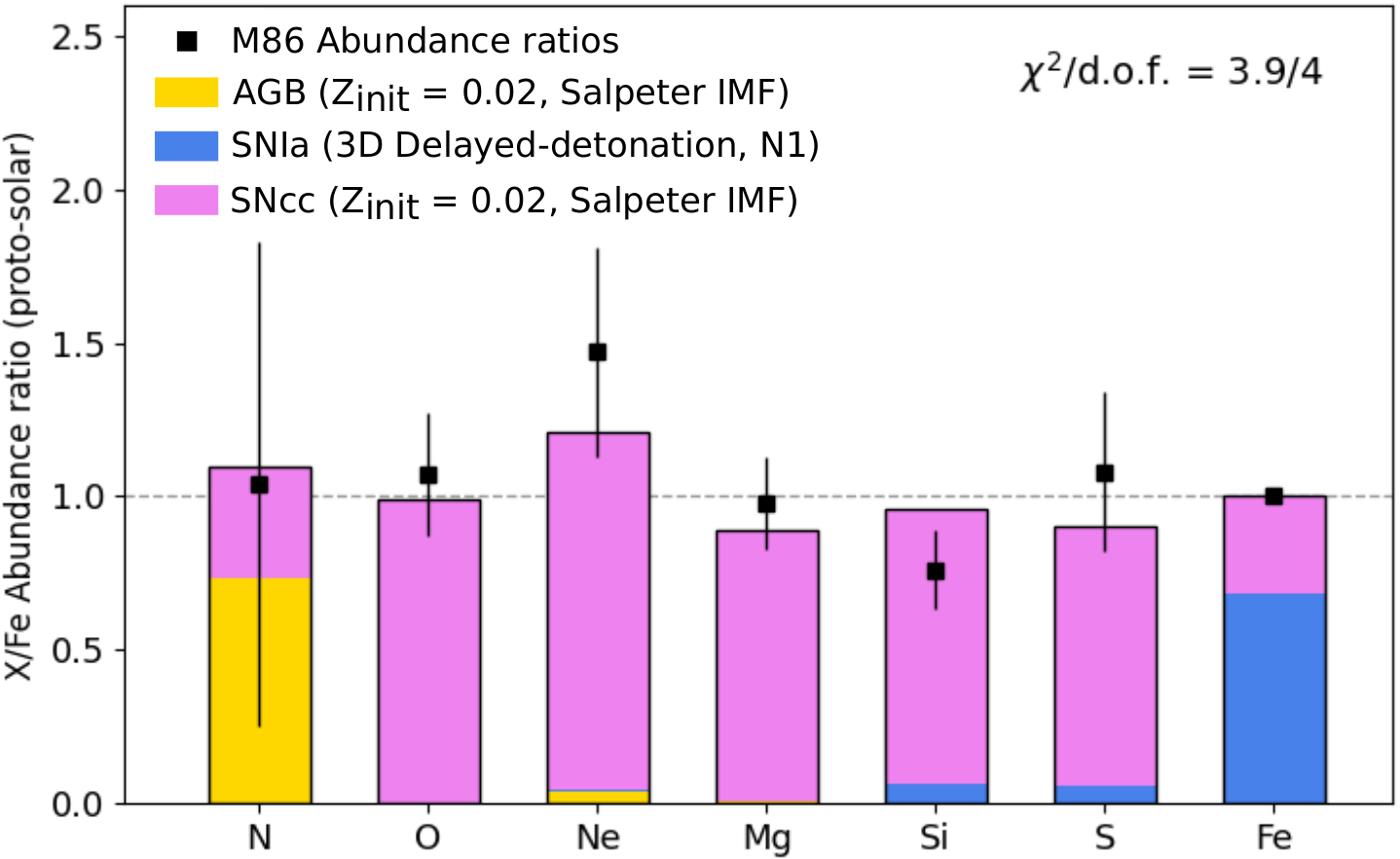}
    \caption{SNe and AGB model fit of the M86 abundance pattern. For SNIa, the Seitenzahl N1 (single ignition spot) model is used, while for SNcc and AGB, we use Nomoto models with Solar initial metallicities ($\text{Z}_\text{init}=0.02$).}
    \label{fig:SN_bars}
\end{figure}




After fitting the SNIa and SNcc models to our abundance pattern of SNe products, we incorporated asymptotic giant branch (AGB) star models to fit our N measurement as well. AGB stars produce light elements like N at the end of their life and enrich their surroundings through stellar winds. To account for the combined contributions of AGB stars, SNcc, and SNIa, we used AGB models from \cite{Nomoto_2013}, adapted from \cite{Karakas_2010}, and refit our data with the inclusion of the N/Fe ratio.

We assumed the initial metallically is the same for SNcc progenitors and AGB stars. In our best-fitting combination, we used Nomoto SNcc model, Salpeter IMF; and Solar initial metallicity ($\text{Z}_{\text{init}} =0.02$) for both SNcc progenitors and AGB stars. For the SNIa component, we used the Seitenzahl model with a single ignition spot ($\text{N}_{\text{s}}=1$); the result is presented in Fig. \ref{fig:SN_bars}, where the colored bars represent the contributions of AGB or SNe production based on the given combination of models, while the black data points correspond to our measured M86 abundance ratios. We caution the reader that our approximations assume that the SNe models with fixed configurations from our best-fits dominate the overall enrichment, while in reality, it may involve different generations of stars and configurations. Nonetheless, our best-fitting results suggest a high SNcc enrichment in M86 galaxy core with a SNcc contribution of 75--85\% based on 3D delayed-detonation SNIa models.





\subsection{Origins of the emission structures: ram-pressure stripping or galaxy-galaxy interaction?}
In M86, we observe an (i) X-ray emitting tail with a true length of >380 kpc, which is almost twice the projected length, according to the orbital estimations of \citet{Randall_2008}, (ii) a plume at the north without any stellar counterparts which has the highest absolute abundances in our analysis, and (iii) a northeastern structure with the lowest absolute abundances. Formation histories of these emission structures in M86 are still in debate. With our elemental abundance results, we aim to constrain the possible scenarios in this section. The ram-pressure stripping occurs when the ram-pressure surpasses the thermal pressure. M86 group enters into Virgo Cluster with $-1551\ \mathrm{km\,s^{-1}}$ velocity and the Virgo ICM has electron density at 350 kpc from the center of Virgo Cluster of $n_e \approx 2.0\times10^{-4}\,\mathrm{cm^{-3}}$ \citep{Urban_2011}. Therefore we estimate the ram-pressure applied to M86 to be $P_{\text{ram}}\approx10^{-11}\,\text{dyne}\,\text{cm}^{-2}$. Similarly, assuming the electron density of M86 core  $n_e \approx 6.2\times10^{-3}\,\mathrm{cm^{-3}}$ \citep{Randall_2008}, the thermal pressure is found to be $P_{\text{th}}\approx3.8\times10^{-13}\,\text{dyne}\,\text{cm}^{-2}$. Since $P_{\text{ram}}\gg P_{\text{th}}$, the hot halo of M86 experiences strong ram-pressure stripping as it moves supersonically in Virgo ICM. 


Furthermore, the neighboring galaxy NGC\,4438, located east of M86 with a projected separation of 120 kpc, is a large spiral galaxy that also exhibits strong morphological disturbances, particularly in its outer stellar disk, where extraplanar tidal arms indicate a past gravitational interaction. Both NGC\,4438 and M86 are blueshifted with respect to M87, infalling toward the Virgo cluster from the far side, with clustercentric line-of-sight velocities of approximately $-1237\ \mathrm{km\,s^{-1}}$ and $-1551\ \mathrm{km\,s^{-1}}$, respectively. \cite{Kenney_2008} reported a projected 120 kpc H$\alpha$ bridge between NGC\,4438 and M86, and using this distance they estimate an average plane-of-sky velocity of $800\ \mathrm{km\,s^{-1}}$ for NGC\,4438, and a timescale of $\sim\!100$ Myr since the closest encounter with M86. In the rest frame of M86, they propose a collision scenario in which NGC\,4438 enters the M86 system from the front with a relative line-of-sight velocity of  $\sim\!314\ \mathrm{km\,s^{-1}}$ and a transverse velocity of $\sim\!800\ \mathrm{km\,s^{-1}}$ toward the east.


Based on the ram-pressure stripping and galaxy-galaxy collision histories of M86, combined with our results on the thermal and chemical structures in M86, we constrain the formation and evolution scenarios of the three emission structures in the following sections.

\subsubsection{Tail}



The X‑ray tail of M86 extends to the northwest—exactly opposite the direction implied by any collision with NGC\,4438. Furthermore, from the true length of the tail and velocity of M86, \citet{Randall_2008} estimate a lower limit of the formation timescale of $\gtrsim230$ Myr.  Since the encounter with NGC\,4438 occurred only $\sim100$ Myr ago, the tail must have been in place long before that collision. Thus, the tail is most naturally explained by ram‑pressure stripping of M86’s hot halo as it moves through the Virgo ICM. Nevertheless, a subsequent interaction between the ram-pressure stripped tail of M86 and NGC\,4438 may explain why the X-ray morphology of this tail differs from that of other elliptical galaxies in the Virgo and Fornax clusters undergoing ram-pressure stripping, which typically exhibit sharp surface brightness edges aligned with centrally trailing tails \citep{Machacek_2005,Machacek_2006a}. Furthermore, \citet{Finoguenov_2004} showed that the tail is low-entropy. Similarly, we showed that the tail regions, despite their considerably large extraction areas, are reasonably isothermal, with a Gaussian temperature distribution characterized by a width $\sigma_\text{kT}$ of $\sim5\%$ of the mean temperature $\text{kT}_\text{mean}$. The narrow temperature distributions suggest that the tail has had sufficient time to reach thermal equilibrium.

The facts that the Mg/Fe and S/Fe ratios in the tail are significantly different from those in the core, along with the absence of temperature variations expected from the mixing of continuously stripped core gas, seem to contradict the simple stripping scenario. However, we remind the reader that based on the orbital calculations in \citet{Randall_2008} in which M86 enters into Virgo cluster towards the observer and from northwestern direction, the lower limit on the actual length of the extended tail is more than twice the projected length. Therefore, tail sections that are spatially closer to the core—where mixing is expected—are very difficult to constrain in the projected image of M86, in which the plume is also another dominant emission source. Moreover, we note that it is also possible that, in addition to the galaxy core, the plume is being stripped and contributes to the formation of the visible tail. Thus, our Tail 1 and Tail 2 regions, which have large extraction areas, represent integrated stripped gas over a large physical volume and a long formation timescale, allowing them to reach thermal equilibrium.

Notably, the tail regions show increased Mg/Fe ratios compared to the plume and the galaxy core. Furthermore, we observe that the more distant tail region, Tail 2, shows enhanced Mg/Fe, Si/Fe and S/Fe ratios compared to the closer tail region, Tail 1. This result is in agreement with the ram-pressure stripping of gas and the consequent increase in star formation in the turbulent tail regions, accompanied by the increased SNcc explosions and $\alpha$ elements.

\subsubsection{Plume}
 

In this study, we have shown that the plume in M86 has the highest absolute abundances for all elements in our measurements and the Mg/Fe ratio in the plume is in excellent agreement with that of the outer, less dense group gas and the Virgo ICM, while it significantly deviates from the Mg/Fe ratio in the core. Since M86 does not harbour an AGN, the possible mechanisms to create the plume structure are the ram-pressure stripping, the galaxy-galaxy interaction, the SNe explosions inside the galaxy core due to a starburst, and/or the combinations of these hot gas perturbations. A more exact computation of the impact of these mechanisms would require 3D simulations, which is out of scope of this paper. Instead, we perform simpler order of magnitude comparisons and together with our elemental abundance results we aim to constrain these scenarios. First, let us assume a scenario in which the gas is uplifted to a distance of 20 kpc from the galaxy core. The energy required for a such lifting is,
\begin{equation}
E_{\rm lift} \;=\; \Delta E_{\rm binding} \;=\;
G\,M_{\rm plume}\,M_{\rm M86}\,
\left(
      \frac{1}{r_{0}}
      -\frac{1}{r_{\rm plume}}
\right),
\end{equation}
where  $r_{0}$ is the initial average distance to the galaxy center of the uplifting gas. Note that this energy equation is oversimplified for an order of magnitude comparison and assumes all uplifting gas is initially concentrated as a shell at  $r_{0}$ and the enclosed mass is constant. If we assume the half of the effective radius of M86 as the average initial distance, which is $\sim\!5$ kpc, we find $\text{E}_{\text{lift}}$ = \(\sim\!4\times10^{57}\,\mathrm{erg}\) using the total M86 enclosed mass at effective radius from \citet{Cappellari2013} and the hot gas mass of the plume from \citet{Randall_2008}. The other energy components of such displacement are the thermal heating energy due to slightly higher temperature of plume compared to the core, and the kinetic energy in the rest frame of M86 galaxy core based on the 20 kpc distance and the fact that there is no shock fronts related to the plume and the motion is subsonic. However, we find these components to be one order of magnitude less than $\text{E}_{\text{lift}}$, therefore we ignore these terms. 

If we assume the galaxy-galaxy collision or ram-pressure stripping induced a starburst in the galaxy core, it is expected that SNcc events and stellar winds eject material into the surroundings. In order to test the pure SNe scenario of the plume origin, let us assume the most extreme case, in which all the ejected material is placed at north of M86 core. We adopt the mechanical energy injection rate from SNcc and stellar winds as given by \citet{Veilleux_2005} to compare this feedback energy with the gas lifting requirement. Assuming a star formation rate of 1\(M_\odot\,\mathrm{yr}^{-1}\), 
a duration of 100\,Myr, and a thermalization efficiency of 10\%, we find that the total mechanical energy output of such a starburst is \(\sim\!2 \times 10^{56}\,\mathrm{erg}\), which is not comparable with the uplifting energy. We see that the origin of the plume cannot be explained by such a starburst alone. Nevertheless, M86 orbits toward the southeast direction, and this gas flow can displace the significant amount of ejected material to the north, enriching the environment. Therefore, this ram-pressure and/or galaxy-galaxy interaction induced starburst can explain the high absolute abundances in the plume.

As for the ram-pressure stripping scenario of the plume, \citet{Randall_2008} report that stripping such a single blob is not possible with a spherical gravitational potential of M86, but it is only possible with an aspherical potential. In our study we find that, similar to the tail, the plume is remarkably isothermal and low in entropy. If the X-ray bright plume had recently formed from the stripping of the dense galaxy core and continued to receive inflow from the core, its temperature structure would be expected to show variations due to ongoing mixing of the core gas. Moreover, the Mg/Fe ratios in the plume and core are significantly different. Therefore, we conclude that the plume has not originated recently from the stripping of the dense core. Nevertheless, if the plume structure formation started at the same time as the tail as M86 enters into Virgo cluster as \citet{Randall_2008} suggested, it should be formed from the same CGM gas of M86 as the tail 200--300 Myr ago. Therefore, if the plume has originated from the CGM of that period, it is expected to preserve its chemical composition while the galaxy core has continuously been enriched by SNIa products. The Mg/Fe ratios in the Virgo ICM, outer M86 hot gas, and the plume are $\sim\!1.39$ Solar. The ratio in the core, on the other hand, is $\sim\!1.03$ Solar. Therefore, in principle, the ongoing Fe production in the galaxy core might change the "initial" Mg/Fe ratio of $\sim\!1.39$ Solar to a $\sim\!1.03$ Solar over time. However, this scenario is entirely based on a "Fe excess" in the galaxy core due to ongoing SNIa events. The facts that the plume and the core has comparable hot gas masses \citep{Randall_2008} and that we observe $\sim\!1.7$ times more Fe absolute abundance in the Plume compared to the Core is in contrast with this scenario\footnote{We note that our Core region is considerably smaller than the core region used by \citet{Randall_2008} for their mass determination. However, our Region 3, which is a circular core region that surrounds and excludes the inner Core region, is roughly similar in size to theirs, and the Fe abundances in our Region 3 and Core region are consistent within 1$\sigma$ ($0.74\pm0.06$ and $0.66\pm0.04$ respectively). Therefore, the interpretation that the core contains less Fe mass than the plume remains valid.}. 

The galaxy–galaxy collision scenario provides a natural explanation for the increased Mg/Fe ratio observed in the plume. A strong perturbation from the close passage of NGC\,4438, combined with the turbulent conditions already present due to ram-pressure stripping, could effectively displace and compress M86’s CGM. \citet{Kenney_2008} estimate the kinetic energy of the gas stripped from NGC\,4438 during its interaction with M86’s hot halo to be \(5\times10^{58}\,\mathrm{erg}\), a value comparable to the total thermal energy of the M86 CGM. This level of energy, combined with ram-pressure flows, can facilitate the removal of the outer halo gas with super-Solar Mg/Fe ratio. Therefore, the observed chemical composition of the plume can be understood as the displaced, Mg rich outer halo gas. The caveat of this scenario is also the high absolute abundances in the plume, which is not observed in the outer halo gas. However, high absolute abundances can be explained by the gas loss from NGC\,4438 during its passage and an increased starburst. We can speculate that the interaction of M86 halo with NGC\,4438 ISM can induce a starburst and a subsequent SNcc enrichment in both NGC\,4438 and M86, which can enrich the M86 halo medium during the passage. If the galaxy-galaxy collision and ram-pressure induced turbulent environment caused the formation of the plume, the plume might act like a repository for these ejected metals. Our high SNcc contribution result in Sect. \ref{subsection:SNE_head} supports this argument. 

In conclusion, a pure ram-pressure stripping scenario is unable to explain the high absolute abundances in the plume. The combination of ram-pressure stripping and galaxy-galaxy collision with NGC\,4438 can explain the high absolute abundances and elevated Mg/Fe ratios in the plume, under the assumptions of the plume has originated from the perturbation of the outer CGM of M86 and further enriched by the starbursts in M86 and NGC\,4438 due to ram-pressure, which is supported by our high SNcc contribution results. We note that tailored simulations are necessary to completely understand the plume phenomena in M86.

\subsubsection{The north-eastern arm}
The north-eastern arm spatially corresponds with the H$\alpha$ emission structures and the M86 edge of the H$\alpha$ bridge between NGC\,4438. The absolute abundances and abundance ratios within the north-eastern arm are quite peculiar. We observe that this structure has the lowest absolute abundances for all elements. It shows a Mg/Fe ratio comparable to that in the core but a Si/Fe ratio more than $1\sigma$ higher than in the core. The S/Fe ratio in the north-eastern arm agrees with the S/Fe ratio everywhere except in the core. It is possible that this region, which is not isothermal like the plume and tail, is highly disturbed due to both galaxy-galaxy collision and ram-pressure stripping, and continuously mixed with the surrounding material.

\section{Conclusions}
\label{sec:conclusion}

In this work, we used a deep (85.6 ks) \textit{XMM-Newton} observation to derive the abundances of N, O, Ne, Mg, Si, S, and Fe in the core, as well as the radial distribution of Mg, Si, S, and Fe abundances and their variations within the emission structures of the M86 galaxy group for the first time. In addition to the group halo \citep{Bohringer_1994}, M86 exhibits an X-ray-emitting plume to the north \citep{Forman_1979}, an arm to the northeast \citep{Rangarajan_1995}, and a long tail of emission \citep{Randall_2008}. The origins of these structures have been debated, with possible explanations including ram-pressure stripping \citep[e.g.,][]{Forman_1979,Randall_2008} and/or a galaxy-galaxy collision with NGC\,4438 \citep[e.g.,][]{Gomez_2010,Finoguenov_2004}. Through careful modeling, we utilized the EPIC MOS, EPIC pn, and RGS instruments to study the chemical enrichment of this system. Our results can be summarized as follows.

\begin{itemize}

    \item We observe an overall Solar abundance ratios of $\alpha$-elements with respect to Fe in the M86 galaxy core, similar to the rest of the hot gaseous content of the Universe than to its stellar population. This is in contrast with  ram-pressure stripped and AGN-hosting elliptical galaxy M89 \citep{Kara_2024}, and in line with the ram-pressure stripped NGC\,1404 \citep{Mernier2022b} which does not harbor an AGN. We conclude that, the loss of the "primordial" galactic hot atmosphere and its replenishment via stellar mass-loss products may require AGN activity with radio-mechanical energy comparable to the binding energy of the host system in addition to ram-pressure stripping.
    

    \item  Our analysis of supernova explosion models, based on the abundance pattern in the core of the M86 galaxy, suggests that the SNcc contribution to the hot gas enrichment is $\gtrsim70\%$, higher than the closed-box galaxy clusters. Our SNcc contribution result suggests that the ram-pressure stripping and galaxy-galaxy interaction possibly increased the star formation rate and the consequent SNcc explosions.

    \item  The emission structures of plume and tail have significantly different Mg/Fe ratio compared to the galaxy core. The plume structure has a Mg/Fe ratio 3.3$\sigma$ higher than the core, while our two tail regions show 2.5$\sigma$ and 4.3$\sigma$ higher Mg/Fe than the core. On the other hand, the Mg/Fe in the plume is remarkably comparable with that of the outskirts of the M86 group and also with the Virgo ICM. Therefore, it is possible to conclude that the plume structure did not originate from the hot gas in the galaxy core through stripping, but rather from the group gas due to galaxy-galaxy interaction with NGC\,4438. We note that the second tail region shows higher Mg/Fe ratios compared to the Virgo ICM and M86 group halo, possibly due to the dynamical interactions induced enhanced star-formation and SNcc events.


    \item We observe that in the RGS data of low-temperature plasma, the O/Fe and Ne/Fe measurements from single-temperature models are $> 2\sigma$ higher than those from a multi-temperature fit.

    \item We conclude that, based on their chemical composition, neither a pure galaxy–galaxy interaction nor pure ram-pressure stripping alone can explain the emission structures. While the tail possibly originates from ram-pressure stripping, the plume may result from the disruption of low-entropy outer hot gas during the collision of NGC\,4438 combined with ram-pressure flows, rather than from the ram-pressure stripping of the dense galaxy core gas.

\end{itemize}

This work shows that the chemical enrichment patterns in a morphologically disturbed, open system can be very distinct from those of closed-box galaxy clusters. Our results illustrate the importance of studying chemical enrichment in lower-mass systems, both for revealing their dynamical histories and for achieving a more complete understanding of the chemical enrichment history of the Universe.

\section*{Acknowledgements}

The authors sincerely thank the anonymous referee for their thoughtful and constructive feedback, which improved the quality of this paper. SK would like to thank Boğaziçi University BAP, project code 19951, for their support. NW is supported by the GACR EXPRO grant No. 21-13491X. ENE would like to thank Boğaziçi University BAP, project code 13760, for their support.

\section*{Data Availability}

The original data discussed in this article can be accessed publicly from the \textit{XMM-Newton} Science Archive (https://nxsa.esac.esa.int/nxsa-web/). The \textit{XMM-Newton} data are processed using the \texttt{SAS} software (https://www.cosmos.esa.int/web/xmm-newton/sas). Additional derived data products can be obtained from the main author upon request.



\bibliographystyle{mnras}
\bibliography{mnras_template} 

\begin{thebibliography}{}
\makeatletter
\relax
\def\mn@urlcharsother{\let\do\@makeother \do\$\do\&\do\#\do\^\do\_\do\%\do\~}
\def\mn@doi{\begingroup\mn@urlcharsother \@ifnextchar [ {\mn@doi@} {\mn@doi@[]}}
\def\mn@doi@[#1]#2{\def\@tempa{#1}\ifx\@tempa\@empty \href {http://dx.doi.org/#2} {doi:#2}\else \href {http://dx.doi.org/#2} {#1}\fi \endgroup}
\def\mn@eprint#1#2{\mn@eprint@#1:#2::\@nil}
\def\mn@eprint@arXiv#1{\href {http://arxiv.org/abs/#1} {{\tt arXiv:#1}}}
\def\mn@eprint@dblp#1{\href {http://dblp.uni-trier.de/rec/bibtex/#1.xml} {dblp:#1}}
\def\mn@eprint@#1:#2:#3:#4\@nil{\def\@tempa {#1}\def\@tempb {#2}\def\@tempc {#3}\ifx \@tempc \@empty \let \@tempc \@tempb \let \@tempb \@tempa \fi \ifx \@tempb \@empty \def\@tempb {arXiv}\fi \@ifundefined {mn@eprint@\@tempb}{\@tempb:\@tempc}{\expandafter \expandafter \csname mn@eprint@\@tempb\endcsname \expandafter{\@tempc}}}

\bibitem[\protect\citeauthoryear{{Bekki} \& {Couch}}{{Bekki} \& {Couch}}{2003}]{sf_2003}
{Bekki} K.,  {Couch} W.~J.,  2003, \mn@doi [\apjl] {10.1086/379054}, \href {https://ui.adsabs.harvard.edu/abs/2003ApJ...596L..13B} {596, L13}

\bibitem[\protect\citeauthoryear{Biffi et~al.,}{Biffi et~al.}{2017}]{Biffi2017}
Biffi V.,  et~al., 2017, \mn@doi [\mnras] {10.1093/mnras/stx444}, 468, 531

\bibitem[\protect\citeauthoryear{{Binggeli}, {Popescu}  \& {Tammann}}{{Binggeli} et~al.}{1993}]{Binggeli_1993}
{Binggeli} B.,  {Popescu} C.~C.,   {Tammann} G.~A.,  1993, \aaps, \href {https://ui.adsabs.harvard.edu/abs/1993A&AS...98..275B} {98, 275}

\bibitem[\protect\citeauthoryear{B{\"o}hringer, Briel, Schwarz, Voges, Hartner  \& Tr{\"u}mper}{B{\"o}hringer et~al.}{1994}]{Bohringer_1994}
B{\"o}hringer H.,  Briel U.~G.,  Schwarz R.~A.,  Voges W.,  Hartner G.,   Tr{\"u}mper J.,  1994, \mn@doi [Nature] {10.1038/368828a0}, 368, 828

\bibitem[\protect\citeauthoryear{{Bulbul}, {Foster}, {Cottam}, {Loewenstein}, {Mushotzki}  \& {Shafer}}{{Bulbul} et~al.}{2011}]{Bulbul_2012}
{Bulbul} Randall E.~S.,  {Foster} A.,  {Cottam} J.,  {Loewenstein} M.,  {Mushotzki} R.,   {Shafer} R.,  2011, in {Vrtilek} J.,  {Green} P.~J.,  eds, Structure in Clusters and Groups of Galaxies in the Chandra Era. p.~3

\bibitem[\protect\citeauthoryear{{Buote}}{{Buote}}{2000}]{Buote2000}
{Buote} D.~A.,  2000, \mn@doi [\apj] {10.1086/309224}, \href {https://ui.adsabs.harvard.edu/abs/2000ApJ...539..172B} {539, 172}

\bibitem[\protect\citeauthoryear{{Buote} \& {Fabian}}{{Buote} \& {Fabian}}{1998}]{Buote1998}
{Buote} D.~A.,  {Fabian} A.~C.,  1998, \mn@doi [\mnras] {10.1046/j.1365-8711.1998.01478.x}, \href {https://ui.adsabs.harvard.edu/abs/1998MNRAS.296..977B} {296, 977}

\bibitem[\protect\citeauthoryear{{Cappellari} et~al.,}{{Cappellari} et~al.}{2013}]{Cappellari2013}
{Cappellari} M.,  et~al., 2013, \mn@doi [\mnras] {10.1093/mnras/stt562}, \href {https://ui.adsabs.harvard.edu/abs/2013MNRAS.432.1709C} {432, 1709}

\bibitem[\protect\citeauthoryear{{Cash}}{{Cash}}{1979}]{Cash_1979}
{Cash} W.,  1979, \mn@doi [\apj] {10.1086/156922}, \href {https://ui.adsabs.harvard.edu/abs/1979ApJ...228..939C} {228, 939}

\bibitem[\protect\citeauthoryear{{Condon} \& {Dressel}}{{Condon} \& {Dressel}}{1978}]{Dressel_Condon_1978}
{Condon} J.~J.,  {Dressel} L.~L.,  1978, \mn@doi [\apj] {10.1086/156047}, \href {https://ui.adsabs.harvard.edu/abs/1978ApJ...221..456C} {221, 456}

\bibitem[\protect\citeauthoryear{Conroy, Graves  \& van Dokkum}{Conroy et~al.}{2013}]{Conroy_2013}
Conroy C.,  Graves G.~J.,   van Dokkum P.~G.,  2013, \mn@doi [\apj] {10.1088/0004-637x/780/1/33}, 780, 33

\bibitem[\protect\citeauthoryear{{Cox}}{{Cox}}{1998}]{Cox_1998}
{Cox} D.~P.,  1998, in {Breitschwerdt} D.,  {Freyberg} M.~J.,   {Truemper} J.,  eds, , Vol.~506, IAU Colloq. 166: The Local Bubble and Beyond.
pp 121--131, \mn@doi{10.1007/BFb0104706}

\bibitem[\protect\citeauthoryear{{Cravens}, {Robertson}  \& {Snowden}}{{Cravens} et~al.}{2001}]{Cravens_2001}
{Cravens} T.~E.,  {Robertson} I.~P.,   {Snowden} S.~L.,  2001, \mn@doi [\jgr] {10.1029/2000JA000461}, \href {https://ui.adsabs.harvard.edu/abs/2001JGR...10624883C} {106, 24883}

\bibitem[\protect\citeauthoryear{{Crowl}, {Kenney}, {van Gorkom}, {Chung}  \& {Rose}}{{Crowl} et~al.}{2006}]{sf_2006}
{Crowl} H.~H.,  {Kenney} J.~D.,  {van Gorkom} J.~H.,  {Chung} A.,   {Rose} J.~A.,  2006, in American Astronomical Society Meeting Abstracts. p. 211.11

\bibitem[\protect\citeauthoryear{{Das}, {Mathur}, {Gupta}, {Nicastro}  \& {Krongold}}{{Das} et~al.}{2019}]{Sanskriti_2019}
{Das} S.,  {Mathur} S.,  {Gupta} A.,  {Nicastro} F.,   {Krongold} Y.,  2019, \mn@doi [\apj] {10.3847/1538-4357/ab5846}, \href {https://ui.adsabs.harvard.edu/abs/2019ApJ...887..257D} {887, 257}

\bibitem[\protect\citeauthoryear{{De Luca} \& {Molendi}}{{De Luca} \& {Molendi}}{2004}]{deLuca_2004}
{De Luca} {Molendi} 2004, \mn@doi [A\&A] {10.1051/0004-6361:20034421}, 419, 837

\bibitem[\protect\citeauthoryear{{Donzelli} \& {Pastoriza}}{{Donzelli} \& {Pastoriza}}{1997}]{SF_Donzelli_1997}
{Donzelli} C.~J.,  {Pastoriza} M.~G.,  1997, \mn@doi [\apjs] {10.1086/313012}, \href {https://ui.adsabs.harvard.edu/abs/1997ApJS..111..181D} {111, 181}

\bibitem[\protect\citeauthoryear{{Eddington}}{{Eddington}}{1926}]{Eddington_rot}
{Eddington} A.~S.,  1926, {The Internal Constitution of the Stars}

\bibitem[\protect\citeauthoryear{{Ehlert}, {Werner}, {Simionescu}, {Allen}, {Kenney}, {Million}  \& {Finoguenov}}{{Ehlert} et~al.}{2013}]{Ehlert_2013}
{Ehlert} S.,  {Werner} N.,  {Simionescu} A.,  {Allen} S.~W.,  {Kenney} J.~D.~P.,  {Million} E.~T.,   {Finoguenov} A.,  2013, \mn@doi [\mnras] {10.1093/mnras/stt060}, \href {https://ui.adsabs.harvard.edu/abs/2013MNRAS.430.2401E} {430, 2401}

\bibitem[\protect\citeauthoryear{{Fabbiano}, {Gioia}  \& {Trinchieri}}{{Fabbiano} et~al.}{1989}]{Fabbiano_1989}
{Fabbiano} G.,  {Gioia} I.~M.,   {Trinchieri} G.,  1989, \mn@doi [\apj] {10.1086/168103}, \href {https://ui.adsabs.harvard.edu/abs/1989ApJ...347..127F} {347, 127}

\bibitem[\protect\citeauthoryear{{Fabian}, {Schwarz}  \& {Forman}}{{Fabian} et~al.}{1980}]{Fabian_1980}
{Fabian} A.~C.,  {Schwarz} J.,   {Forman} W.,  1980, \mn@doi [\mnras] {10.1093/mnras/192.2.135}, \href {https://ui.adsabs.harvard.edu/abs/1980MNRAS.192..135F} {192, 135}

\bibitem[\protect\citeauthoryear{Fink et~al.,}{Fink et~al.}{2014}]{Fink_2014}
Fink M.,  et~al., 2014, \mn@doi [\mnras] {10.1093/mnras/stt2315}, 438, 1762

\bibitem[\protect\citeauthoryear{{Finoguenov}, {Pietsch}, {Aschenbach}  \& {Miniati}}{{Finoguenov} et~al.}{2004}]{Finoguenov_2004}
{Finoguenov} A.,  {Pietsch} W.,  {Aschenbach} B.,   {Miniati} F.,  2004, \mn@doi [\aap] {10.1051/0004-6361:20031672}, \href {https://ui.adsabs.harvard.edu/abs/2004A&A...415..415F} {415, 415}

\bibitem[\protect\citeauthoryear{{Forman}, {Schwarz}, {Jones}, {Liller}  \& {Fabian}}{{Forman} et~al.}{1979}]{Forman_1979}
{Forman} W.,  {Schwarz} J.,  {Jones} C.,  {Liller} W.,   {Fabian} A.~C.,  1979, \mn@doi [\apjl] {10.1086/183103}, \href {https://ui.adsabs.harvard.edu/abs/1979ApJ...234L..27F} {234, L27}

\bibitem[\protect\citeauthoryear{{Foster}, {Ji}, {Smith}  \& {Brickhouse}}{{Foster} et~al.}{2012}]{Foster2012}
{Foster} A.~R.,  {Ji} L.,  {Smith} R.~K.,   {Brickhouse} N.~S.,  2012, \mn@doi [\apj] {10.1088/0004-637X/756/2/128}, \href {https://ui.adsabs.harvard.edu/abs/2012ApJ...756..128F} {756, 128}

\bibitem[\protect\citeauthoryear{{Gastaldello}, {Simionescu}, {Mernier}, {Biffi}, {Gaspari}, {Sato}  \& {Matsushita}}{{Gastaldello} et~al.}{2021}]{Gastaldello2021}
{Gastaldello} F.,  {Simionescu} A.,  {Mernier} F.,  {Biffi} V.,  {Gaspari} M.,  {Sato} K.,   {Matsushita} K.,  2021, \mn@doi [Universe] {10.3390/universe7070208}, \href {https://ui.adsabs.harvard.edu/abs/2021Univ....7..208G} {7, 208}

\bibitem[\protect\citeauthoryear{Gatuzz et~al.,}{Gatuzz et~al.}{2023a}]{Gatuzz2023}
Gatuzz E.,  et~al., 2023a, \mn@doi [\mnras] {10.1093/mnras/stad447}, 520, 4793

\bibitem[\protect\citeauthoryear{{Gatuzz} et~al.,}{{Gatuzz} et~al.}{2023b}]{Gatuzz_2023}
{Gatuzz} E.,  et~al., 2023b, \mn@doi [\mnras] {10.1093/mnras/stad2716}, \href {https://ui.adsabs.harvard.edu/abs/2023MNRAS.525.6394G} {525, 6394}

\bibitem[\protect\citeauthoryear{{Gil de Paz} et~al.,}{{Gil de Paz} et~al.}{2007}]{Armando_2017}
{Gil de Paz} A.,  et~al., 2007, \mn@doi [\apjs] {10.1086/516636}, \href {https://ui.adsabs.harvard.edu/abs/2007ApJS..173..185G} {173, 185}

\bibitem[\protect\citeauthoryear{{Gomez} et~al.,}{{Gomez} et~al.}{2010}]{Gomez_2010}
{Gomez} H.~L.,  et~al., 2010, \mn@doi [\aap] {10.1051/0004-6361/201014530}, \href {https://ui.adsabs.harvard.edu/abs/2010A&A...518L..45G} {518, L45}

\bibitem[\protect\citeauthoryear{{Gunn} \& {Gott}}{{Gunn} \& {Gott}}{1972}]{Gunn_1972}
{Gunn} J.~E.,  {Gott} J.~Richard I.,  1972, \mn@doi [\apj] {10.1086/151605}, 176, 1

\bibitem[\protect\citeauthoryear{{Hickox} \& {Alexander}}{{Hickox} \& {Alexander}}{2018}]{Hickox2018}
{Hickox} R.~C.,  {Alexander} D.~M.,  2018, \mn@doi [\araa] {10.1146/annurev-astro-081817-051803}, \href {https://ui.adsabs.harvard.edu/abs/2018ARA&A..56..625H} {56, 625}

\bibitem[\protect\citeauthoryear{{Hishi}, {Fujimoto}, {Kotake}, {Ito}, {Tanaka}, {Kai}  \& {Kinoshita}}{{Hishi} et~al.}{2017}]{Hishi_2017}
{Hishi} U.,  {Fujimoto} R.,  {Kotake} M.,  {Ito} H.,  {Tanaka} K.,  {Kai} Y.,   {Kinoshita} Y.,  2017, \mn@doi [\pasj] {10.1093/pasj/psx014}, \href {https://ui.adsabs.harvard.edu/abs/2017PASJ...69...42H} {69, 42}

\bibitem[\protect\citeauthoryear{{Hummel}}{{Hummel}}{2003}]{Hummel_1980}
{Hummel} E.,  2003, {VizieR Online Data Catalog: 1415MHz Survey of Bright Galaxies (Hummel, 1980)}, VizieR On-line Data Catalog: J/A+AS/41/151. Originally published in: 1980A\&AS...41..151H

\bibitem[\protect\citeauthoryear{Irwin, Athey  \& Bregman}{Irwin et~al.}{2003}]{Irwin_2003}
Irwin J.~A.,  Athey A.~E.,   Bregman J.~N.,  2003, \mn@doi [\apj] {10.1086/368179}, 587, 356

\bibitem[\protect\citeauthoryear{{Kaastra}}{{Kaastra}}{2017}]{Kaastra_2017}
{Kaastra} J.~S.,  2017, \mn@doi [\aap] {10.1051/0004-6361/201629319}, \href {https://ui.adsabs.harvard.edu/abs/2017A&A...605A..51K} {605, A51}

\bibitem[\protect\citeauthoryear{{Kaastra} \& {Bleeker}}{{Kaastra} \& {Bleeker}}{2016}]{Kaastra_2016}
{Kaastra} {Bleeker} 2016, \mn@doi [A\&A] {10.1051/0004-6361/201527395}, 587, A151

\bibitem[\protect\citeauthoryear{{Kaastra}, {Raassen}, {de Plaa}  \& {Gu}}{{Kaastra} et~al.}{2024}]{Kaastra_2024}
{Kaastra} J.~S.,  {Raassen} A.~J.~J.,  {de Plaa} J.,   {Gu} L.,  2024, {SPEX X-ray spectral fitting package}, \mn@doi{10.5281/zenodo.10822753}

\bibitem[\protect\citeauthoryear{{Kara}, {Pl{\v{s}}ek}, {Protu{\v{s}}ov{\'a}}, {Breuer}, {Werner}, {Mernier}  \& {Ercan}}{{Kara} et~al.}{2024}]{Kara_2024}
{Kara} S.,  {Pl{\v{s}}ek} T.,  {Protu{\v{s}}ov{\'a}} K.,  {Breuer} J.-P.,  {Werner} N.,  {Mernier} F.,   {Ercan} E.~N.,  2024, \mn@doi [\mnras] {10.1093/mnras/stae065}, \href {https://ui.adsabs.harvard.edu/abs/2024MNRAS.528.1500K} {528, 1500}

\bibitem[\protect\citeauthoryear{{Karakas}}{{Karakas}}{2010}]{Karakas_2010}
{Karakas} A.~I.,  2010, {VizieR Online Data Catalog: Updated stellar yields from AGB models (Karakas, 2010)}, VizieR On-line Data Catalog: J/MNRAS/403/1413. Originally published in: 2010MNRAS.403.1413K

\bibitem[\protect\citeauthoryear{{Kenney}, {Tal}, {Crowl}, {Feldmeier}  \& {Jacoby}}{{Kenney} et~al.}{2008}]{Kenney_2008}
{Kenney} J. D.~P.,  {Tal} T.,  {Crowl} H.~H.,  {Feldmeier} J.,   {Jacoby} G.~H.,  2008, \mn@doi [\apjl] {10.1086/593300}, \href {https://ui.adsabs.harvard.edu/abs/2008ApJ...687L..69K} {687, L69}

\bibitem[\protect\citeauthoryear{{Kennicutt}, {Keel}, {van der Hulst}, {Hummel}  \& {Roettiger}}{{Kennicutt} et~al.}{1987}]{SF_Kennicutt_1987}
{Kennicutt} Jr. R.~C.,  {Keel} W.~C.,  {van der Hulst} J.~M.,  {Hummel} E.,   {Roettiger} K.~A.,  1987, \mn@doi [\aj] {10.1086/114384}, \href {https://ui.adsabs.harvard.edu/abs/1987AJ.....93.1011K} {93, 1011}

\bibitem[\protect\citeauthoryear{{Koutroumpa}, {Smith}, {Edgar}, {Kuntz}, {Plucinsky}  \& {Snowden}}{{Koutroumpa} et~al.}{2011}]{Koutroumpa_2011}
{Koutroumpa} D.,  {Smith} R.~K.,  {Edgar} R.~J.,  {Kuntz} K.~D.,  {Plucinsky} P.~P.,   {Snowden} S.~L.,  2011, \mn@doi [\apj] {10.1088/0004-637X/726/2/91}, \href {https://ui.adsabs.harvard.edu/abs/2011ApJ...726...91K} {726, 91}

\bibitem[\protect\citeauthoryear{{Kuntz}}{{Kuntz}}{2019}]{Kuntz_2019}
{Kuntz} K.~D.,  2019, \mn@doi [\aapr] {10.1007/s00159-018-0114-0}, \href {https://ui.adsabs.harvard.edu/abs/2019A&ARv..27....1K} {27, 1}

\bibitem[\protect\citeauthoryear{{Kuntz} \& {Snowden}}{{Kuntz} \& {Snowden}}{2008}]{Kuntz_2008}
{Kuntz} K.~D.,  {Snowden} S.~L.,  2008, \mn@doi [\apj] {10.1086/524719}, \href {https://ui.adsabs.harvard.edu/abs/2008ApJ...674..209K} {674, 209}

\bibitem[\protect\citeauthoryear{{Lakhchaura}, {Mernier}  \& {Werner}}{{Lakhchaura} et~al.}{2019}]{Lakhchaura_2019}
{Lakhchaura} K.,  {Mernier} F.,   {Werner} N.,  2019, \mn@doi [\aap] {10.1051/0004-6361/201834755}, \href {https://ui.adsabs.harvard.edu/abs/2019A&A...623A..17L} {623, A17}

\bibitem[\protect\citeauthoryear{{Larson} \& {Tinsley}}{{Larson} \& {Tinsley}}{1978}]{SF_Larson_1978}
{Larson} R.~B.,  {Tinsley} B.~M.,  1978, \mn@doi [\apj] {10.1086/155753}, \href {https://ui.adsabs.harvard.edu/abs/1978ApJ...219...46L} {219, 46}

\bibitem[\protect\citeauthoryear{{Lehmer} et~al.,}{{Lehmer} et~al.}{2012}]{Lehmer_2012}
{Lehmer} B.~D.,  et~al., 2012, \mn@doi [\apj] {10.1088/0004-637X/752/1/46}, \href {https://ui.adsabs.harvard.edu/abs/2012ApJ...752...46L} {752, 46}

\bibitem[\protect\citeauthoryear{{Limongi} \& {Chieffi}}{{Limongi} \& {Chieffi}}{2018}]{Limongi_2018}
{Limongi} M.,  {Chieffi} A.,  2018, \mn@doi [\apjs] {10.3847/1538-4365/aacb24}, \href {https://ui.adsabs.harvard.edu/abs/2018ApJS..237...13L} {237, 13}

\bibitem[\protect\citeauthoryear{{Liu}, {Zhai}  \& {Tozzi}}{{Liu} et~al.}{2019}]{Liu_2019}
{Liu} A.,  {Zhai} M.,   {Tozzi} P.,  2019, \mn@doi [\mnras] {10.1093/mnras/stz533}, \href {https://ui.adsabs.harvard.edu/abs/2019MNRAS.485.1651L} {485, 1651}

\bibitem[\protect\citeauthoryear{{Lodders}, {Palme}  \& {Gail}}{{Lodders} et~al.}{2009}]{Lodders2009}
{Lodders} K.,  {Palme} H.,   {Gail} H.~P.,  2009, \mn@doi [Landolt Börnstein] {10.1007/978-3-540-88055-4_34}, \href {https://ui.adsabs.harvard.edu/abs/2009LanB...4B..712L} {4B, 712}

\bibitem[\protect\citeauthoryear{{Loewenstein}}{{Loewenstein}}{2004}]{Loewenstein_2004}
{Loewenstein} M.,  2004, in {McWilliam} A.,  {Rauch} M.,  eds, Origin and Evolution of the Elements. p.~422 (\mn@eprint {arXiv} {astro-ph/0310557}), \mn@doi{10.48550/arXiv.astro-ph/0310557}

\bibitem[\protect\citeauthoryear{{Machacek}, {Dosaj}, {Forman}, {Jones}, {Markevitch}, {Vikhlinin}, {Warmflash}  \& {Kraft}}{{Machacek} et~al.}{2005}]{Machacek_2005}
{Machacek} M.,  {Dosaj} A.,  {Forman} W.,  {Jones} C.,  {Markevitch} M.,  {Vikhlinin} A.,  {Warmflash} A.,   {Kraft} R.,  2005, \mn@doi [\apj] {10.1086/427548}, \href {https://ui.adsabs.harvard.edu/abs/2005ApJ...621..663M} {621, 663}

\bibitem[\protect\citeauthoryear{Machacek, Jones, Forman  \& Nulsen}{Machacek et~al.}{2006}]{Machacek_2006a}
Machacek M.,  Jones C.,  Forman W.~R.,   Nulsen P.,  2006, \mn@doi [\apj] {10.1086/503350}, 644, 155–166

\bibitem[\protect\citeauthoryear{Madau \& Dickinson}{Madau \& Dickinson}{2014}]{MadauDickinson2014}
Madau P.,  Dickinson M.,  2014, \mn@doi [\araa] {10.1146/annurev-astro-081811-125615}, 52, 415

\bibitem[\protect\citeauthoryear{{Maeder}}{{Maeder}}{2009}]{Maeder_2009}
{Maeder} A.,  2009, {Physics, Formation and Evolution of Rotating Stars}, \mn@doi{10.1007/978-3-540-76949-1.
}

\bibitem[\protect\citeauthoryear{{Mao} et~al.,}{{Mao} et~al.}{2019}]{Mao_2019}
{Mao} J.,  et~al., 2019, \mn@doi [\aap] {10.1051/0004-6361/201730931}, \href {https://ui.adsabs.harvard.edu/abs/2019A&A...621A...9M} {621, A9}

\bibitem[\protect\citeauthoryear{{Mathews}}{{Mathews}}{1990}]{Mathews1990}
{Mathews} W.~G.,  1990, \mn@doi [\apj] {10.1086/168708}, \href {https://ui.adsabs.harvard.edu/abs/1990ApJ...354..468M} {354, 468}

\bibitem[\protect\citeauthoryear{Mathews \& Brighenti}{Mathews \& Brighenti}{2003}]{MathewsBrighetni2003}
Mathews W.~G.,  Brighenti F.,  2003, \mn@doi [\araa] {10.1146/annurev.astro.41.090401.094542}, 41, 191

\bibitem[\protect\citeauthoryear{{McCammon} et~al.,}{{McCammon} et~al.}{2002}]{McCammon_2002}
{McCammon} D.,  et~al., 2002, \mn@doi [\apj] {10.1086/341727}, \href {https://ui.adsabs.harvard.edu/abs/2002ApJ...576..188M} {576, 188}

\bibitem[\protect\citeauthoryear{{Mei} et~al.,}{{Mei} et~al.}{2007}]{Mei_2007}
{Mei} S.,  et~al., 2007, \mn@doi [\apj] {10.1086/509598}, \href {https://ui.adsabs.harvard.edu/abs/2007ApJ...655..144M} {655, 144}

\bibitem[\protect\citeauthoryear{{Merluzzi}, {Busarello}, {Dopita}, {Haines}, {Steinhauser}, {Bourdin}  \& {Mazzotta}}{{Merluzzi} et~al.}{2016}]{sf_2016}
{Merluzzi} P.,  {Busarello} G.,  {Dopita} M.~A.,  {Haines} C.~P.,  {Steinhauser} D.,  {Bourdin} H.,   {Mazzotta} P.,  2016, \mn@doi [\mnras] {10.1093/mnras/stw1198}, \href {https://ui.adsabs.harvard.edu/abs/2016MNRAS.460.3345M} {460, 3345}

\bibitem[\protect\citeauthoryear{{Mernier} \& {Biffi}}{{Mernier} \& {Biffi}}{2022}]{Mernier2022}
{Mernier} F.,  {Biffi} V.,  2022, arXiv e-prints, \href {https://ui.adsabs.harvard.edu/abs/2022arXiv220207097M} {p. arXiv:2202.07097}

\bibitem[\protect\citeauthoryear{{Mernier}, {de Plaa, J.}, {Lovisari, L.}, {Pinto, C.}, {Zhang, Y.-Y.}, {Kaastra, J. S.}, {Werner, N.}  \& {Simionescu, A.}}{{Mernier} et~al.}{2015}]{Mernier_2015}
{Mernier} {de Plaa, J.} {Lovisari, L.} {Pinto, C.} {Zhang, Y.-Y.} {Kaastra, J. S.} {Werner, N.}  {Simionescu, A.} 2015, \mn@doi [A\&A] {10.1051/0004-6361/201425282}, 575, A37

\bibitem[\protect\citeauthoryear{{Mernier}, {de Plaa}, {Pinto}, {Kaastra}, {Kosec}, {Zhang}, {Mao}  \& {Werner}}{{Mernier} et~al.}{2016a}]{Mernier2016a}
{Mernier} F.,  {de Plaa} J.,  {Pinto} C.,  {Kaastra} J.~S.,  {Kosec} P.,  {Zhang} Y.~Y.,  {Mao} J.,   {Werner} N.,  2016a, \mn@doi [\aap] {10.1051/0004-6361/201527824}, \href {https://ui.adsabs.harvard.edu/abs/2016A&A...592A.157M} {592, A157}

\bibitem[\protect\citeauthoryear{{Mernier} et~al.,}{{Mernier} et~al.}{2016b}]{Mernier_SNe}
{Mernier} F.,  et~al., 2016b, \mn@doi [\aap] {10.1051/0004-6361/201628765}, \href {https://ui.adsabs.harvard.edu/abs/2016A&A...595A.126M} {595, A126}

\bibitem[\protect\citeauthoryear{{Mernier} et~al.,}{{Mernier} et~al.}{2017}]{Mernier_2017}
{Mernier} F.,  et~al., 2017, in {Ness} J.-U.,  {Migliari} S.,  eds, The X-ray Universe 2017. p.~148

\bibitem[\protect\citeauthoryear{{Mernier} et~al.,}{{Mernier} et~al.}{2018a}]{Mernier2018m}
{Mernier} F.,  et~al., 2018a, \mn@doi [\mnras] {10.1093/mnrasl/sly080}, \href {https://ui.adsabs.harvard.edu/abs/2018MNRAS.478L.116M} {478, L116}

\bibitem[\protect\citeauthoryear{{Mernier} et~al.,}{{Mernier} et~al.}{2018b}]{Mernier2018b}
{Mernier} F.,  et~al., 2018b, \mn@doi [\mnras] {10.1093/mnrasl/sly134}, \href {https://ui.adsabs.harvard.edu/abs/2018MNRAS.480L..95M} {480, L95}

\bibitem[\protect\citeauthoryear{Mernier et~al.,}{Mernier et~al.}{2022}]{Mernier2022b}
Mernier F.,  et~al., 2022, \mn@doi [\mnras] {10.1093/mnras/stac253}, 511, 3159

\bibitem[\protect\citeauthoryear{{Mihos} \& {Hernquist}}{{Mihos} \& {Hernquist}}{1996}]{SF_Mihos_1996}
{Mihos} J.~C.,  {Hernquist} L.,  1996, \mn@doi [\apj] {10.1086/177353}, \href {https://ui.adsabs.harvard.edu/abs/1996ApJ...464..641M} {464, 641}

\bibitem[\protect\citeauthoryear{{Million}, {Werner}, {Simionescu}  \& {Allen}}{{Million} et~al.}{2011}]{Million_2011}
{Million} E.~T.,  {Werner} N.,  {Simionescu} A.,   {Allen} S.~W.,  2011, \mn@doi [\mnras] {10.1111/j.1365-2966.2011.19664.x}, \href {https://ui.adsabs.harvard.edu/abs/2011MNRAS.418.2744M} {418, 2744}

\bibitem[\protect\citeauthoryear{{Mitchell}, {Culhane}, {Davison}  \& {Ives}}{{Mitchell} et~al.}{1976}]{Mitchell1976}
{Mitchell} R.~J.,  {Culhane} J.~L.,  {Davison} P.~J.~N.,   {Ives} J.~C.,  1976, \mn@doi [\mnras] {10.1093/mnras/175.1.29P}, \href {https://ui.adsabs.harvard.edu/abs/1976MNRAS.175P..29M} {175, 29P}

\bibitem[\protect\citeauthoryear{{Moll{\'a}}, {V{\'\i}lchez}, {Gavil{\'a}n}  \& {D{\'\i}az}}{{Moll{\'a}} et~al.}{2006}]{Molla_2006}
{Moll{\'a}} M.,  {V{\'\i}lchez} J.~M.,  {Gavil{\'a}n} M.,   {D{\'\i}az} A.~I.,  2006, \mn@doi [\mnras] {10.1111/j.1365-2966.2006.10892.x}, \href {https://ui.adsabs.harvard.edu/abs/2006MNRAS.372.1069M} {372, 1069}

\bibitem[\protect\citeauthoryear{{Moreno} et~al.,}{{Moreno} et~al.}{2019}]{SF_Moreno_2019}
{Moreno} J.,  et~al., 2019, \mn@doi [\mnras] {10.1093/mnras/stz417}, \href {https://ui.adsabs.harvard.edu/abs/2019MNRAS.485.1320M} {485, 1320}

\bibitem[\protect\citeauthoryear{Nomoto, Kobayashi  \& Tominaga}{Nomoto et~al.}{2013}]{Nomoto_2013}
Nomoto K.,  Kobayashi C.,   Tominaga N.,  2013, \mn@doi [\araa] {10.1146/annurev-astro-082812-140956}, 51, 457

\bibitem[\protect\citeauthoryear{{Panagoulia}, {Fabian}  \& {Sanders}}{{Panagoulia} et~al.}{2013}]{Panagoulia_2013}
{Panagoulia} E.~K.,  {Fabian} A.~C.,   {Sanders} J.~S.,  2013, \mn@doi [\mnras] {10.1093/mnras/stt969}, \href {https://ui.adsabs.harvard.edu/abs/2013MNRAS.433.3290P} {433, 3290}

\bibitem[\protect\citeauthoryear{{Panagoulia}, {Sanders}  \& {Fabian}}{{Panagoulia} et~al.}{2015}]{Panagoulia_2015}
{Panagoulia} E.~K.,  {Sanders} J.~S.,   {Fabian} A.~C.,  2015, \mn@doi [\mnras] {10.1093/mnras/stu2469}, \href {https://ui.adsabs.harvard.edu/abs/2015MNRAS.447..417P} {447, 417}

\bibitem[\protect\citeauthoryear{{Randall}, {Nulsen}, {Forman}, {Jones}, {Machacek}, {Murray}  \& {Maughan}}{{Randall} et~al.}{2008}]{Randall_2008}
{Randall} S.,  {Nulsen} P.,  {Forman} W.~R.,  {Jones} C.,  {Machacek} M.,  {Murray} S.~S.,   {Maughan} B.,  2008, \mn@doi [\apj] {10.1086/592324}, \href {https://ui.adsabs.harvard.edu/abs/2008ApJ...688..208R} {688, 208}

\bibitem[\protect\citeauthoryear{{Rangarajan}, {White}, {Ebeling}  \& {Fabian}}{{Rangarajan} et~al.}{1995}]{Rangarajan_1995}
{Rangarajan} F.~V.~N.,  {White} D.~A.,  {Ebeling} H.,   {Fabian} A.~C.,  1995, \mn@doi [\mnras] {10.1093/mnras/277.3.1047}, \href {https://ui.adsabs.harvard.edu/abs/1995MNRAS.277.1047R} {277, 1047}

\bibitem[\protect\citeauthoryear{{R{\"o}pke} et~al.,}{{R{\"o}pke} et~al.}{2012}]{Ropke_2012}
{R{\"o}pke} F.~K.,  et~al., 2012, \mn@doi [\apjl] {10.1088/2041-8205/750/1/L19}, \href {https://ui.adsabs.harvard.edu/abs/2012ApJ...750L..19R} {750, L19}

\bibitem[\protect\citeauthoryear{{Salpeter}}{{Salpeter}}{1955}]{Salpater1955}
{Salpeter} E.~E.,  1955, \mn@doi [\apj] {10.1086/145971}, \href {https://ui.adsabs.harvard.edu/abs/1955ApJ...121..161S} {121, 161}

\bibitem[\protect\citeauthoryear{{Sanders} \& {Fabian}}{{Sanders} \& {Fabian}}{2011}]{Sanders_Fabian_2011}
{Sanders} J.~S.,  {Fabian} A.~C.,  2011, \mn@doi [\mnras] {10.1111/j.1745-3933.2010.01000.x}, \href {https://ui.adsabs.harvard.edu/abs/2011MNRAS.412L..35S} {412, L35}

\bibitem[\protect\citeauthoryear{{Sanders}, {Fabian}, {Allen}, {Morris}, {Graham}  \& {Johnstone}}{{Sanders} et~al.}{2008}]{Sanders_2008}
{Sanders} J.~S.,  {Fabian} A.~C.,  {Allen} S.~W.,  {Morris} R.~G.,  {Graham} J.,   {Johnstone} R.~M.,  2008, \mn@doi [\mnras] {10.1111/j.1365-2966.2008.12952.x}, \href {https://ui.adsabs.harvard.edu/abs/2008MNRAS.385.1186S} {385, 1186}

\bibitem[\protect\citeauthoryear{{Sanders}, {Fabian}, {Frank}, {Peterson}  \& {Russell}}{{Sanders} et~al.}{2010}]{Sanders_2010}
{Sanders} J.~S.,  {Fabian} A.~C.,  {Frank} K.~A.,  {Peterson} J.~R.,   {Russell} H.~R.,  2010, \mn@doi [\mnras] {10.1111/j.1365-2966.2009.15902.x}, \href {https://ui.adsabs.harvard.edu/abs/2010MNRAS.402..127S} {402, 127}

\bibitem[\protect\citeauthoryear{Seitenzahl et~al.,}{Seitenzahl et~al.}{2013}]{Seitenzahl_2012}
Seitenzahl I.~R.,  et~al., 2013, \mn@doi [\mnras] {10.1093/mnras/sts402}, 429, 1156

\bibitem[\protect\citeauthoryear{{Serlemitsos}, {Smith}, {Boldt}, {Holt}  \& {Swank}}{{Serlemitsos} et~al.}{1977}]{Serlemitsos1977}
{Serlemitsos} P.~J.,  {Smith} B.~W.,  {Boldt} E.~A.,  {Holt} S.~S.,   {Swank} J.~H.,  1977, \mn@doi [\apjl] {10.1086/182342}, \href {https://ui.adsabs.harvard.edu/abs/1977ApJ...211L..63S} {211, L63}

\bibitem[\protect\citeauthoryear{{Sim} et~al.,}{{Sim} et~al.}{2013}]{Sim_2013}
{Sim} S.~A.,  et~al., 2013, \mn@doi [\mnras] {10.1093/mnras/stt1574}, \href {https://ui.adsabs.harvard.edu/abs/2013MNRAS.436..333S} {436, 333}

\bibitem[\protect\citeauthoryear{{Simionescu}, {Werner}, {Urban}, {Allen}, {Ichinohe}  \& {Zhuravleva}}{{Simionescu} et~al.}{2015}]{Simionescu_2015}
{Simionescu} A.,  {Werner} N.,  {Urban} O.,  {Allen} S.~W.,  {Ichinohe} Y.,   {Zhuravleva} I.,  2015, \mn@doi [\apjl] {10.1088/2041-8205/811/2/L25}, \href {https://ui.adsabs.harvard.edu/abs/2015ApJ...811L..25S} {811, L25}

\bibitem[\protect\citeauthoryear{{Simionescu}, {Werner}, {Mantz}, {Allen}  \& {Urban}}{{Simionescu} et~al.}{2017}]{Simionescu_2017}
{Simionescu} A.,  {Werner} N.,  {Mantz} A.,  {Allen} S.~W.,   {Urban} O.,  2017, \mn@doi [\mnras] {10.1093/mnras/stx919}, \href {https://ui.adsabs.harvard.edu/abs/2017MNRAS.469.1476S} {469, 1476}

\bibitem[\protect\citeauthoryear{{Simionescu} et~al.,}{{Simionescu} et~al.}{2019}]{Simionescu2019b}
{Simionescu} A.,  et~al., 2019, \mn@doi [\mnras] {10.1093/mnras/sty3220}, \href {https://ui.adsabs.harvard.edu/abs/2019MNRAS.483.1701S} {483, 1701}

\bibitem[\protect\citeauthoryear{{Snowden}}{{Snowden}}{2009}]{Snowden_2009}
{Snowden} S.~L.,  2009, \mn@doi [\ssr] {10.1007/s11214-008-9343-2}, \href {https://ui.adsabs.harvard.edu/abs/2009SSRv..143..253S} {143, 253}

\bibitem[\protect\citeauthoryear{{Snowden} \& {Kuntz}}{{Snowden} \& {Kuntz}}{2013}]{Snowden_Kuntz_2013}
{Snowden} S.~L.,  {Kuntz} K.~D.,  2013, XMM ESAS cookbook

\bibitem[\protect\citeauthoryear{Su et~al.,}{Su et~al.}{2017}]{Su_2017}
Su Y.,  et~al., 2017, \mn@doi [\apj] {10.3847/1538-4357/834/1/74}, 834, 74

\bibitem[\protect\citeauthoryear{{Sweet}}{{Sweet}}{1950}]{Sweet_rot}
{Sweet} P.~A.,  1950, \mn@doi [\mnras] {10.1093/mnras/110.6.548}, \href {https://ui.adsabs.harvard.edu/abs/1950MNRAS.110..548S} {110, 548}

\bibitem[\protect\citeauthoryear{{Takeda}, {Nulsen}  \& {Fabian}}{{Takeda} et~al.}{1984}]{Takeda_1984}
{Takeda} H.,  {Nulsen} P.~E.~J.,   {Fabian} A.~C.,  1984, \mn@doi [\mnras] {10.1093/mnras/208.2.261}, \href {https://ui.adsabs.harvard.edu/abs/1984MNRAS.208..261T} {298, 261}

\bibitem[\protect\citeauthoryear{Thomas, Maraston, Schawinski, Sarzi  \& Silk}{Thomas et~al.}{2010}]{Thomas2010}
Thomas D.,  Maraston C.,  Schawinski K.,  Sarzi M.,   Silk J.,  2010, \mn@doi [\mnras] {10.1111/j.1365-2966.2010.16427.x}, 404, 1775

\bibitem[\protect\citeauthoryear{{Truong} et~al.,}{{Truong} et~al.}{2019}]{Truong2019}
{Truong} N.,  et~al., 2019, \mn@doi [\mnras] {10.1093/mnras/stz161}, \href {https://ui.adsabs.harvard.edu/abs/2019MNRAS.484.2896T} {484, 2896}

\bibitem[\protect\citeauthoryear{{Urban}, {Werner}, {Simionescu}, {Allen}  \& {B{\"o}hringer}}{{Urban} et~al.}{2011}]{Urban_2011}
{Urban} O.,  {Werner} N.,  {Simionescu} A.,  {Allen} S.~W.,   {B{\"o}hringer} H.,  2011, \mn@doi [\mnras] {10.1111/j.1365-2966.2011.18526.x}, \href {https://ui.adsabs.harvard.edu/abs/2011MNRAS.414.2101U} {414, 2101}

\bibitem[\protect\citeauthoryear{Urban, Werner, Allen, Simionescu  \& Mantz}{Urban et~al.}{2017}]{Urban2017}
Urban O.,  Werner N.,  Allen S.~W.,  Simionescu A.,   Mantz A.,  2017, \mn@doi [\mnras] {10.1093/mnras/stx1542}, 470, 4583

\bibitem[\protect\citeauthoryear{{Urdampilleta}, {Mernier}, {Kaastra}, {Simionescu}, {de Plaa}, {Kara}  \& {Ercan}}{{Urdampilleta} et~al.}{2019}]{Urdampilleta_2019}
{Urdampilleta} I.,  {Mernier} F.,  {Kaastra} J.~S.,  {Simionescu} A.,  {de Plaa} J.,  {Kara} S.,   {Ercan} E.~N.,  2019, \mn@doi [\aap] {10.1051/0004-6361/201935452}, \href {https://ui.adsabs.harvard.edu/abs/2019A&A...629A..31U} {629, A31}

\bibitem[\protect\citeauthoryear{{Veilleux}, {Cecil}  \& {Bland-Hawthorn}}{{Veilleux} et~al.}{2005}]{Veilleux_2005}
{Veilleux} S.,  {Cecil} G.,   {Bland-Hawthorn} J.,  2005, \mn@doi [\araa] {10.1146/annurev.astro.43.072103.150610}, \href {https://ui.adsabs.harvard.edu/abs/2005ARA&A..43..769V} {43, 769}

\bibitem[\protect\citeauthoryear{{Vulcani} et~al.,}{{Vulcani} et~al.}{2018}]{sf_2018}
{Vulcani} B.,  et~al., 2018, \mn@doi [\apjl] {10.3847/2041-8213/aae68b}, \href {https://ui.adsabs.harvard.edu/abs/2018ApJ...866L..25V} {866, L25}

\bibitem[\protect\citeauthoryear{{Werner} \& {Mernier}}{{Werner} \& {Mernier}}{2020}]{Werner_2020}
{Werner} N.,  {Mernier} F.,  2020, in , Reviews in Frontiers of Modern Astrophysics; From Space Debris to Cosmology.
pp 279--310, \mn@doi{10.1007/978-3-030-38509-5_10}

\bibitem[\protect\citeauthoryear{{Werner}, {de Plaa, J.}, {Kaastra, J. S.}, {Vink, Jacco}, {Bleeker, J. A. M.}, {Tamura, T.}, {Peterson, J. R.}  \& {Verbunt, F.}}{{Werner} et~al.}{2006}]{Werner_2006}
{Werner} {de Plaa, J.} {Kaastra, J. S.} {Vink, Jacco} {Bleeker, J. A. M.} {Tamura, T.} {Peterson, J. R.}  {Verbunt, F.} 2006, \mn@doi [A\&A] {10.1051/0004-6361:20053868}, 449, 475

\bibitem[\protect\citeauthoryear{{Werner}, {Zhuravleva}, {Churazov}, {Simionescu}, {Allen}, {Forman}, {Jones}  \& {Kaastra}}{{Werner} et~al.}{2009}]{Werner_2009}
{Werner} N.,  {Zhuravleva} I.,  {Churazov} E.,  {Simionescu} A.,  {Allen} S.~W.,  {Forman} W.,  {Jones} C.,   {Kaastra} J.~S.,  2009, \mn@doi [\mnras] {10.1111/j.1365-2966.2009.14860.x}, \href {https://ui.adsabs.harvard.edu/abs/2009MNRAS.398...23W} {398, 23}

\bibitem[\protect\citeauthoryear{Werner, Urban, Simionescu  \& Allen}{Werner et~al.}{2013}]{Werner2013}
Werner N.,  Urban O.,  Simionescu A.,   Allen S.~W.,  2013, \mn@doi [Nature] {10.1038/nature12646}, 502, 656

\bibitem[\protect\citeauthoryear{{White}, {Fabian}, {Forman}, {Jones}  \& {Stern}}{{White} et~al.}{1990}]{White_1991}
{White} D.~A.,  {Fabian} A.~C.,  {Forman} W.,  {Jones} C.,   {Stern} C.,  1990, in {Hollenbach} D.~J.,  {Thronson} Jr. H.~A.,  eds,  NASA Conference Publication Vol. 3084, NASA Conference Publication. p.~201

\bibitem[\protect\citeauthoryear{Zhang, Churazov, Forman  \& Jones}{Zhang et~al.}{2018}]{Zhang_2018}
Zhang C.,  Churazov E.,  Forman W.~R.,   Jones C.,  2018, \mn@doi [\mnras] {10.1093/mnras/sty2501}, 482, 20–29

\bibitem[\protect\citeauthoryear{{de Plaa}, {Werner}, {Bleeker}, {Vink}, {Kaastra}  \& {M{\'e}ndez}}{{de Plaa} et~al.}{2007}]{dePlaa_2007}
{de Plaa} J.,  {Werner} N.,  {Bleeker} J.~A.~M.,  {Vink} J.,  {Kaastra} J.~S.,   {M{\'e}ndez} M.,  2007, \mn@doi [\aap] {10.1051/0004-6361:20066382}, \href {https://ui.adsabs.harvard.edu/abs/2007A&A...465..345D} {465, 345}

\bibitem[\protect\citeauthoryear{{de Plaa} et~al.,}{{de Plaa} et~al.}{2017a}]{dePlaa2017}
{de Plaa} J.,  et~al., 2017a, \mn@doi [\aap] {10.1051/0004-6361/201629926}, \href {https://ui.adsabs.harvard.edu/abs/2017A&A...607A..98D} {607, A98}

\bibitem[\protect\citeauthoryear{{de Plaa} et~al.,}{{de Plaa} et~al.}{2017b}]{dePlaa__2017}
{de Plaa} J.,  et~al., 2017b, \mn@doi [\aap] {10.1051/0004-6361/201629926}, \href {https://ui.adsabs.harvard.edu/abs/2017A&A...607A..98D} {607, A98}

\makeatother
\end{thebibliography}




\appendix

\section{Measurement biases in the RGS data of low temperature plasma}
\label{subsection:RGS-systematics}

\begin{figure*}
    \centering
    \makebox[\textwidth][c]{
        \includegraphics[height=4.5cm]{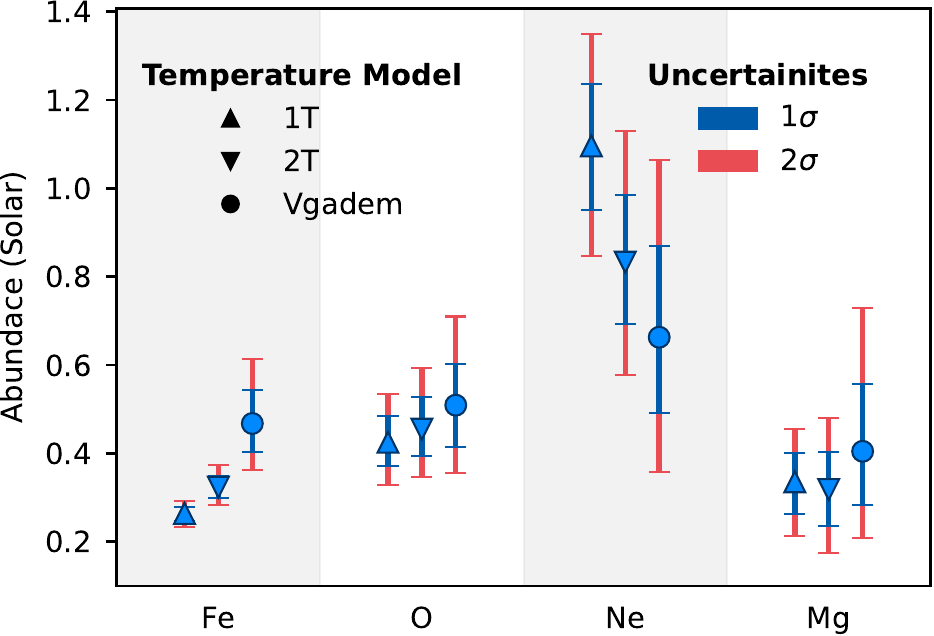}%
        \hspace{5.mm} 
        \includegraphics[height=4.5cm]{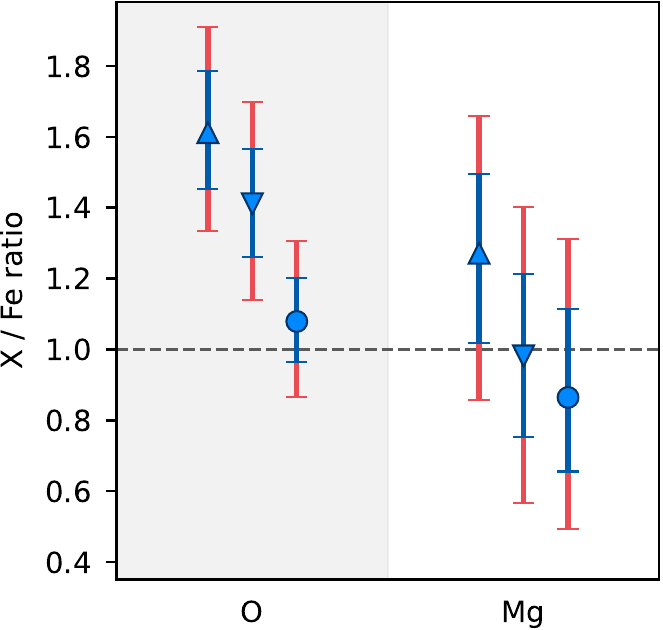}
            \hspace{5.mm} 
        \includegraphics[height=4.5cm]{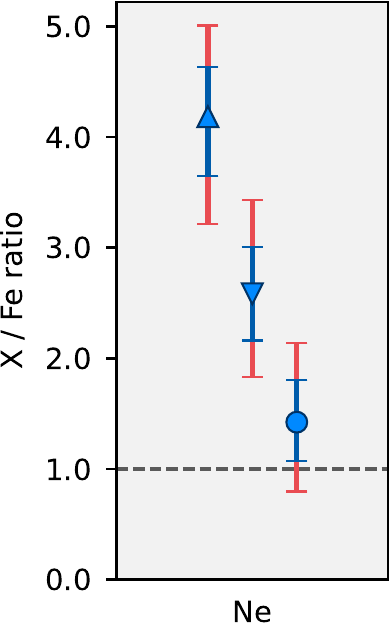}    
    }

    \caption{Measurement differences between different temperature model for the RGS data. The blue bars represents the 1$\sigma$ errors, while the red bars represents the 2$\sigma$.   }
    \label{fig:rgs_syst}
\end{figure*}

\begin{figure*}
    \centering
    \makebox[\textwidth][c]{
        \includegraphics[height=4.5cm]{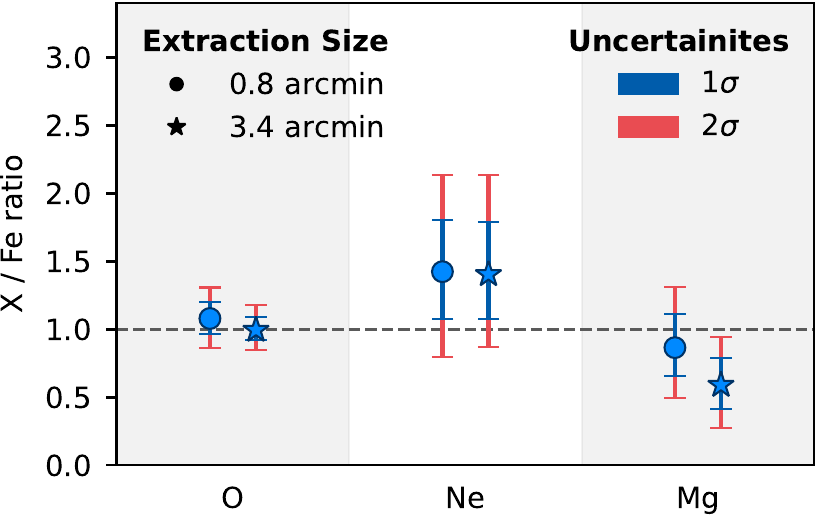}%
        \label{fig:rgs_syst_1}
        \hspace{5.mm} 
        \includegraphics[height=4.5cm]{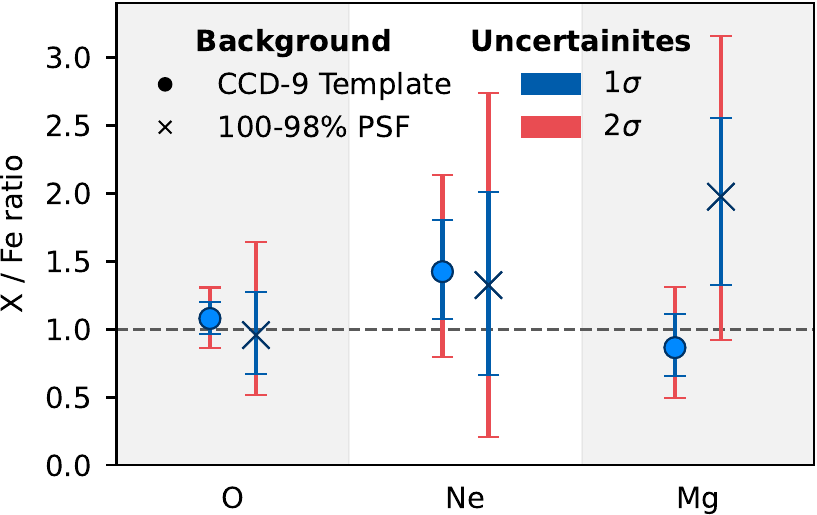}
    }

    \caption{ The blue bars represents the 1$\sigma$ errors, while the red bars represents the 2$\sigma$. \textit{Left:} Measurement differences between different extraction sizes for the RGS data. \textit{Right:} Measurement differences between different background methods for the RGS data. 
  }
    \label{fig:rgs_syst_2}
\end{figure*}

We have investigated the measurement differences of the RGS data of low-temperature plasma ($kT \lesssim 2-3$ keV) based on the (i) applied temperature model, (ii) extraction size, and (iii) background subtraction method. In this section we present the results based on different measurement techniques for the RGS data of low-temperature plasma. Except the comparison between different temperature models, for the rest of the comparisons on the extraction size and background method, we used \texttt{vgadem} model for consistency.

\subsection{Temperature model}

We observe that the Fe abundance do not agree within 1$\sigma$ between 1T, 2T and \texttt{vgadem} models, while 1T and \texttt{vgadem} do not agree within 2$\sigma$. The Ne absolute abundance also do not agree within 1$\sigma$ for 1T and \texttt{vgadem} models. Although not very prominent, we note that we observe a reverse trend with Ne abundance compared to Fe, O, and Mg abundances based on the temperature model used. In Ne measurements \texttt{vgadem} gives the lowest abundance value, which is the reverse for others. The results are presented in Fig. \ref{fig:rgs_syst}.

As for the abundance ratios, we see the same O/Fe bias observed in \citep{Kara_2024} in which O/Fe ratio of RGS data of low-temperature plasma is lower in \texttt{vgadem} model with more than 1$\sigma$ uncertainty. We also show here that the O/Fe ratio do not agree within 2$\sigma$ error between these models, and additionally 2T model do not agree with \texttt{vgadem} for 1$\sigma$. Unlike O/Fe, we see that for RGS measurements Mg/Fe agrees with results obtained from different temperature models, although a decreasing trend is also prominent for 1T to \texttt{vgadem}. The most prominent measurement bias due to imperfections in constraining the temperature structure, on the other hand, is Ne/Fe ratio. First, we observe that Ne/Fe for 1T model is highly unusual with >4 Solar. None of the temperature models agree with each other within 1$\sigma$. We see that the \texttt{vgadem} model clearly gives the most realistic abundance ratio for the Ne/Fe in the RGS data of low-temperature plasma with a value $1.47\pm0.41$. For M86, we conclude that \texttt{vgadem} gives the most reliable measurements for RGS data.

\subsection{Extraction size}

We extracted RGS spectra first with a 0.8 arcmin extraction size, which minimizes the line broadening, and also with a 3.4 arcmin size that consists of more photon counts with enhanced instrumental broadening. Reassuringly, we observe that all abundance ratios of O/Fe, Ne/Fe and Mg/Fe are agrees with each other within 1$\sigma$ uncertainty. This also confirms that the applied correction of line broadening is not significantly biased.

\subsection{Background subtraction}

Conventionally, two background subtraction methods can be applied in RGS data analysis. One method uses a template background file based on the count rate in CCD 9, which is assumed to contain no source counts. The other method involves extracting a spectrum by excluding a given percentage of the cross-dispersion PSF. For this analysis, we excluded 98\% of the PSF—meaning we extracted the spectra outside of 98\% of the PSF—for testing purposes. In Fig. \ref{fig:rgs_syst_2}, we see that for O/Fe and Ne/Fe, abundance ratios obtained from different backgrounds agree within 1$\sigma$. The Mg/Fe measurements, on the other hand, do not agree with each other within 1$\sigma$, although they do agree within 2$\sigma$. We note that for the spectra with the PSF background, the background-subtracted photon count is roughly half of that in the spectra with the template background. The low statistics and disagreement in Mg/Fe values might originate from imperfections in constraining the temperature structure. Considering the irregular surface brightness profile of M86, the background spectra obtained outside 98\% of the PSF likely include photons from X-ray structures such as the North-Eastern Arm and the edges of the Plume itself. As a result, this background might not accurately represent the emission from the outskirts of the given source, which could lead to an actual deprojection of the data. We conclude that the disagreement between the Mg/Fe values most likely originates from systematics introduced by selecting a background spectrum containing photon counts from different X-ray structures. Nevertheless, the discrepancy in abundance ratios obtained using different background methods highlights that the choice of background subtraction in the RGS data of low-temperature plasma is not a trivial factor in abundance measurements.

\subsection{AtomDB vs. SPEXACT}

\begin{figure}
    \centering
    \includegraphics[width=0.8\columnwidth]{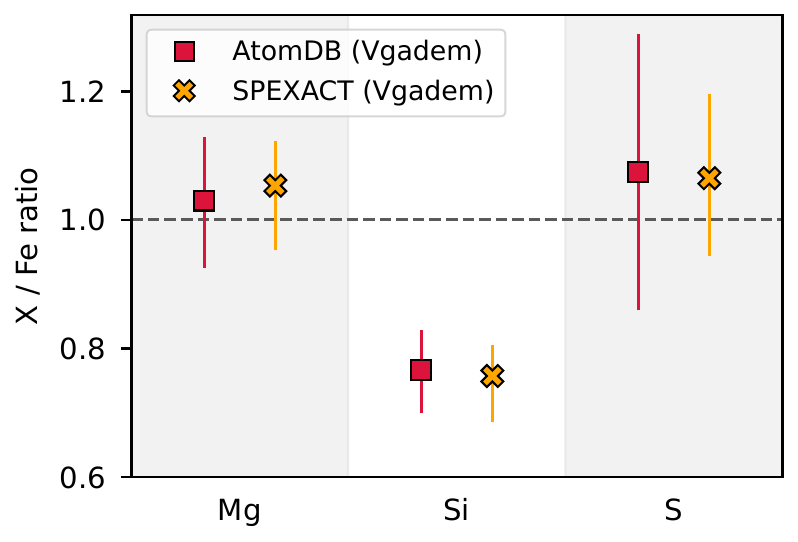}
    \caption{\textit{XMM-Newton}/EPIC results obtained by \texttt{vgadem} model using AtomDB (red squares) and SPEXACT (yellow crosses).}
    \label{fig:spec}
\end{figure}

Additionally, we compared the results obtained from different atomic databases, AtomDB \citep{Foster2012} and SPEXACT \citep{Kaastra_2024}. The \texttt{vgadem} model allows \texttt{XSPEC} users to change the used atomic database while performing the fitting. After performing our test, we observe that the results are in good agreement, presented in Fig. \ref{fig:spec}.

\bsp	
\label{lastpage}
\end{document}